\documentclass[structabstract]{aa}  
\usepackage{graphicx} 
\usepackage{txfonts}
\usepackage{natbib}
\usepackage{breqn}
\usepackage{lscape}
\usepackage{hyperref}

\newcommand{\HeI}[1]{\mbox{He\,{\sc i}~$\lambda${#1}}}
\newcommand{\HeII}[1]{\mbox{He\,{\sc ii}~$\lambda${#1}}}

\newcommand{\CIII}[1]{\mbox{C\,{\sc iii}~$\lambda${#1}}}
\newcommand{\CIV}[1]{\mbox{C\,{\sc iv}~$\lambda${#1}}}

\newcommand{\NIII}[1]{\mbox{N\,{\sc iii}~$\lambda${#1}}}

\newcommand{\SiIII}[1]{\mbox{Si\,{\sc iii}~$\lambda${#1}}}

\begin{document}

   \title{MONOS: Multiplicity Of Northern O-type Spectroscopic systems.}
   \subtitle{I. Project description and spectral classifications and \linebreak
             visual multiplicity of previously known objects.}
   \titlerunning{MONOS I. Project description, spectral classifications, and visual multiplicity.}

   \author{J. Ma\'{\i}z Apell\'aniz \inst{1}
           \and E. Trigueros P\'aez \inst{1, 2}
           \and I. Negueruela \inst{2}
           \and R. H. Barb\'a \inst{3} 
           \and S. Sim\'on-D\'{\i}az \inst{4,5}
           \and J. Lorenzo \inst{2}
           \and A. Sota \inst{6} 
           \and 
     \linebreak R. C. Gamen \inst{7}
           \and C. Fari\~na \inst{4,8}
           \and J. Salas \inst{9}
           \and J. A. Caballero \inst{1}
           \and N. I. Morrell \inst{10}
           \and A. Pellerin \inst{11}
           \and E. J. Alfaro \inst{6} 
           \and 
     \linebreak A. Herrero \inst{4,5}
           \and J. I. Arias \inst{3}
           \and A. Marco \inst{12}
           }

   \institute{Centro de Astrobiolog\'{\i}a, CSIC-INTA. Campus ESAC. 
              Camino bajo del castillo s/n. 
              E-28\,692 Vill. de la Ca\~nada, Madrid, Spain\linebreak
              \email{jmaiz@cab.inta-csic.es}
         \and
              Departamento de F\'{\i}sica Aplicada. Universidad de Alicante. 
              Ctra. S. Vicente del Raspeig. 
              E-03\,690 S. Vicente del Raspeig, Spain
         \and
			          Departamento de F\'{\i}sica y Astronom\'{\i}a. Universidad de La Serena. 
             Av. cisternas 1200 norte. 
             La Serena, Chile.
         \and
             Instituto de Astrof\'{\i}sica de Canarias. 
             E-38\,200 La Laguna, Tenerife, Spain
         \and
             Departamento de Astrof\'{\i}sica. Universidad de La Laguna.
             E-38\,205 La Laguna, Tenerife, Spain
         \and
         			 Instituto de Astrof{\'\i}sica de Andaluc{\'\i}a-CSIC. 
         			 Glorieta de la astronom\'{\i}a s/n. 
         			 E-18\,008 Granada, Spain
         \and
             Instituto de Astrof\'{\i}sica de La Plata (CONICET, UNLP). 
             Paseo del bosque s/n. 
             1900 La Plata, Argentina
         \and
             Isaac Newton Group of Telescopes.                           
             Apartado de correos 321.
             E-38\,700 Santa Cruz de La Palma, La Palma, Spain
         \and
             Agrupaci\'on Astron\'omica de Huesca.
             Parque Tecnol\'ogico Walqa, parcela 13.
             E-22\,197 Huesca, Spain
         \and
             Las Campanas Observatory. Carnegie Observatories. 
             Casilla 601. 
             La Serena, Chile
         \and
             Department of Physics and Astronomy. State University of New York at Geneseo.
             1 College Circle.
             Geneseo, NY 14\,454, U.S.A.
         \and
              Departamento de F\'{\i}sica, Ingenier\'{\i}a de Sistemas y Teor\'{\i}a de la Se\~nal. Universidad de Alicante. 
              Ctra. S. Vicente del Raspeig. 
              E-03\,690 S. Vicente del Raspeig, Spain
             }

   \date{Received 25 Feb 2019 / Accepted 25 April 2019}

 
  \abstract
   {Multiplicity in massive stars is a key element to understand the chemical and dynamical evolution of galaxies. Among massive stars, 
    those of O type play a crucial role due to their high masses and short lifetimes.}
   {MONOS (Multiplicity Of Northern O-type Spectroscopic systems) is a project designed to collect information and study O-type 
    spectroscopic binaries with $\delta > -20\degr$. In this first paper we describe the sample and provide spectral classifications and 
    additional information for objects with previous spectroscopic and/or eclipsing binary orbits. In future papers we will test the 
    validity of previous solutions and calculate new spectroscopic orbits.}
   {The spectra in this paper have two sources: the Galactic O-Star Spectroscopic Survey (GOSSS), a project that is obtaining blue-violet
    $R\sim 2500$ spectroscopy of thousands of massive stars, and LiLiMaRlin, a library of libraries of high-resolution spectroscopy of 
    massive stars obtained from four different surveys (CAF\'E-BEANS, OWN, IACOB, and NoMaDS) and additional data from our own observing 
    programs and public archives. We also use lucky images obtained with AstraLux.}
   {We present homogeneous spectral classifications for 92 O-type spectroscopic multiple systems and ten optical companions, many of them original.
    We discuss the visual multiplicity of each system with the support of AstraLux images and additional sources.
    For eleven O-type objects and for six B-type objects we present their first GOSSS spectral classifications. 
    For two known eclipsing binaries we detect double absorption lines (SB2) 
    or a single moving line (SB1) for the first time, to which we add a third system already reported by us recently. 
    For two previous 
    SB1 systems we detect their SB2 nature for the first time and give their first separate spectral classifications, something we also do for
    a third object just recently identified as a SB2. We also detect nine new astrometric companions and provide updated information on 
    several others. We emphasize the results for two stars: for $\sigma$~Ori~AaAbB we 
    provide spectral classifications for the three components with a single observation for the first time thanks to a lucky spectroscopy
    observation obtained close to the Aa,Ab periastron and for $\theta^1$~Ori~CaCb we add it to the class of Galactic Of?p stars, raising the number of
    its members to six. Our sample of O-type spectroscopic binaries contains more triple- or higher-order systems than double systems.}
   {}
   \maketitle
%

\section{Introduction}

$\,\!$\indent Massive stars are key components of galaxies and one of the most important factors that determine their chemical and 
dynamical evolution. However our knowledge of them is still quite incomplete. Among massive stars multiplicity (both visual and 
spectroscopic) is very high \citep{DuchKrau13,Sotaetal14}. This effect may be related to their formation mechanisms \citep{ZinnYork07} and is not 
fully studied since there are many hidden or poorly studied systems (\citealt{Masoetal98}, from now on M98, \citealt{Masoetal09}). 
As a large fraction of them are part of short-period
systems \citep{SanaEvan11}, an accurate knowledge of their binary properties is crucial to understand the role of massive stars as a 
population \citep{Langetal08}. This strong preference for close systems with short periods implies that almost a third of 
the systems will interact while both components are still on the main sequence and nearly 70\% of all massive stars will exchange mass 
with the companion \citep{Sanaetal12a}.


We have started an ambitious project that aims to bring homogeneity to the extensive but diverse literature data on Galactic O-type 
spectroscopic binaries. The logical division into two hemispheres produced by the samples available from different observatories prompted us 
to do two subprojects, one for the south and one for the north. In the southern hemisphere the OWN subproject \citep{Barbetal10,Barbetal17} 
is obtaining spectroscopic orbits for a large number of Galactic O-type (plus WN) systems to do a systematic analysis of their 
multiplicity for periods in the range from $\sim$1~day to a few years. The northern equivalent, MONOS (Multiplicity Of Northern O-type 
Spectroscopic systems) is presented in this paper, with the division between the two established at $\delta = -20\degr$ (e.g. we include 
the Orion and the M16 stars in the northern sample) to leave similar numbers of Galactic O stars in each subproject. The main data basis for 
MONOS is LiLiMaRlin \citep{Maizetal19a}, a 
{\bf Li}brary of {\bf Li}braries of {\bf Ma}ssive-Star High-{\bf R}eso{\bf l}ut{\bf i}o{\bf n} Spectra built by collecting data from four 
different surveys (CAF\'E-BEANS, \citealt{Neguetal15a}; IACOB, \citealt{SimDetal15b}; NoMaDS, \citealt{Maizetal12}; and OWN itself) plus 
additional spectra from other programs led by us and from public archives. Currently LiLiMaRlin has 18\,077 epochs for 1665 stars, of 
which 549 are O stars. 

In this first MONOS paper we select as our sample spectroscopic and/or eclipsing O+OBcc binaries (by OBcc we mean a star of spectral type 
O or B or a compact object) with previously published orbits and $\delta > -20\degr$. We specifically exclude systems that have been 
tentatively identified as spectroscopic binaries but that have no published orbits, see below for the case of some OWN targets.  
We present spectral classifications following the Galactic O-Star Spectroscopic Survey (GOSSS, \citealt{Maizetal11}) 
methodology and spectrograms for the systems that had not appeared or had different spectral
classifications in previous GOSSS papers. For each system we discuss its visual multiplicity, in some cases with the help of AstraLux lucky 
images. We have also compiled literature spectral classifications for those SB2/SB3 (double-lined/triple-lined spectroscopic systems) targets 
that had been previously separated into kinematic components, and for some targets we give new spectral classifications based on LiLiMaRlin spectra. 
In paper II we will start presenting our compilation 
of literature orbits and compare our LiLiMaRlin radial velocity measurements with their predictions. In subsequent MONOS papers we will publish new 
spectroscopic orbits.

Our plans for OWN include three papers in the near future. Two of them will be similar to MONOS-I (spectral classification and multiplicity)
and MONOS-II (a compilation of published orbital solutions) but for targets with $\delta < -20\degr$. The third paper will present a large
number of new orbits, many of them for systems that currently have none, and will be referred to here as the OWN orbit paper. A few of the 
MONOS-I targets in the equatorial region have orbits from that future OWN paper.





\section{Methods}

\subsection{Building the sample}

$\,\!$\indent To collect the sample, we started with the Galactic O-Star Catalog (GOSC, \citealt{Maizetal04b,Maizetal12,Maizetal17c,Sotaetal08},
\url{http://gosc.cab.inta-csic.es}), which is a repository of data for massive stars compiled from different projects, most notably GOSSS.
GOSC has a public version that includes the previously published GOSSS 
spectral types and a private version that adds those targets that we have already classified as being of O type using GOSSS data but that we 
have not published yet. We have used the current private version of GOSC to select O-type systems with $\delta >-20\degr$ and we have found 520 
of them. For those targets we have thoroughly searched the literature for spectroscopic or eclipsing orbits where the companion is an O star, a B star, or 
an unseen object (that is, we exclude cases where the companion may be a non-OB supergiant or a Wolf-Rayet star such as WR~113, WR~127, WR~133, WR~139, 
WR~140, WR~151, or WR 153ab) and found 92 systems. Of those, five have 
only eclipsing orbits published but the references have no information on the radial velocity amplitudes (so there are no published spectroscopic orbits).
One of the sources we have used is SB9, the ninth catalog of spectroscopic binary orbits \citep{Pouretal04}, but we note that there is a significant number of
orbits missing there. We do not include in our sample systems where only an indication of variability in radial velocity (as opposed to a full published orbit) is available, 
as other mechanisms such as pulsations can masquerade as spectroscopic binaries, especially when only a few epochs are available and no clear period 
is observed \citep{SimDetal17}. Indeed, some of the objects presented here have published orbits but we have strong suspicions that they are not really 
spectroscopic binaries but are instead misidentifications in the literature. We have left those cases in the sample for completeness but in MONOS-II we 
will present a table with the published orbits, compare their predictions with our data, and discuss which targets should be removed from the list.

\subsection{GOSSS data}

$\,\!$\indent GOSSS is obtaining $R\sim$2500 blue-violet spectroscopy with a high signal-to-noise ratio (S/N) of all 
optically accessible Galactic O stars.  To this date, three survey papers (\citealt{Sotaetal11a,Sotaetal14,Maizetal16}, from now on, 
GOSSS I+II+III) have been published and along with other recent papers \citep{Maizetal18a,Maizetal18b} GOSSS has produced spectral types for 
a total of 594 O-type, 24 non-O early-type, and 11 late-type systems. The GOSSS spectra are being gathered with six facilities: 
the 1.5~m Telescope at the Observatorio de Sierra Nevada (OSN),
the 2.5~m du Pont Telescope at Las Campanas Observatory (LCO), 
the 3.5~m Telescope at the Observatorio de Calar Alto (CAHA), 
and the 2.0~m Liverpool Telescope (LT),
the 4.2~m William Herschel Telescope (WHT), 
and the 10.4~m Gran Telescopio Canarias (GTC) at the Observatorio del Roque de los Muchachos (ORM). 
Additionally, several hundreds of O stars and several thousands of B- and later-type 
stars have been observed, and their data will be published in the near future. The reduction of the GOSSS data is explained in the three 
survey papers above. The spectral classification is performed with MGB and the GOSSS spectral classification grid OB2500~v3.0 
\citep{Maizetal12,Maizetal15b}. MGB is used in this paper for the spectral classification of SB2 systems because the software allows for the 
generation of synthetic linear combinations of standards with different spectral classifications, radial velocity separations, magnitude differences, 
and rotation indices. The GOSSS spectra and spectral types can be accessed through the GOSC web site.

The GOSSS spectrograms and spectral types for the 92 systems (as well as for ten optical companions) described in the previous subsection are the 
main scientific content of this paper. The new spectrograms are shown in Fig.~\ref{GOSSS} and in Table~\ref{spclas} we give the names, GOSC IDs, 
coordinates, and spectral types for the objects in our sample sorted by GOSC ID (or, equivalently, by Galactic longitude). In the rest of 
this subsection we describe some general information about the GOSSS data in that table and in the next section we analyze it star by star.

We have been recently successful \citep{Maizetal18a} in applying a new technique, lucky spectroscopy, to spatially separate close visual components
in GOSSS data, which is used for five of the systems in this paper (MY~Cam~A+B, LY~Aur~A+B, $\delta$~Ori~Aa+Ab, $\zeta$~Ori~AaAb+B, and 
$\sigma$~Ori~AaAb+B). The first two of those systems did not appear in \citet{Maizetal18a} and for the last one we have repeated the observations and 
obtained new data. LY~Aur~A and MY~Cam~A are fainter targets than the other three and that led us to observe them with a different setup.
Instead of using EEV12, the standard CCD for the ISIS spectrograph at the WHT, we selected the alternative CCD, QUCAM3. QUCAM3 has the advantage
of having a much faster readout time than EEV12, allowing us to obtain a spectrum per second with little dead time between exposures, as opposed to
EEV12, for which the one-exposure cycle lasts for $\sim$15 seconds. On the other hand, QUCAM3 spans a significantly lower wavelength range than EEV12 at
the same spectral resolution, so we had to do three different central wavelengths per star to cover the whole GOSSS spectral range. Also, QUCAM3 does 
not allow to do 0.1~s exposures, as EEV12 does. Therefore, we used QUCAM3 for faint stars and EEV12 for bright ones.

Regarding the components included in the name of the star, we follow the general strategy for GOSSS, namely, we always try to spatially 
deconvolve visual components and give the spectral classification for the brightest visual star (or combination of if no deconvolving of 
all of them is possible). In those cases where the number of components in the GOSSS spectrum is different from the number in the high-resolution
aperture, we indicate it (see section on spectroscopic nomenclature below).
If more than one visual component is left in the final spectrum we only name those that can significantly alter the 
spectral classification, for which we establish a dividing criterion at a $\Delta B$ of 2 magnitudes. For example, the A and B components 
in HD~193\,443 are separated by 0\farcs1 and have a $\Delta B$ of 0.3 (too close even for lucky spectroscopy), so we cannot resolve them 
and both have a significant effect on the (combined) spectral type and we leave ``AB'' in the name. On the other hand, $\iota$~Ori also 
has two visual components (Aa~and~Ab) separated by 0\farcs1 but in that case $\Delta B$ is larger than 3 magnitudes, so the secondary 
does not contribute significantly to the spectral classification and its name is left out (note that for $\iota$~Ori there are two additional 
dim B and C 
components located further away). Whenever possible we follow the Washington Double Star Catalog (WDS, \citealt{Masoetal01}) component 
nomenclature and we use our AstraLux lucky images, both previously published \citep{Maiz10a} and newly presented here, to add 
information about visual companions. 

The GOSSS spectral classifications in Table~\ref{spclas} can be divided in four categories:

\begin{itemize}
 \item Spectral classifications of spectroscopic binaries previously published and not modified here. There are 58 of them and in each 
       case the reference is given. Note that in those cases we do not show the spectrograms in Fig.~\ref{GOSSS}, as they are available
       in the previous references.
 \item Objects that had previously appeared in GOSSS papers but for which we provide a new classification. That can be because we have 
       obtained a new epoch in which we have caught the two spectroscopic components with a large enough radial velocity separation to provide 
       separate classifications or because we have been able to reclassify an old epoch with new standards. They are marked as ``mod'' (for 
       modified) and there are 17 of them.
 \item Spectroscopic binaries that had no previous GOSSS spectral classification. They are marked as ``new'' and there are 17 of them.
 \item Visual companions to spectroscopic binaries that were observed by placing them on the GOSSS slit at the same time as the primary 
       target. They are marked as ``vis'' (for visual companion) and there are ten of them.   
\end{itemize}


\begin{figure*}
\centerline{\includegraphics*[width=\linewidth]{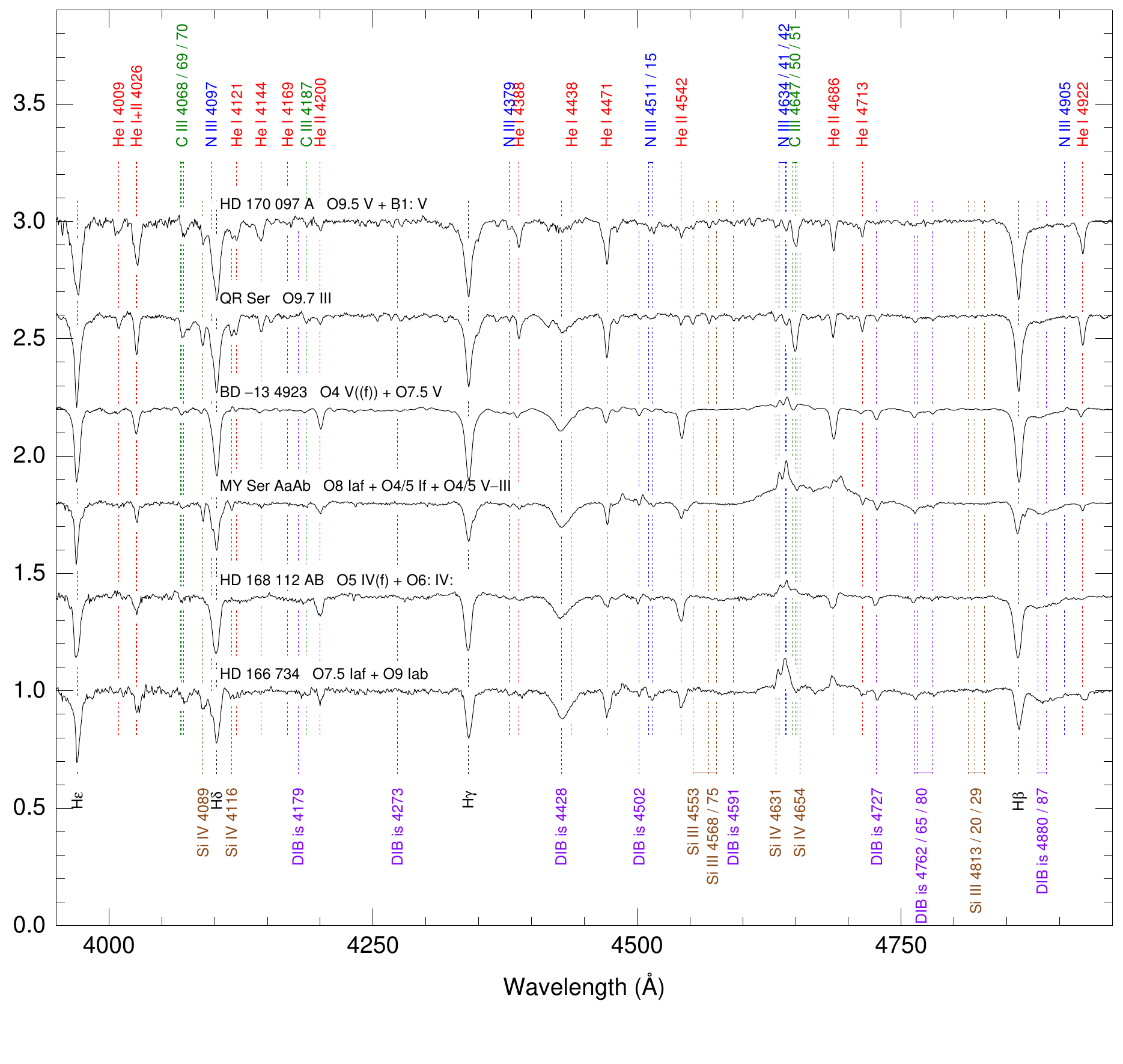}}
\caption{New GOSSS spectrograms. The targets are sorted by Galactic O-Star (GOS) ID.}
\label{GOSSS}
\end{figure*}	

\addtocounter{figure}{-1}

\begin{figure*}
\centerline{\includegraphics*[width=\linewidth]{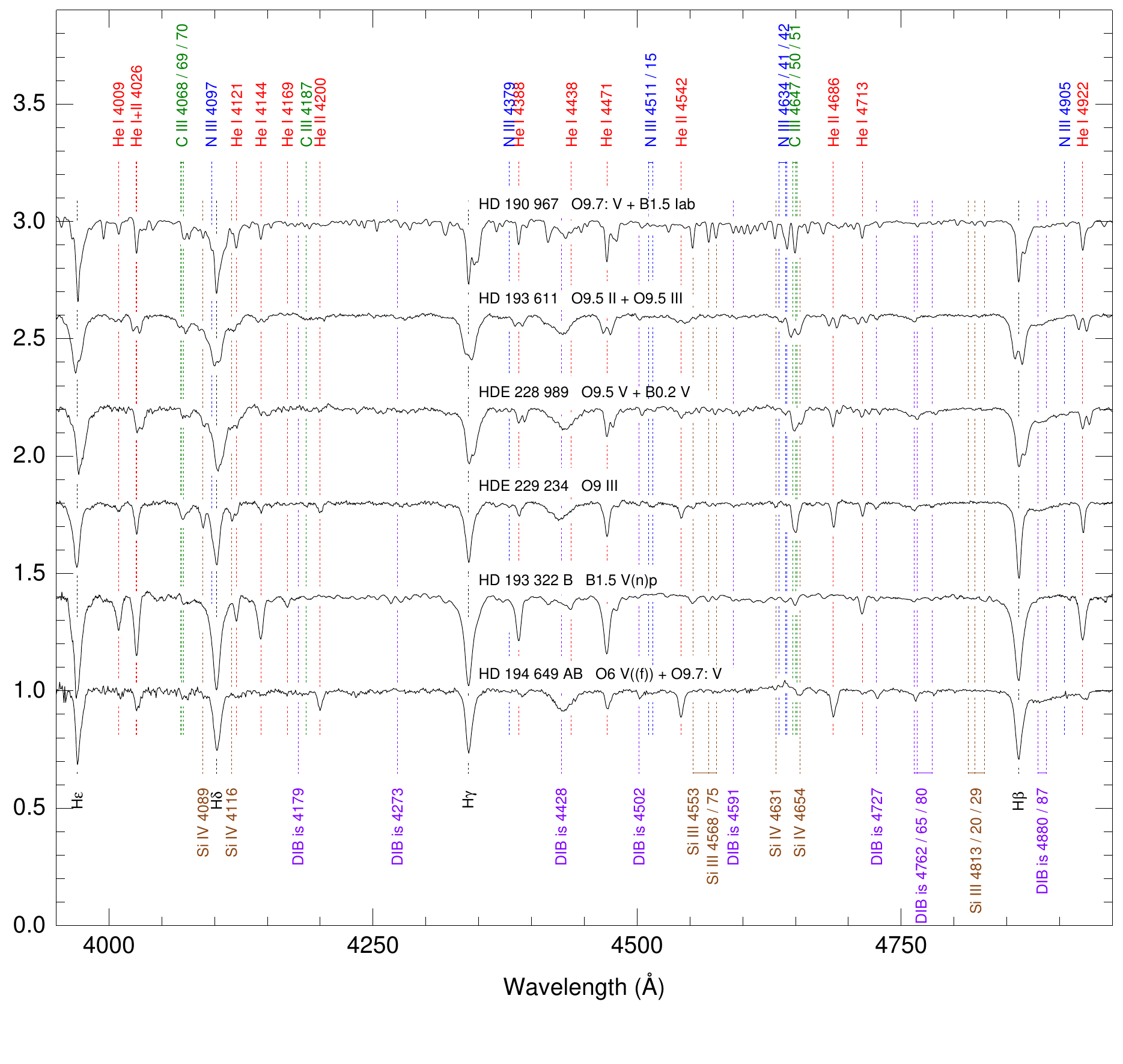}}
\caption{(continued).}
\end{figure*}	

\addtocounter{figure}{-1}

\begin{figure*}
\centerline{\includegraphics*[width=\linewidth]{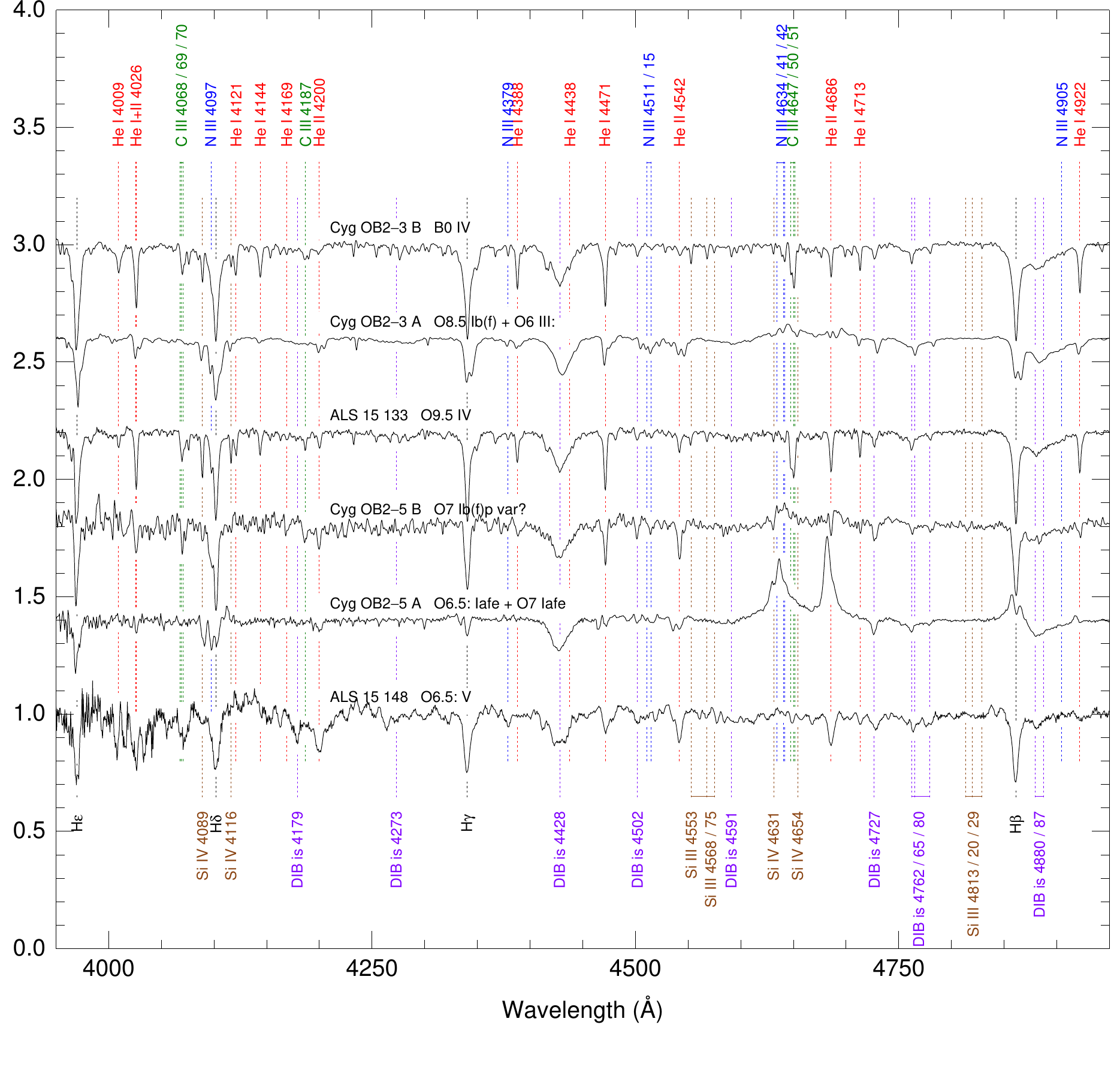}}
\caption{(continued).}
\end{figure*}	

\addtocounter{figure}{-1}

\begin{figure*}
\centerline{\includegraphics*[width=\linewidth]{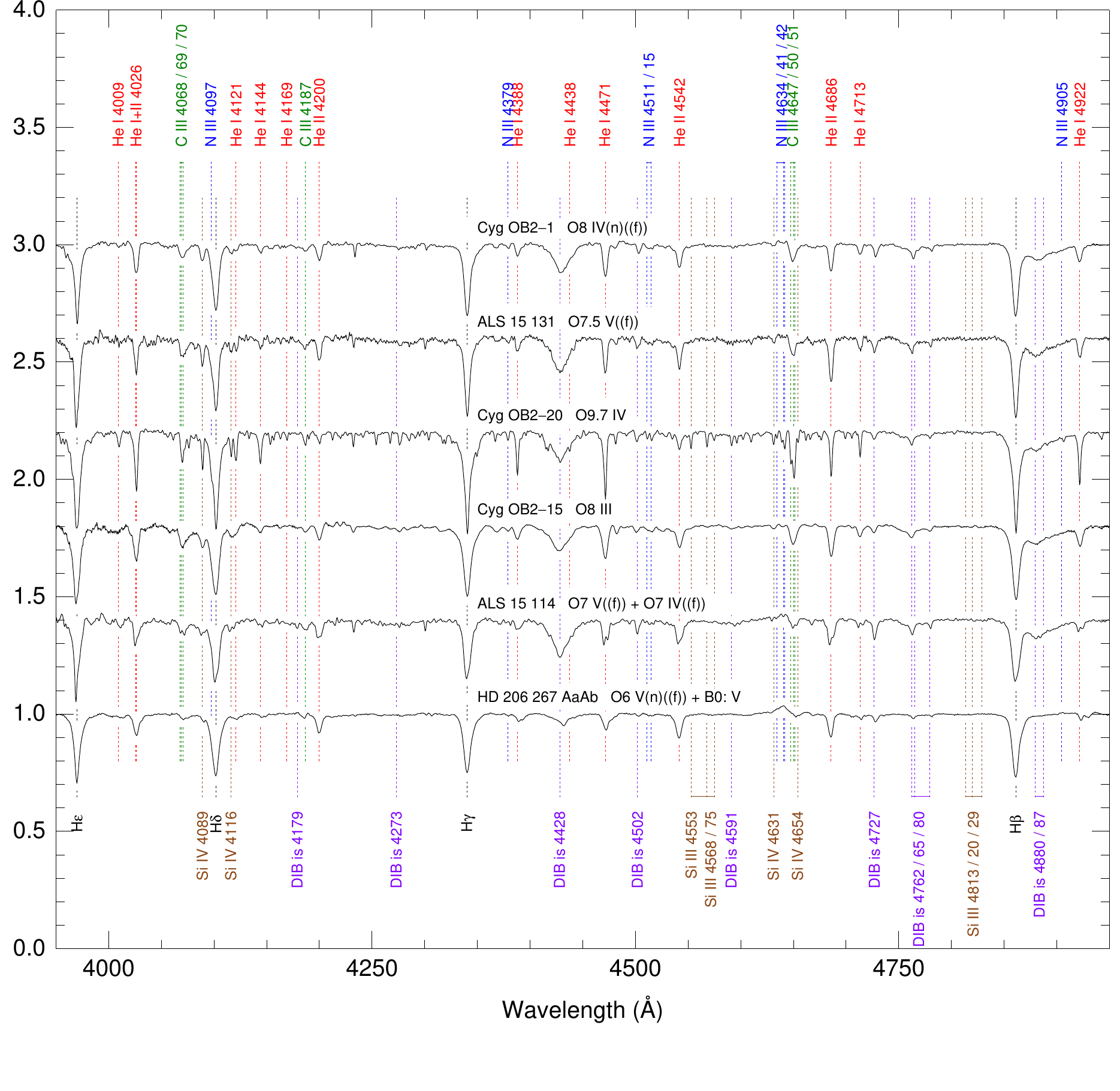}}
\caption{(continued).}
\end{figure*}	

\addtocounter{figure}{-1}

\begin{figure*}
\centerline{\includegraphics*[width=\linewidth]{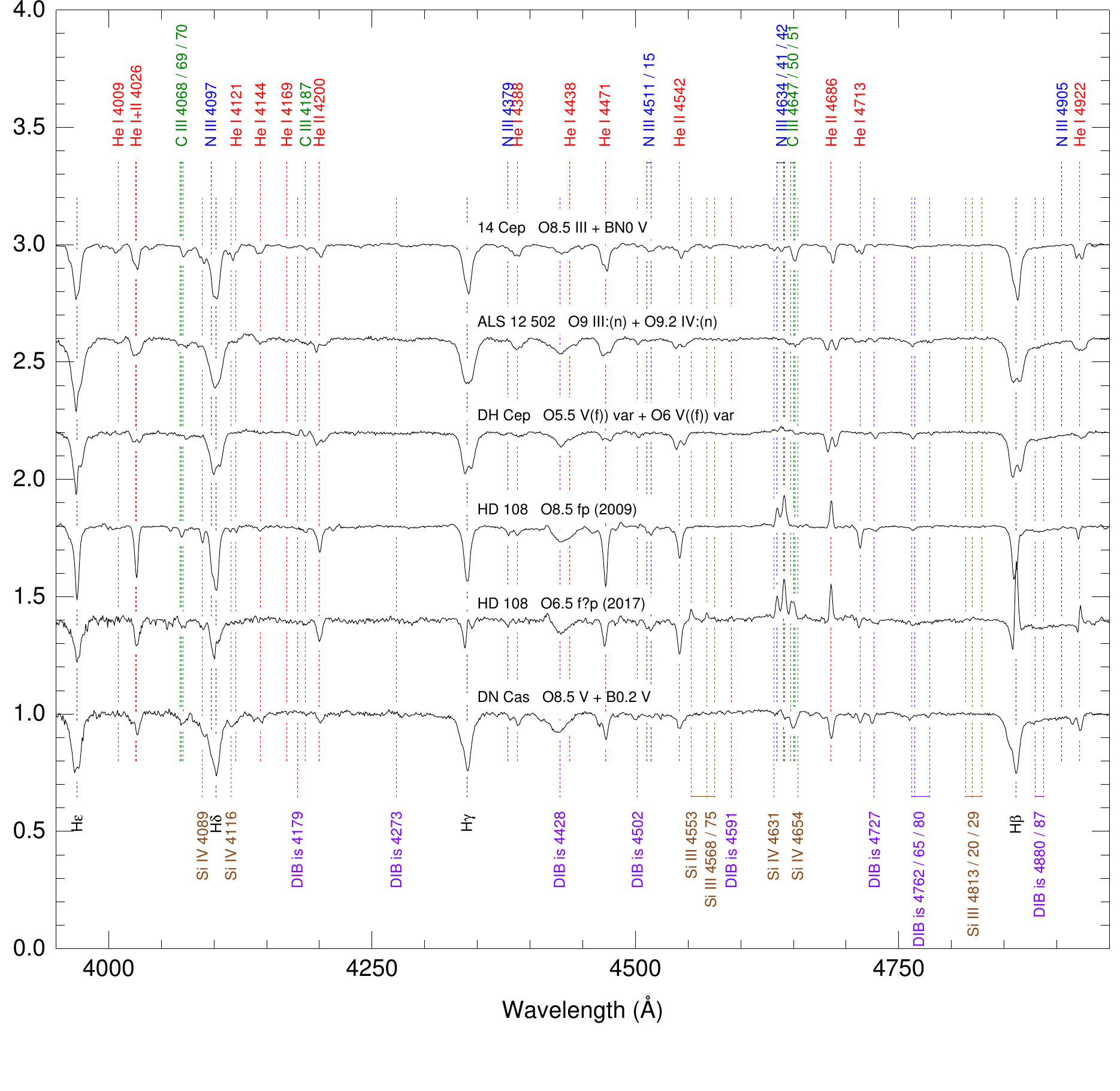}}
\caption{(continued).}
\end{figure*}	

\addtocounter{figure}{-1}

\begin{figure*}
\centerline{\includegraphics*[width=\linewidth]{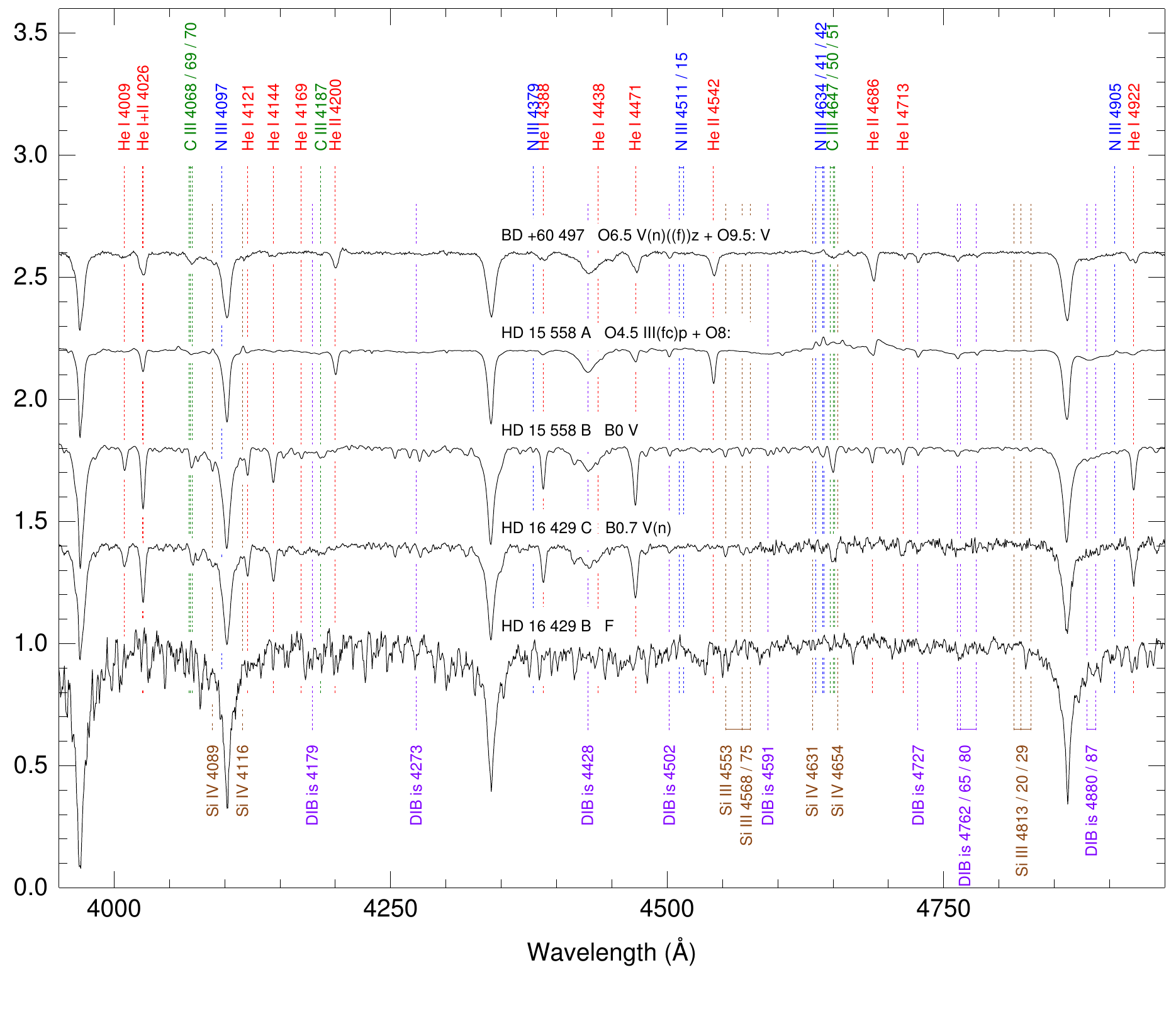}}
\caption{(continued).}
\end{figure*}	

\addtocounter{figure}{-1}

\begin{figure*}
\centerline{\includegraphics*[width=\linewidth]{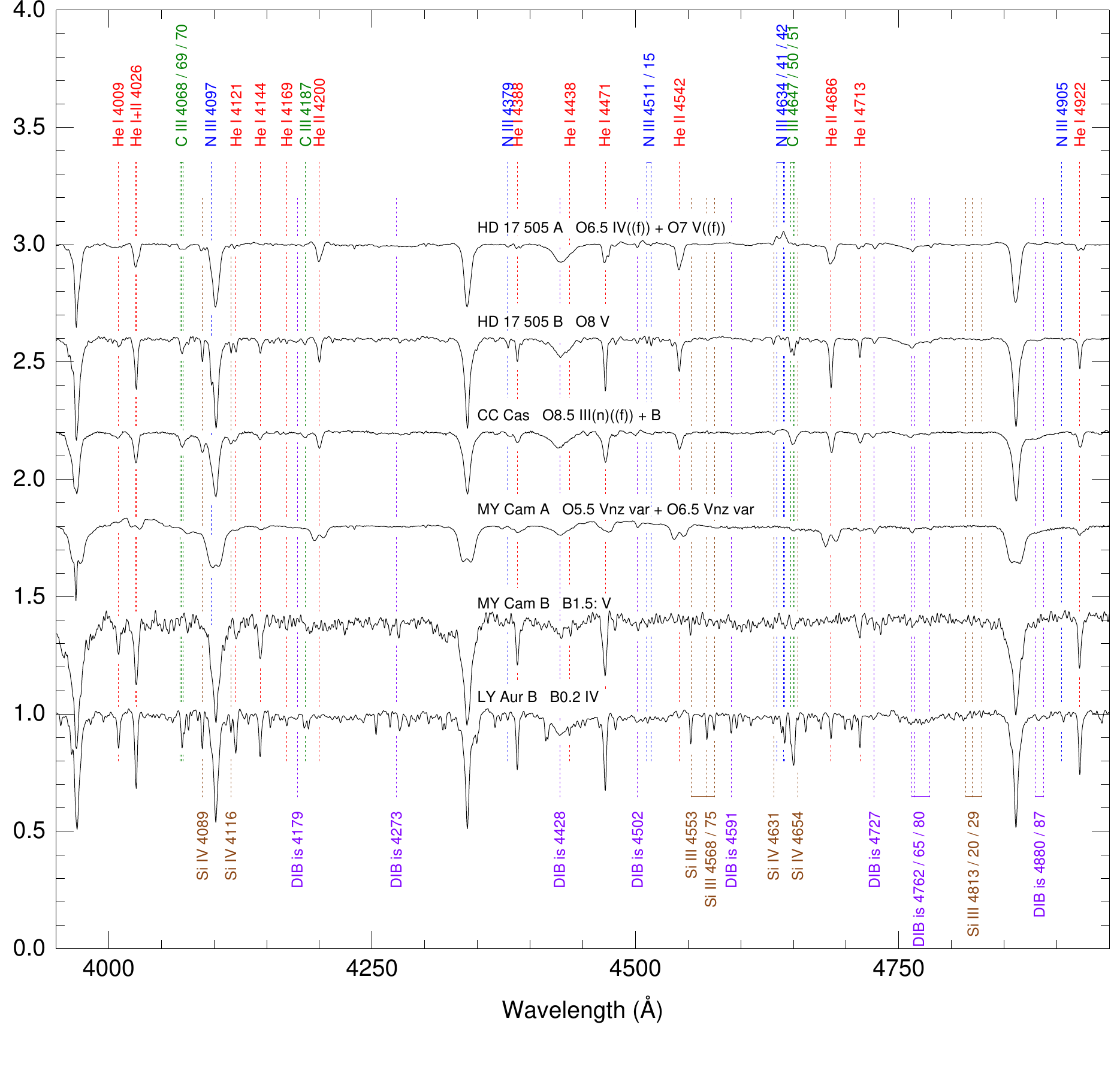}}
\caption{(continued).}
\end{figure*}	

\addtocounter{figure}{-1}

\begin{figure*}
\centerline{\includegraphics*[width=\linewidth]{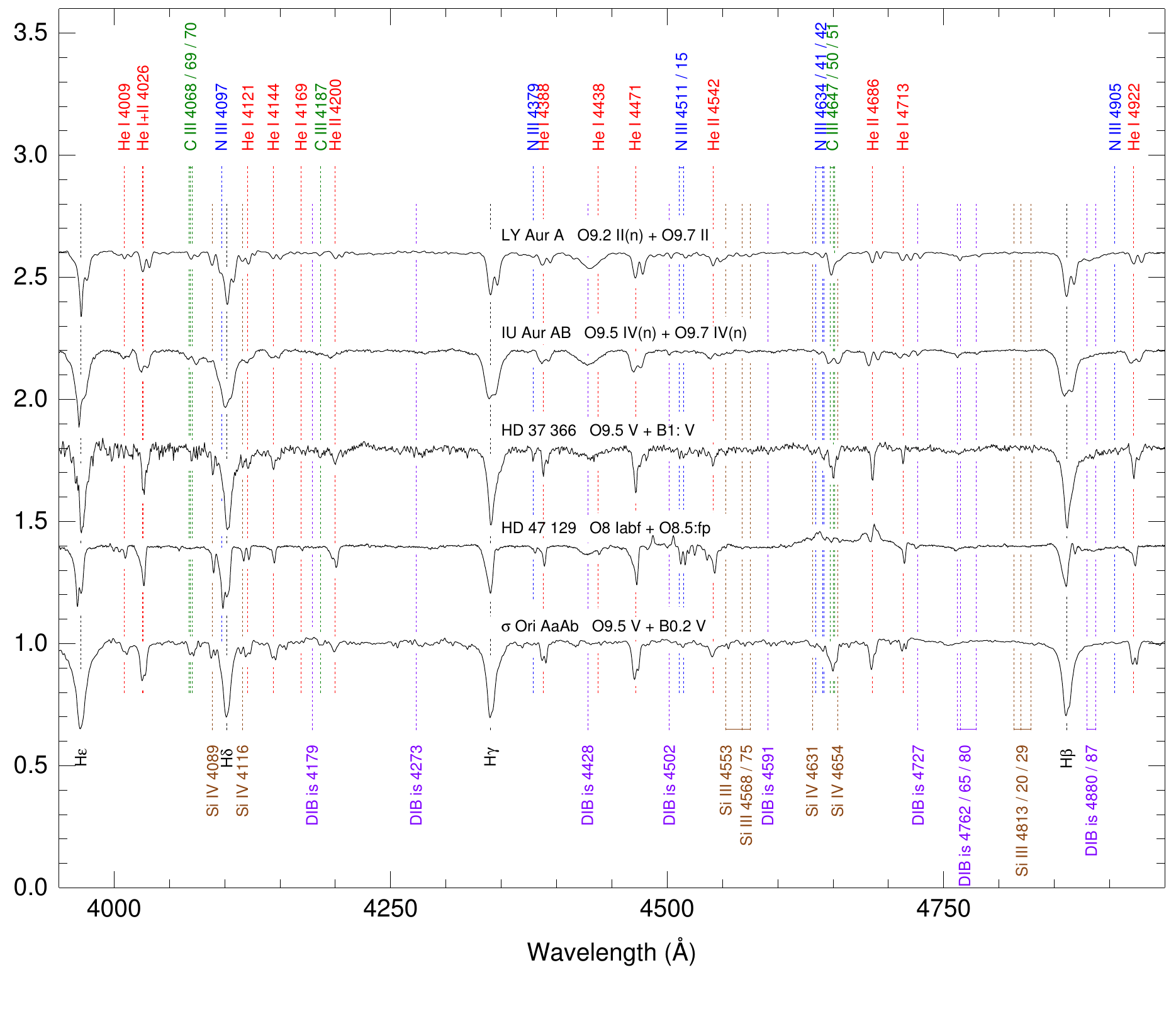}}
\caption{(continued).}
\end{figure*}	

\addtocounter{figure}{-1}

\begin{figure*}
\centerline{\includegraphics*[width=\linewidth]{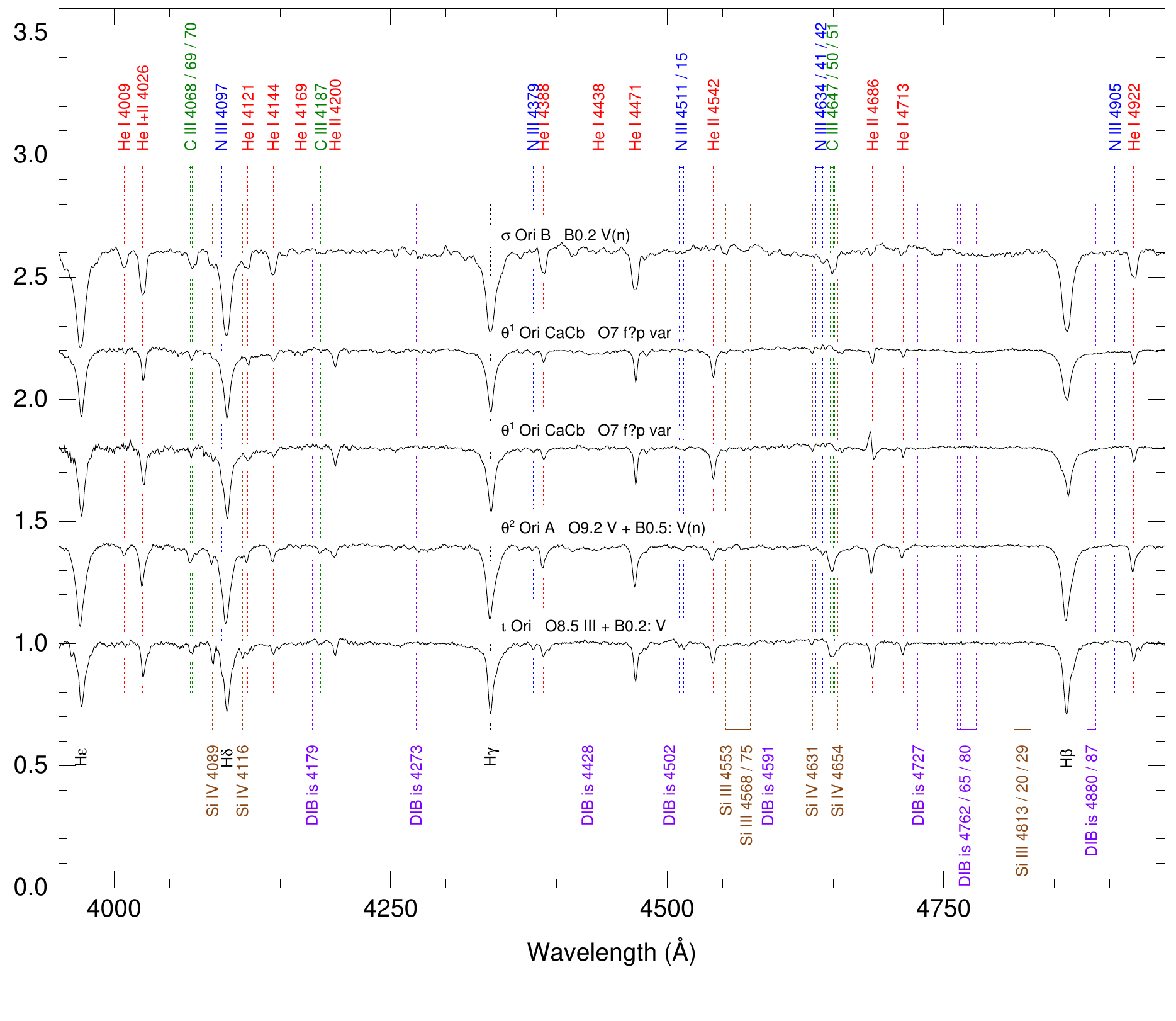}}
\caption{(continued).}
\end{figure*}	

\subsection{Alternate spectral classifications}

$\,\!$\indent In addition to the spectral types derived from GOSSS data, Table~\ref{spclas} gives spectral classifications from two other
types of sources. First, the literature provides separate spectral types for part of the sample, as some of the papers that publish
orbits also publish the spectral types of each component. Most of those types are generated using spectral disentangling techniques \citep{GonzLeva06},
which allow for the decomposition of the individual spectra of a multiple object by using the information in a large number of high-S/N 
high-spectral-resolution spectrograms. Spectral disentangling allows for the separation of multiple components whose presence is not 
immediately apparent in single exposures. When done properly, it is a highly efficient technique but one has to be careful with data selection and 
parameter tuning to avoid the generation of artificial structures in the output.

The second source of spectral types is the LiLiMaRlin data \citep{Maizetal19a} that will be used for future MONOS papers, which consists of high-resolution spectra 
obtained with seven different telescopes: HET~9.2~m, NOT~2.56~m, CAHA~2.2~m, MPG-ESO~2.2~m, OHP~1.93~m, Mercator~1.2~m, and Stella~1.2~m.
Here we use the high-resolution spectra for some 
targets for which we do not have GOSSS epochs at large radial velocity separations. In future papers we will use disentangling techniques but for now we 
simply select the epoch with the best radial velocity separation between components, apply a smoothing algorithm to degrade the resolution to 2500, and use MGB with the 
GOSSS standards to derive the spectral classifications. The LiLiMarlin spectra processed in that way used to derive new spectral classifications are shown in 
Fig.~\ref{LiLiMaRlin}.

The alternate spectral types are also listed in Table~\ref{spclas}, indicating either the reference (for literature values) or ``new'' (for LiLiMarlin data).
Most of the literature and all of the LiLiMaRlin spectra were obtained with echelle spectrographs using circular apertures with diameters of 1\arcsec-3\arcsec. 
Therefore, their spatial resolution is significantly worse than that of GOSSS spectra and may include different components. In the next two subsections
we explain how we deal with this.

We repeat here a comment about spectral classifications we have done in the past but that is important to keep in mind. Why do we present different spectral
classifications and keep updating them? There are three reasons.

\begin{itemize}
 \item Spectral classification is a process by which the spectrogram of a star is compared to those of a grid of standards. Since we published our first OSTAR grid
       in GOSSS~I we have updated it regularly with data of better quality, filled some gaps in the grid, and defined new subtypes. For example, some stars in the original grid 
       have turned out to be SB2 and we have replaced them with others. Therefore, some of our own spectral types need to be revised over time and literature types need
       to be analyzed to see how they were obtained.
 \item We are obtaining new data not only for the standards but also for stars that were previously observed. This can be because the spectra had poor S/N, the 
       spatial resolution could be improved to spatially separate visual components, or the phase at which a SB2 system had been observed was not optimal for the
       kinematic separation of components.
 \item Finally, the spectral types of the stars themselves can change over time. The observed cases with secular evolution are rare (but they exist, see 
       \citealt{Walbetal17}) but more frequent are those that vary periodically or quasi-periodically, such as magnetic stars \citep{Walbetal10a}, Oe/Be stars
       \citep{PortRivi03}, and spectroscopic binaries.  For that reason, it is important to differentiate between ``star X has spectral type Y'' (which implies that it 
       is constant or nearly so) and ``star X shows a spectral type Y at moment Z'' (which does not). For example, Oe/Be stars can lose their disks (to regain them 
       later) so one can say that an Oe/Be star has a non-Oe/Be spectral type at a given moment.
\end{itemize}

It is for the reasons above that we present new and alternate spectral classifications in this paper. SB2 systems are especially subject to these issues, mostly due to the
need of a large radial velocity difference to easily separate components. For close systems, the apparent spectral types of each star throughout the orbit can change as
[a] high rotational velocities induce temperature differences between equator and poles, [b] ellipsoidal deformations vary the effective stellar areas as a function 
of orbital phase, [c] the regions of each star directly exposed to the radiation of the companion are overheated, and [d] different parts of each component are visible or 
blocked by eclipses (in extreme cases this also applies to the emission lines generated in interacting regions close to the stars). 

We also warn about the confusion that exists in some papers about real measurements and estimates of spectral types. It is possible to read in the literature expressions
like ``no signature of the companion is seen in the spectrum and the non-detection of the X line implies that the companion must be of spectral type Y to Z'' or 
``given the $\Delta m$ between the visual A and B components, we deduce that B has to have a spectral type X''. Those comments can be (more or less) reasonable 
estimates of the spectral type but they are not real measurements, as they do not follow the criteria for spectral classifications. It is important to label them as such 
in order not to propagate errors in papers that cite the original works.

\begin{figure*}
\centerline{\includegraphics*[width=\linewidth]{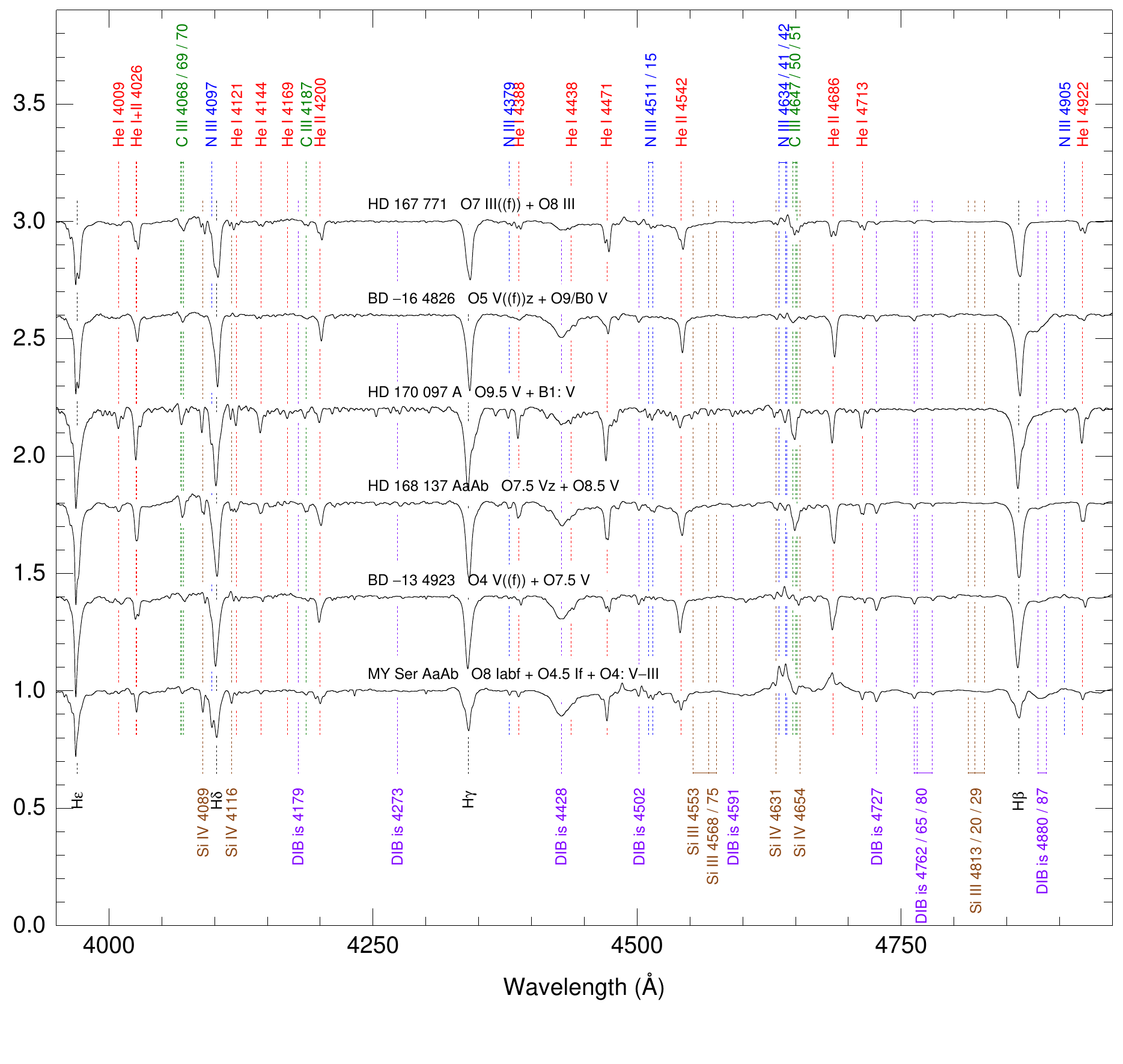}}
\caption{New LiLiMaRlin spectrograms degraded to $R$ = 2500. The targets are sorted by GOS ID. The Cyg~OB2-9 spectrogram does not include the 
         4710-4760~\AA\ region}
\label{LiLiMaRlin}
\end{figure*}	

\addtocounter{figure}{-1}

\begin{figure*}
\centerline{\includegraphics*[width=\linewidth]{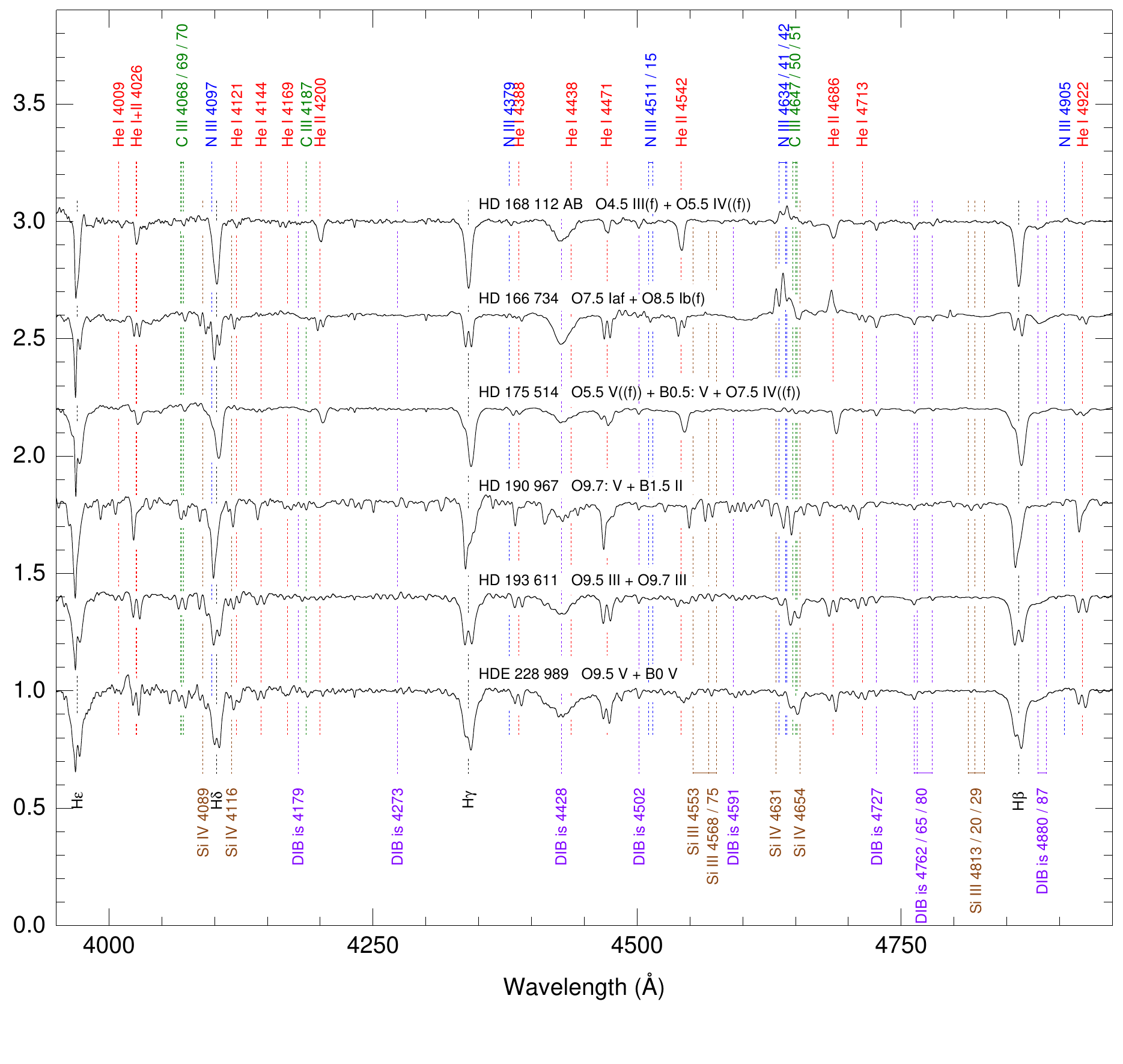}}
\caption{(continued).}
\end{figure*}	

\addtocounter{figure}{-1}

\begin{figure*}
\centerline{\includegraphics*[width=\linewidth]{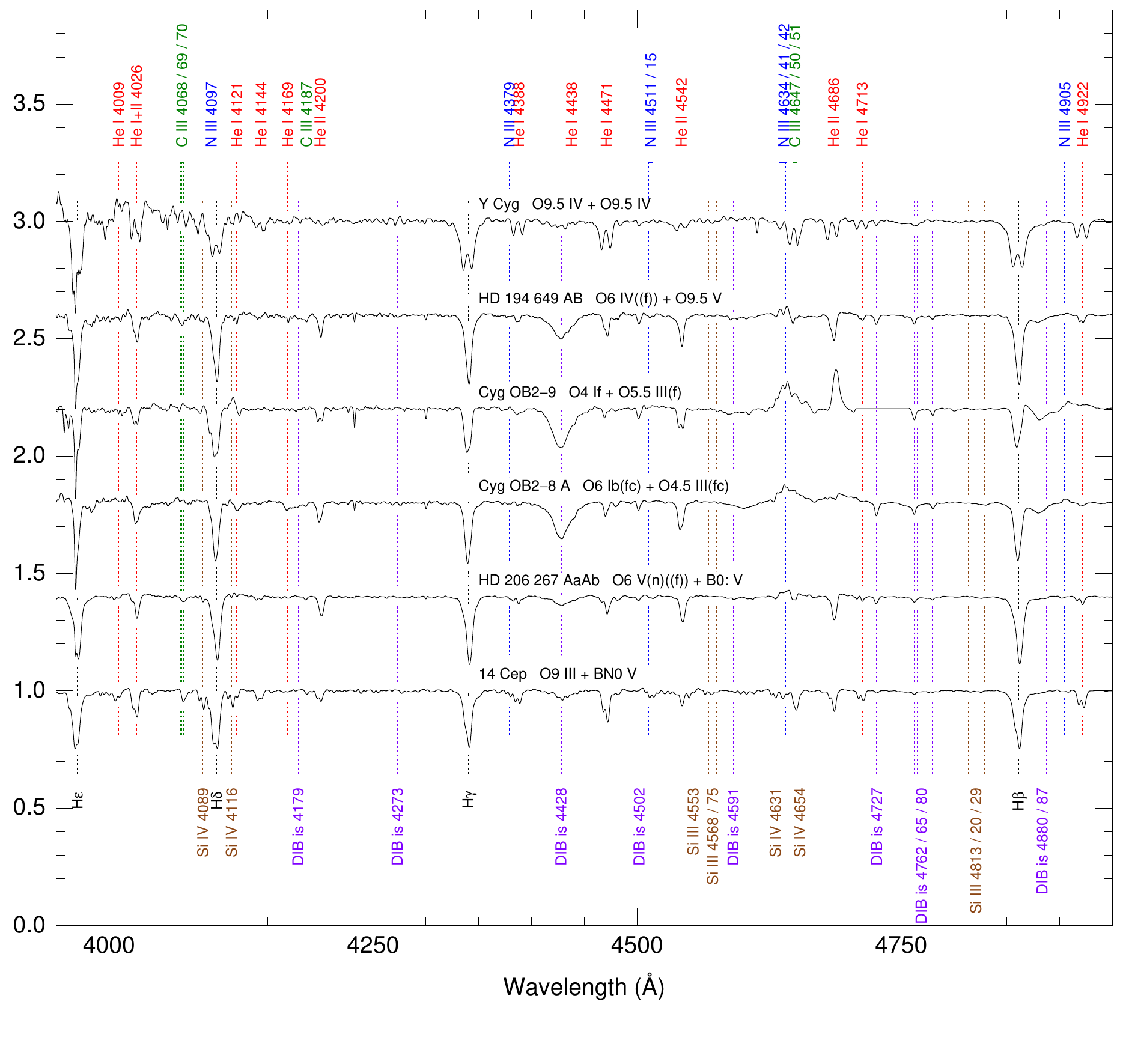}}
\caption{(continued).}
\end{figure*}	

\addtocounter{figure}{-1}

\begin{figure*}
\centerline{\includegraphics*[width=\linewidth]{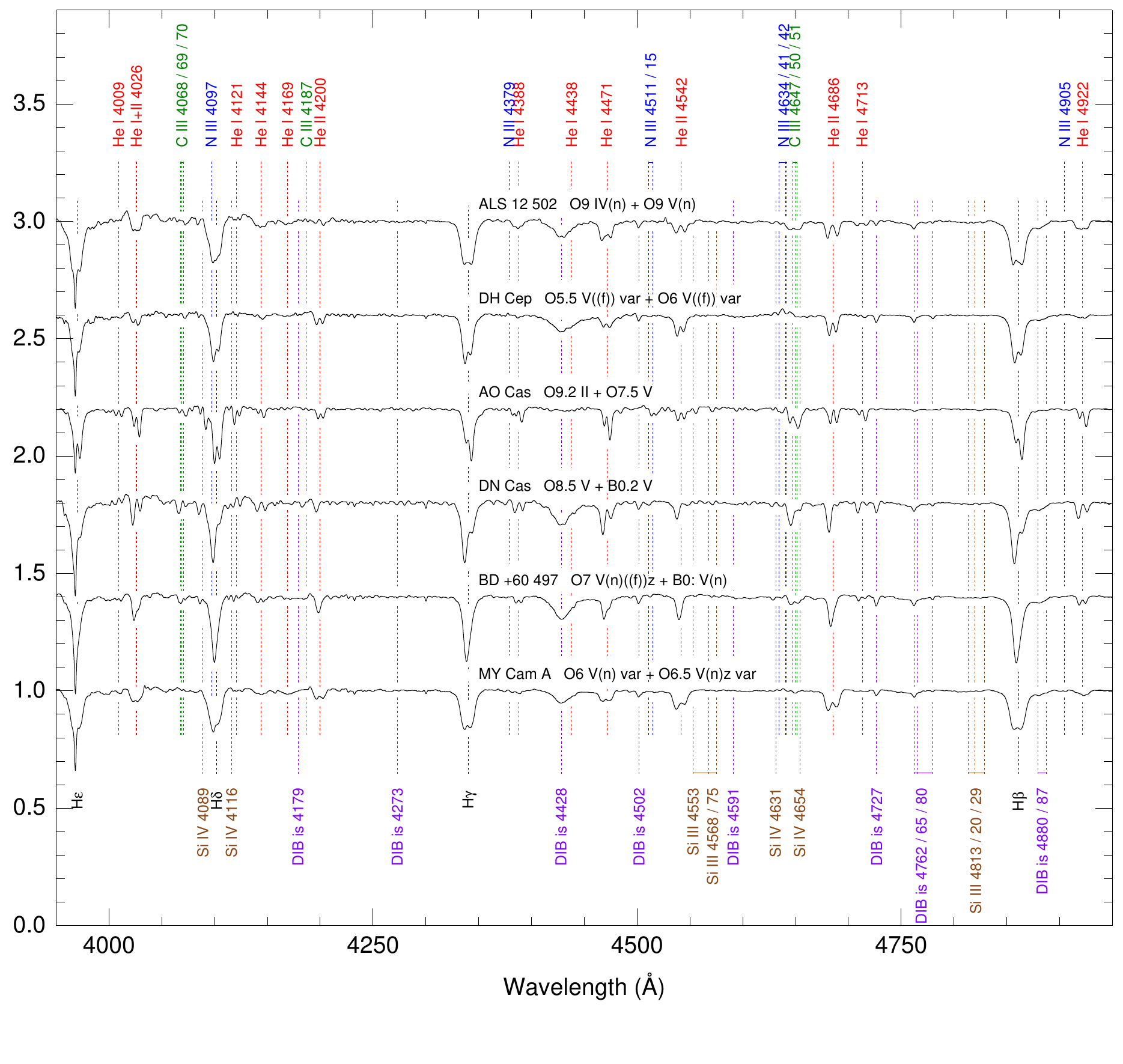}}
\caption{(continued).}
\end{figure*}	

\addtocounter{figure}{-1}

\begin{figure*}
\centerline{\includegraphics*[width=\linewidth]{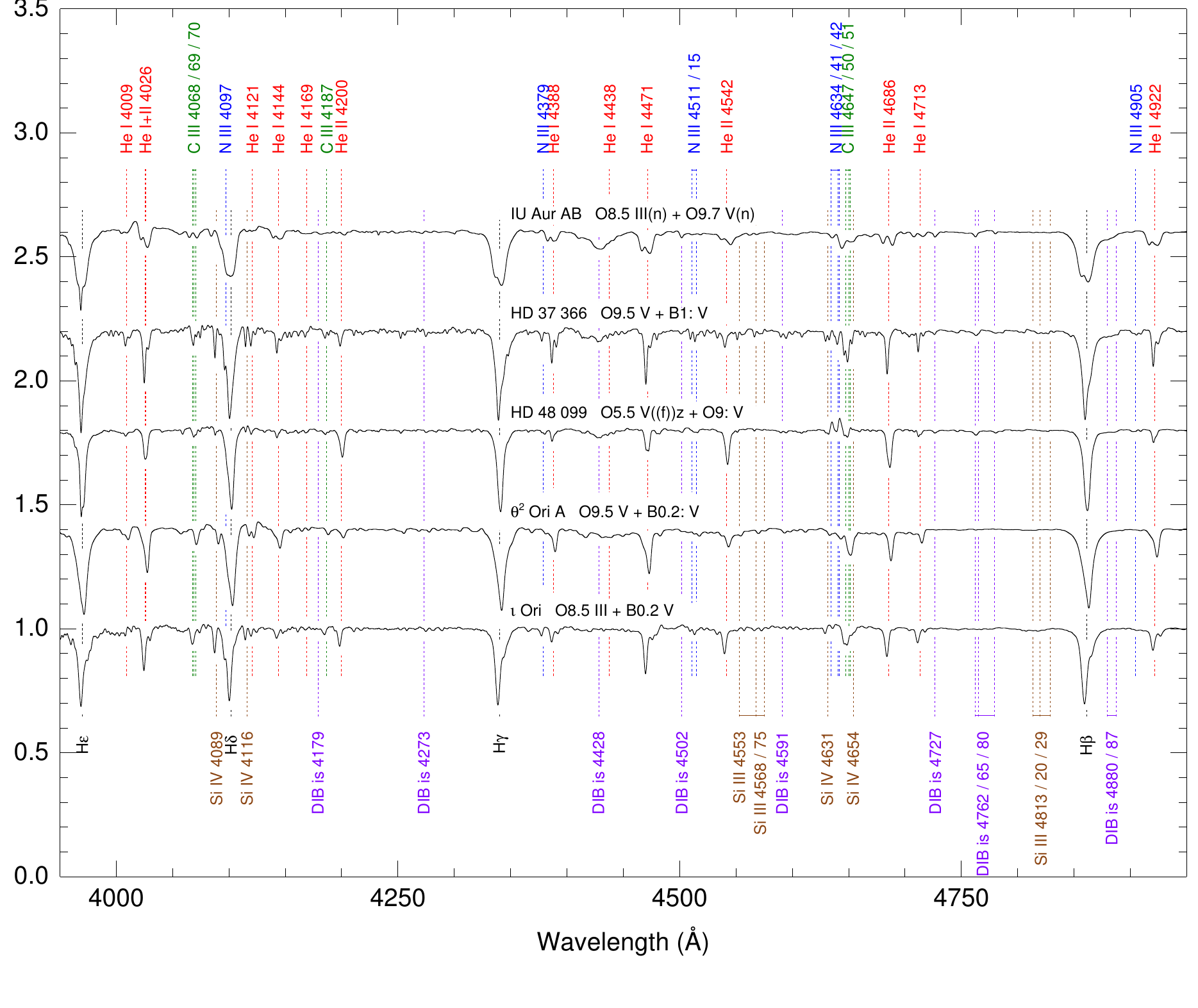}}
\caption{(continued).}
\end{figure*}	

\subsection{Visual multiplicity and AstraLux images}

$\,\!$\indent Even though the main goal of MONOS is to study spectroscopic multiplicity, that is, stellar components detected through radial velocity
differences, we find it necessary to also discuss the visual multiplicity of our sample, that is, stellar components detected through imaging or
interferometric means. There are several reasons for this:

\begin{itemize}
 \item The most important one is that to obtain a complete picture of the dynamics of massive star systems and their formation it is necessary to explore the
       different scales involved, from contact or near-contact systems separated by distances measured in stellar radii to weakly bound systems separated by 
       distances of a fraction of a pc.
 \item It has become clear in recent years that many multiple system are triples or of a higher order \citep{Sotaetal14}. These are typically composed of an inner 
       (spectroscopic) pair orbited at a larger distance by a visual companion. Therefore, many multiple systems have both visual and spectroscopic components.
 \item The line between visual and spectroscopic components has become blurred in the last decade, with an increasing number of cases where a pair is detected 
       through both means \citep{SimDetal15a,Aldoetal15,LeBoetal17,Maizetal17a}. This provides an excellent opportunity to derive the stellar parameters of each 
       component with even better precision.
 \item As a result of the two previous reasons, it is possible to have confusing circumstances where it is not clear whether a newly discovered visual corresponds to
       a previously known spectroscopic one (or viceversa). Therefore, it is a good idea to have all the information accessible in one place, following the pioneering
       work of M98 in this respect.
\end{itemize}

\begin{figure*}
\centerline{\includegraphics*[width=0.240\linewidth]{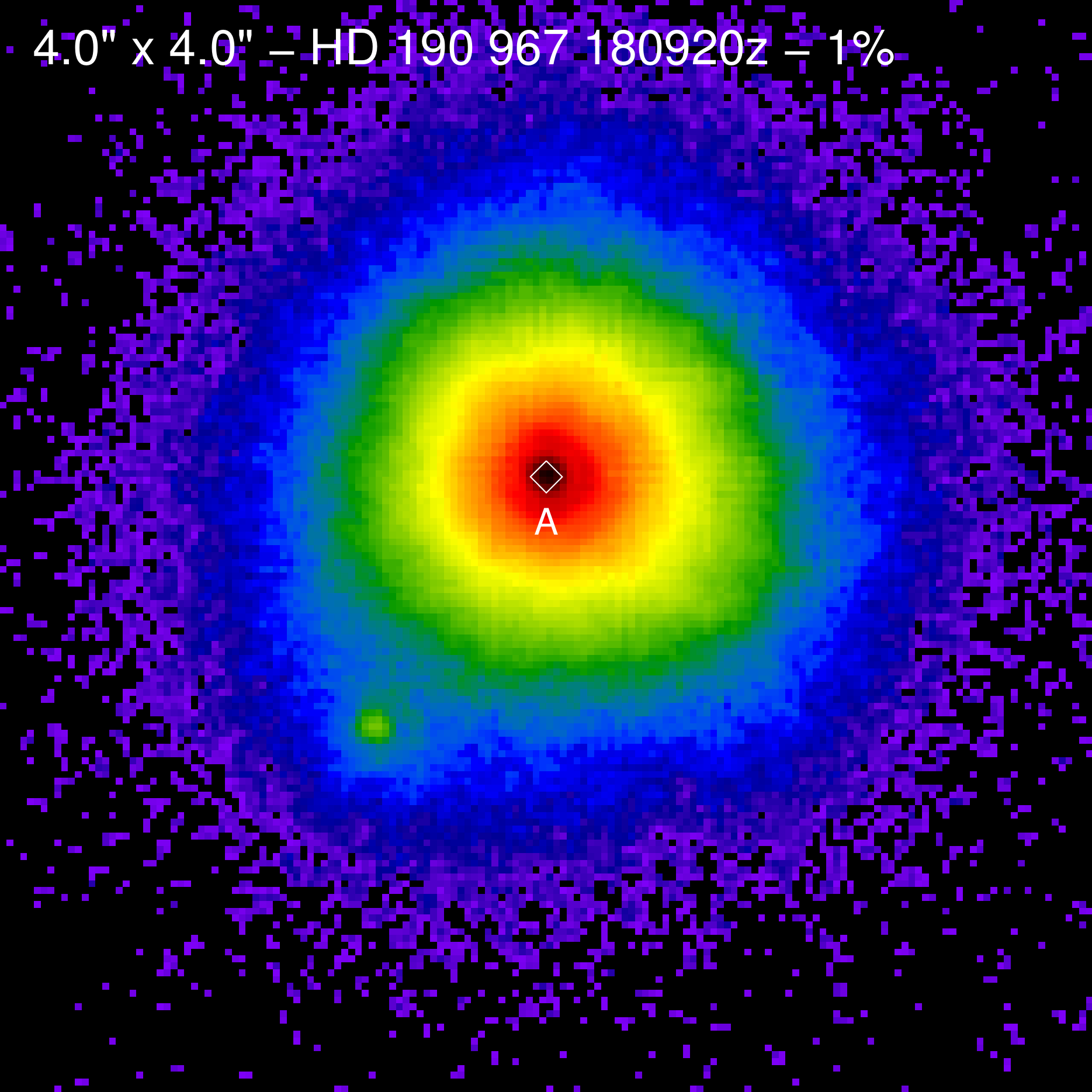} \
            \includegraphics*[width=0.240\linewidth]{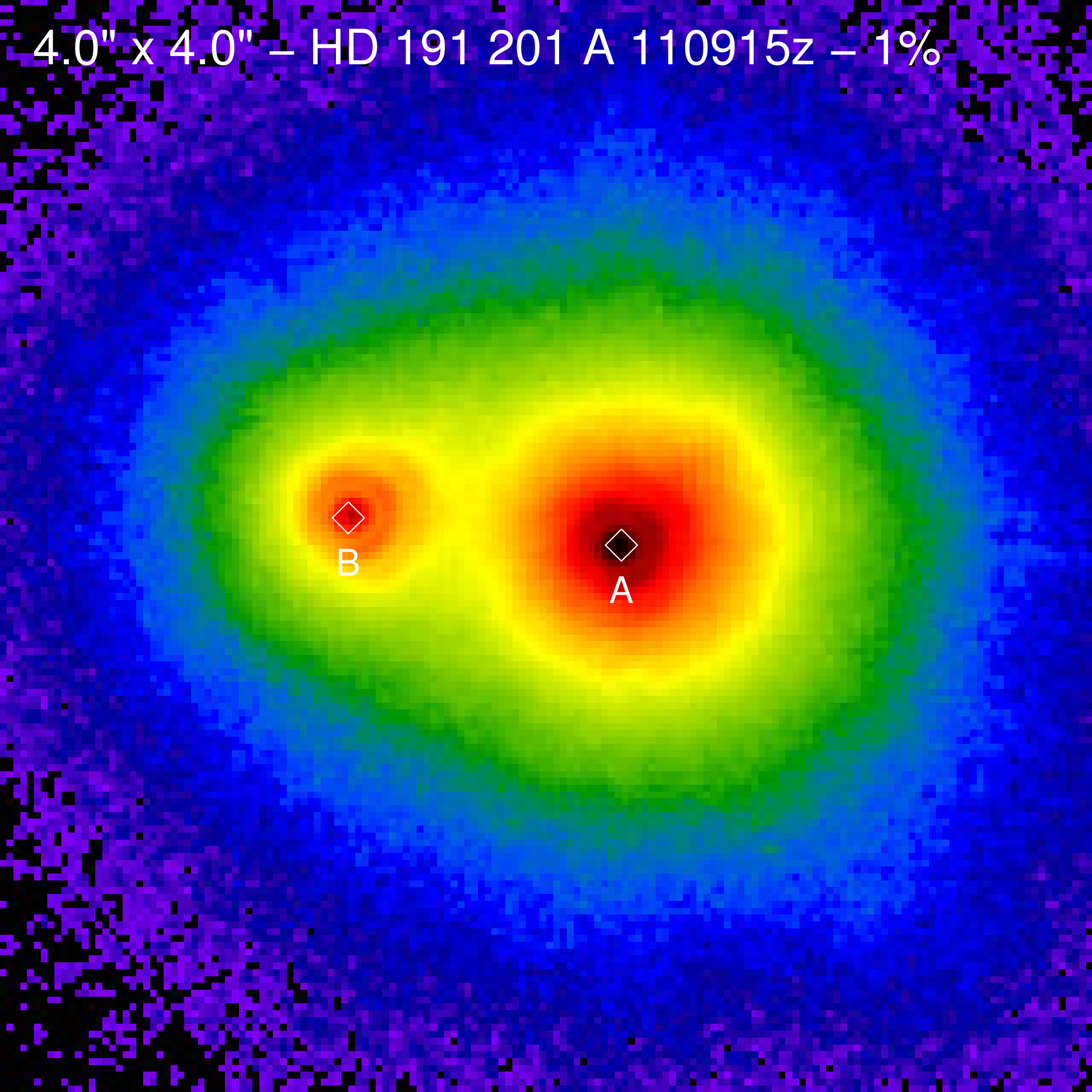} \
            \includegraphics*[width=0.240\linewidth]{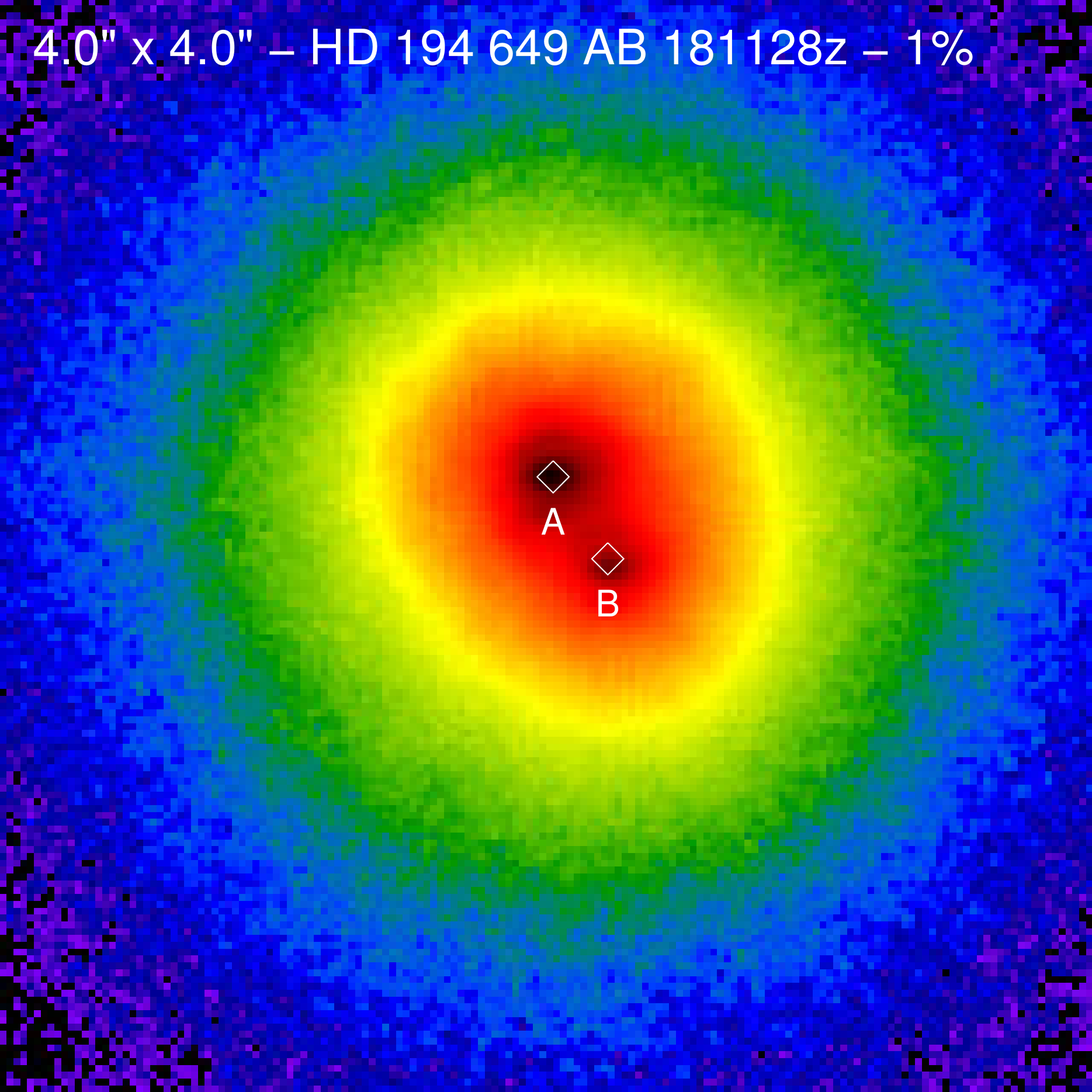} \
            \includegraphics*[width=0.240\linewidth]{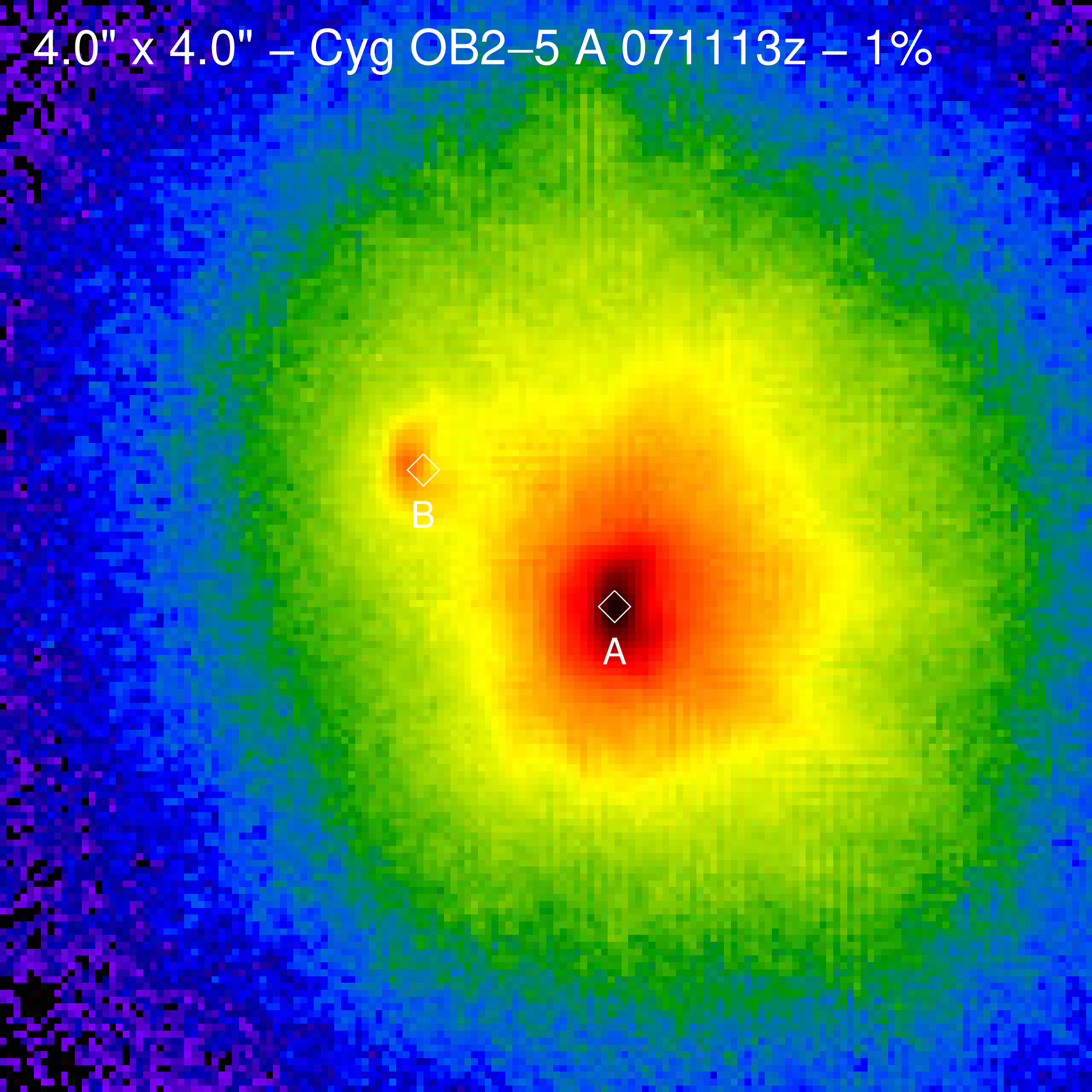}}
\vspace{1mm}
\centerline{\includegraphics*[width=0.240\linewidth]{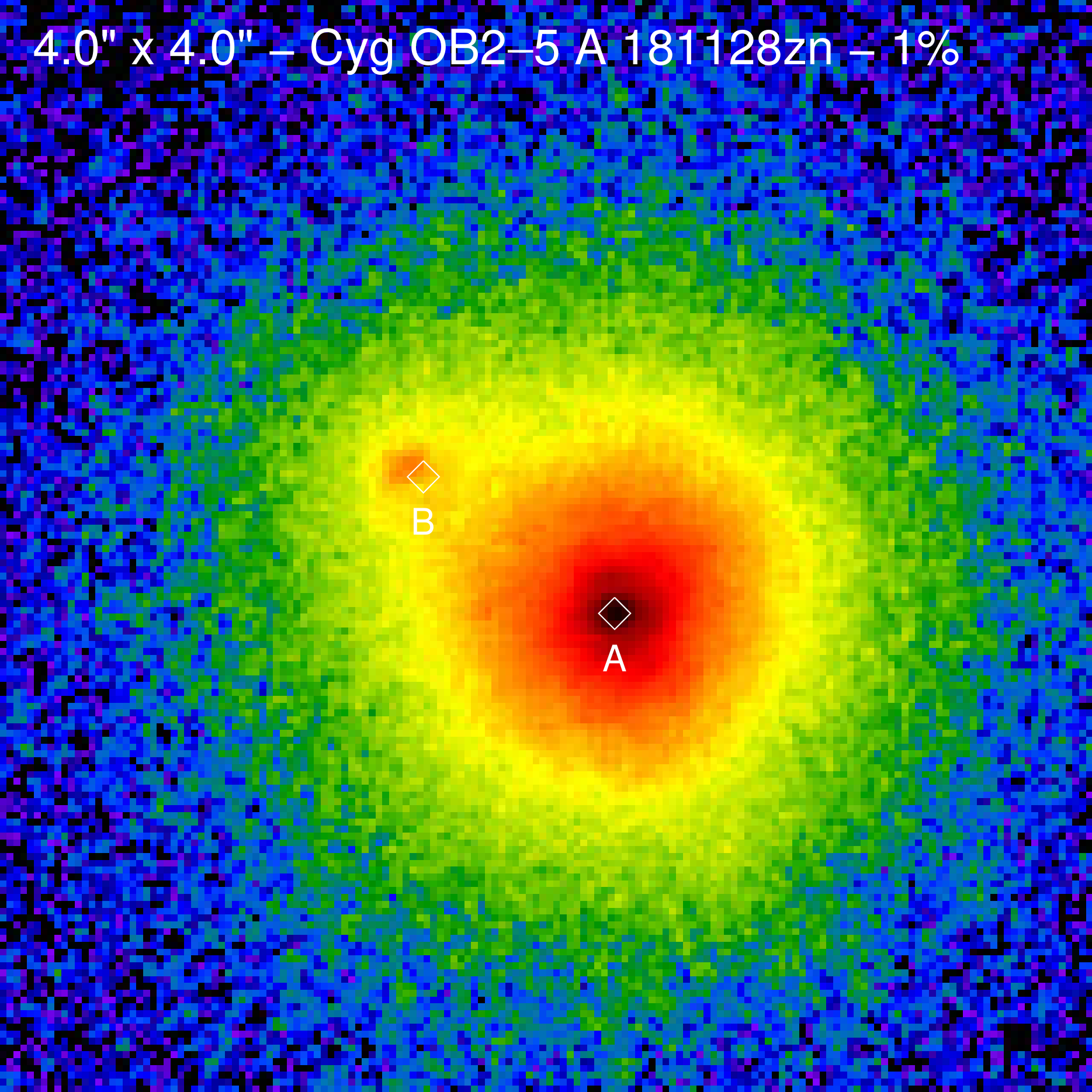} \
            \includegraphics*[width=0.240\linewidth]{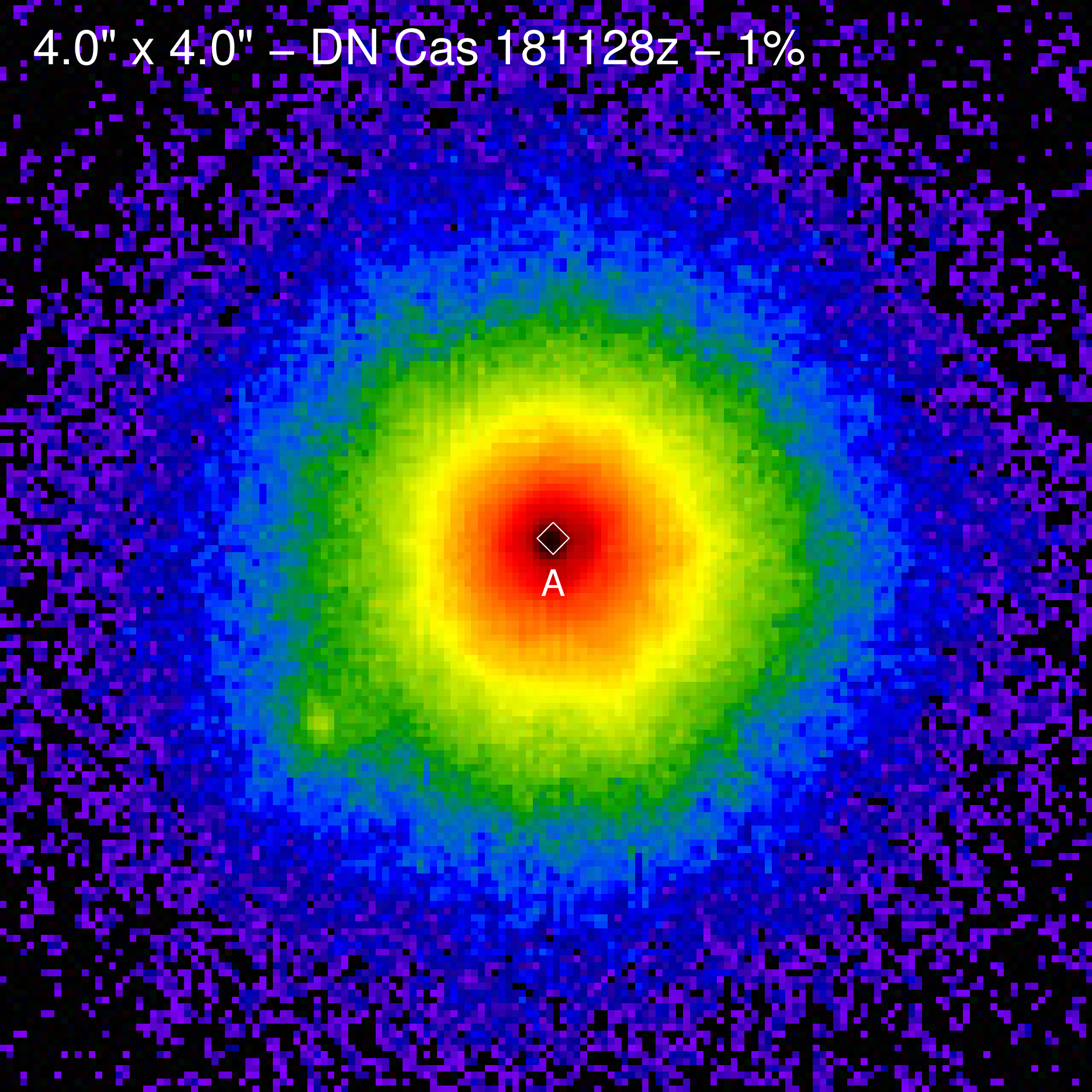} \
            \includegraphics*[width=0.240\linewidth]{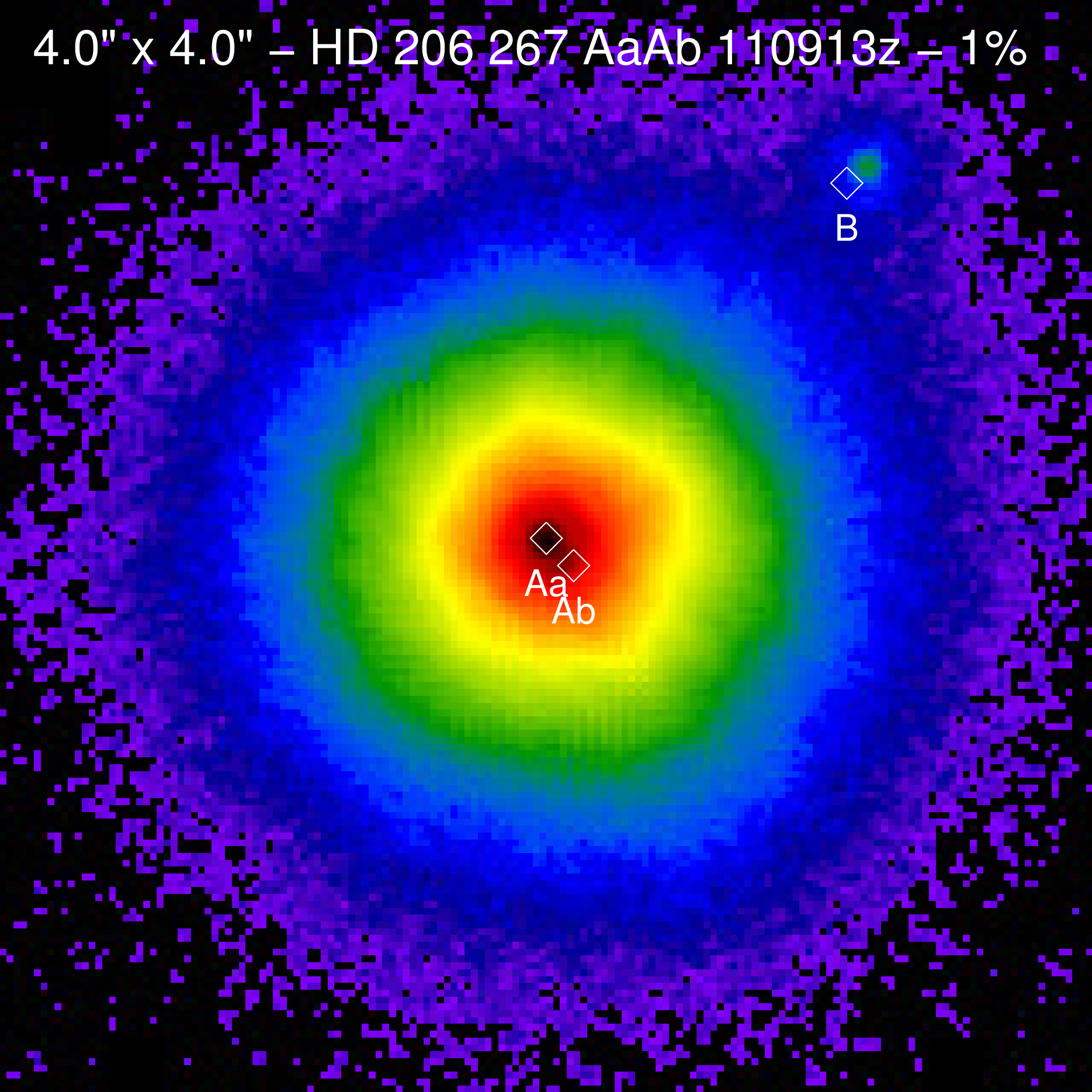} \
            \includegraphics*[width=0.240\linewidth]{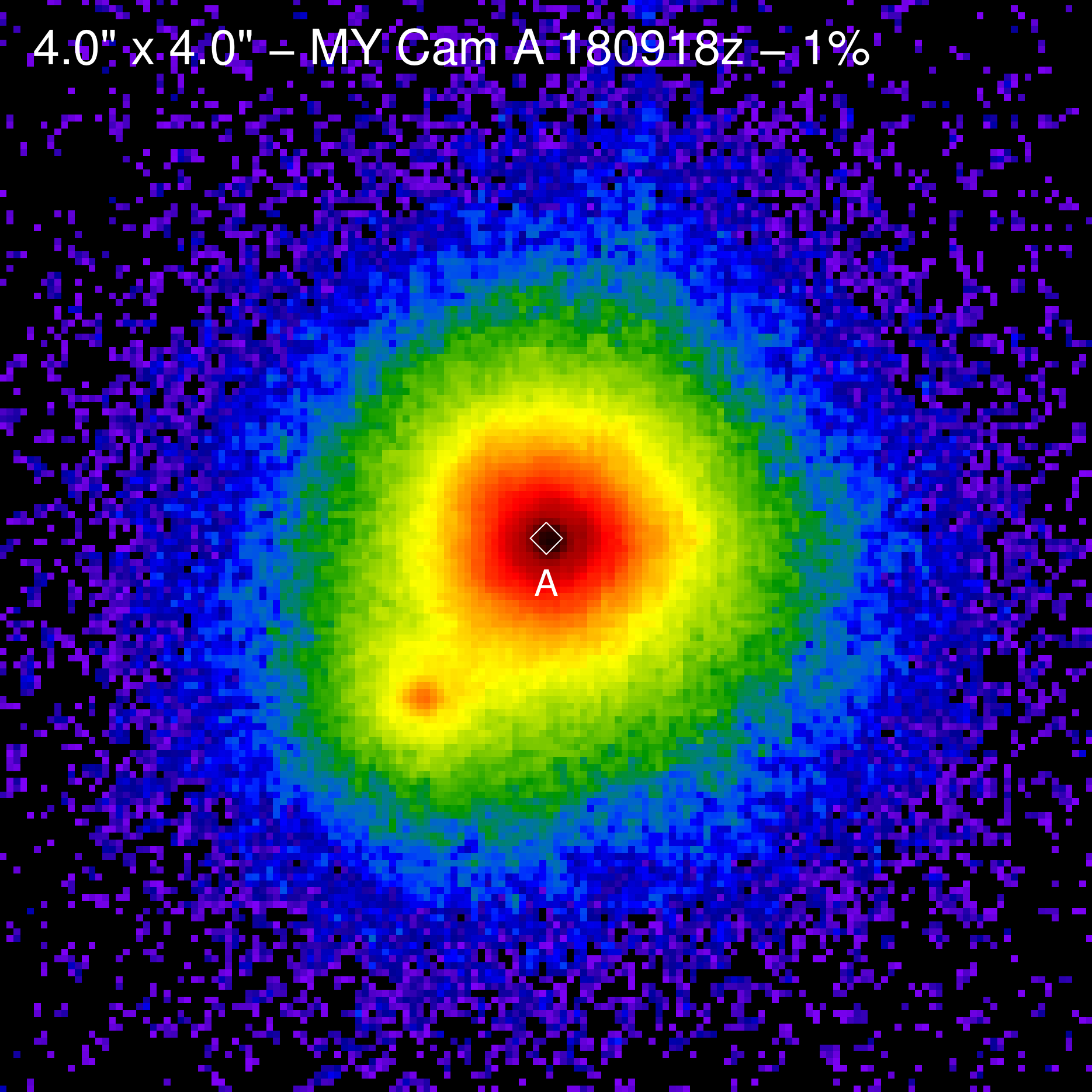}}
\vspace{1mm}
\centerline{\includegraphics*[width=0.240\linewidth]{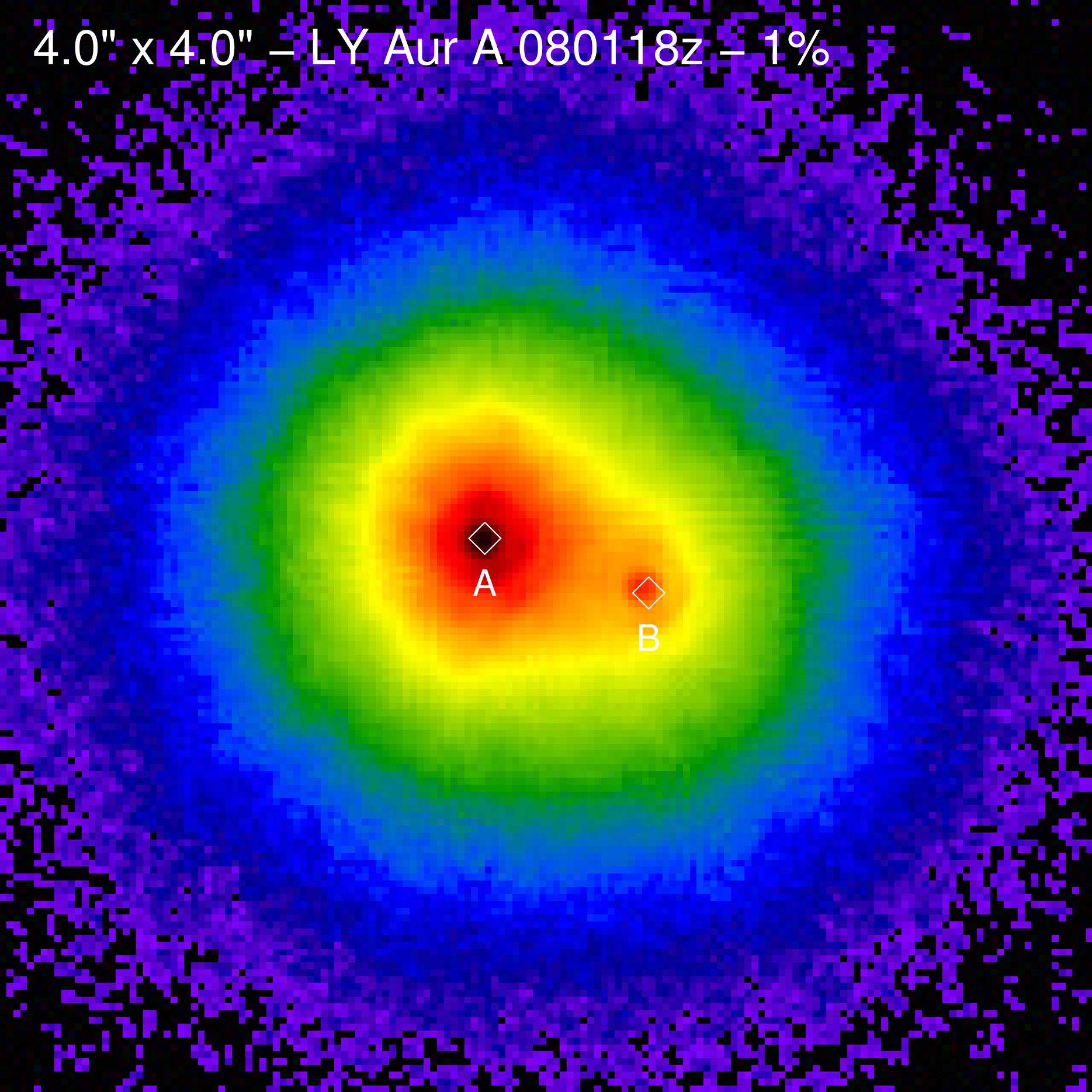} \
            \includegraphics*[width=0.240\linewidth]{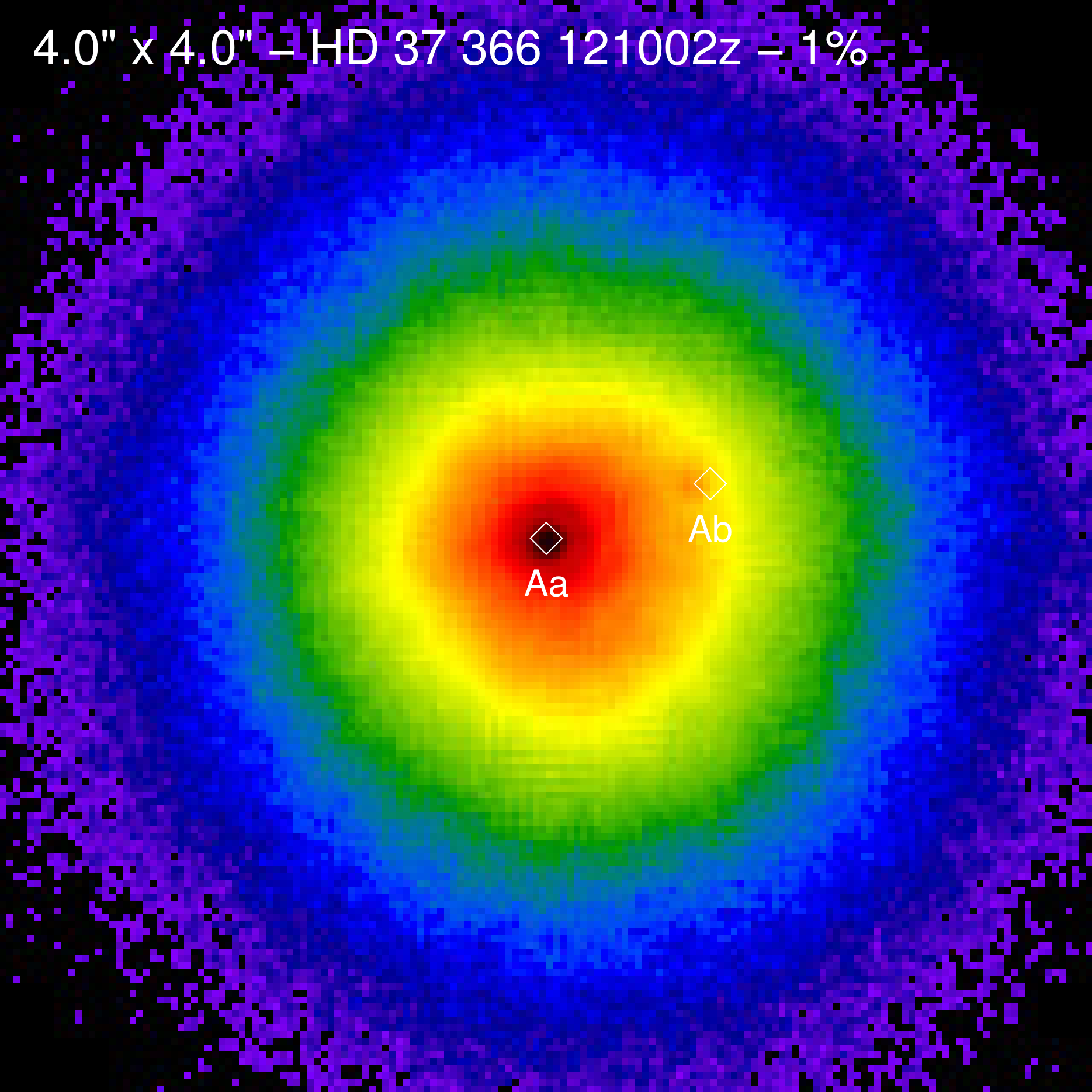} \
            \includegraphics*[width=0.240\linewidth]{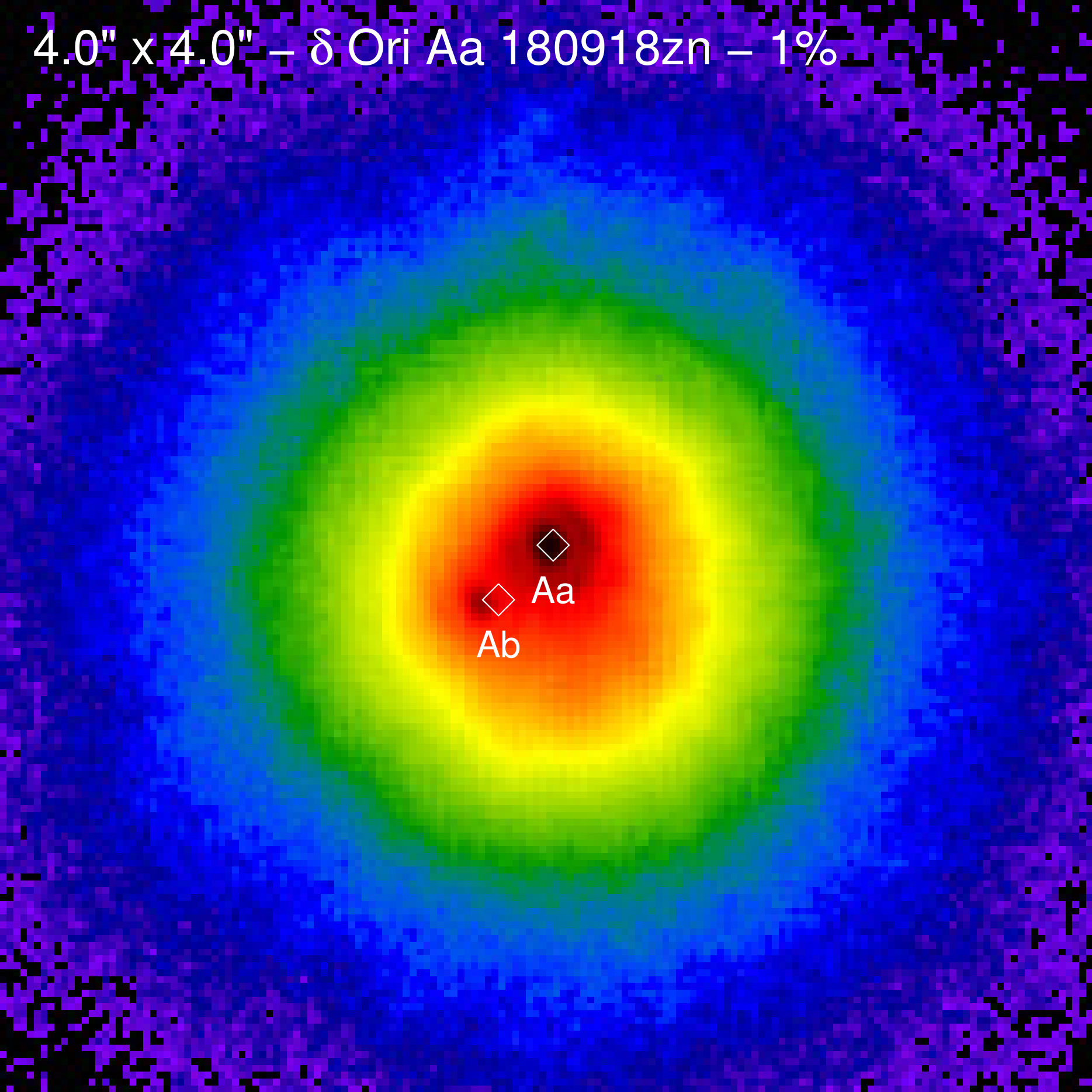} \
            \includegraphics*[width=0.240\linewidth]{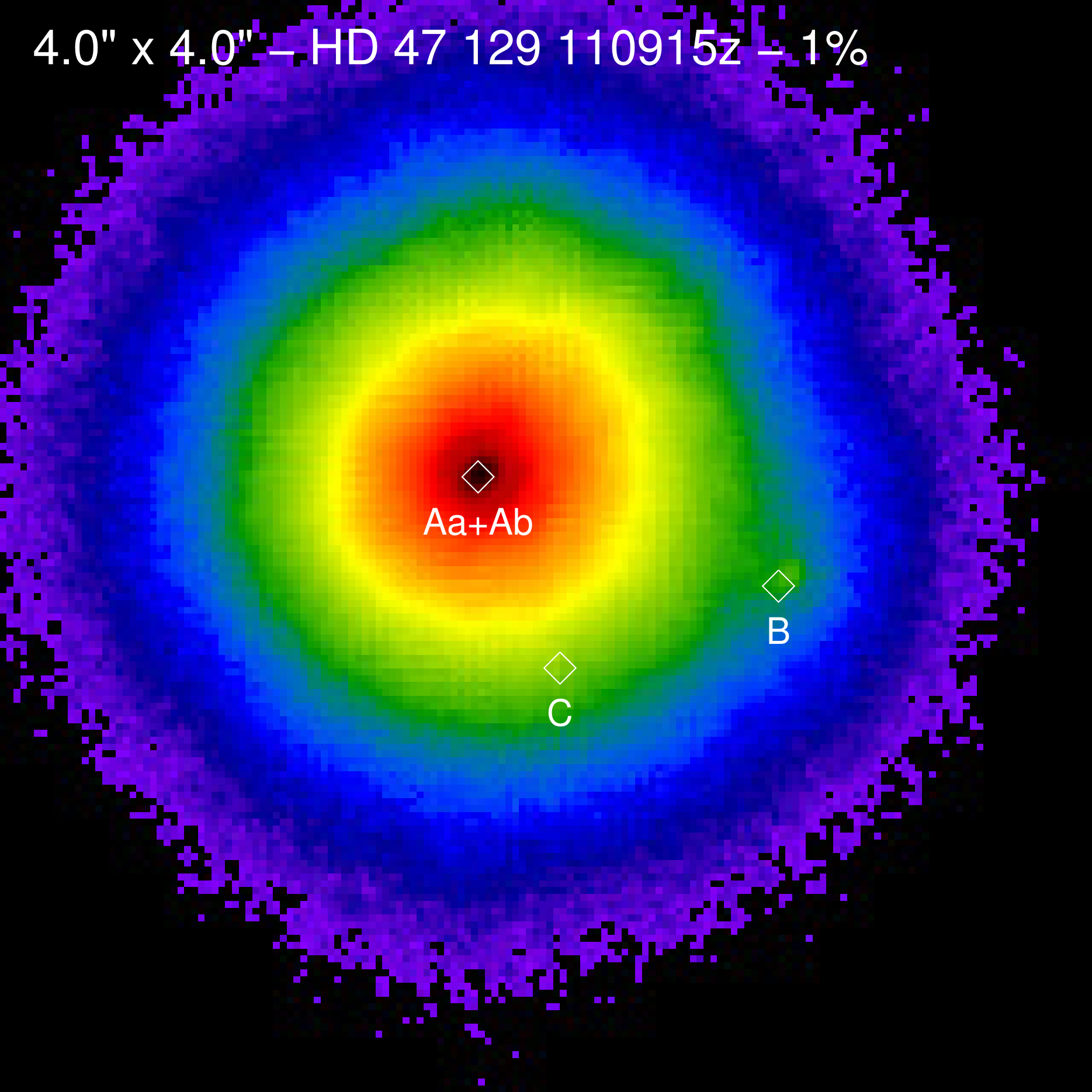}}
\vspace{1mm}
\centerline{\includegraphics*[width=0.240\linewidth]{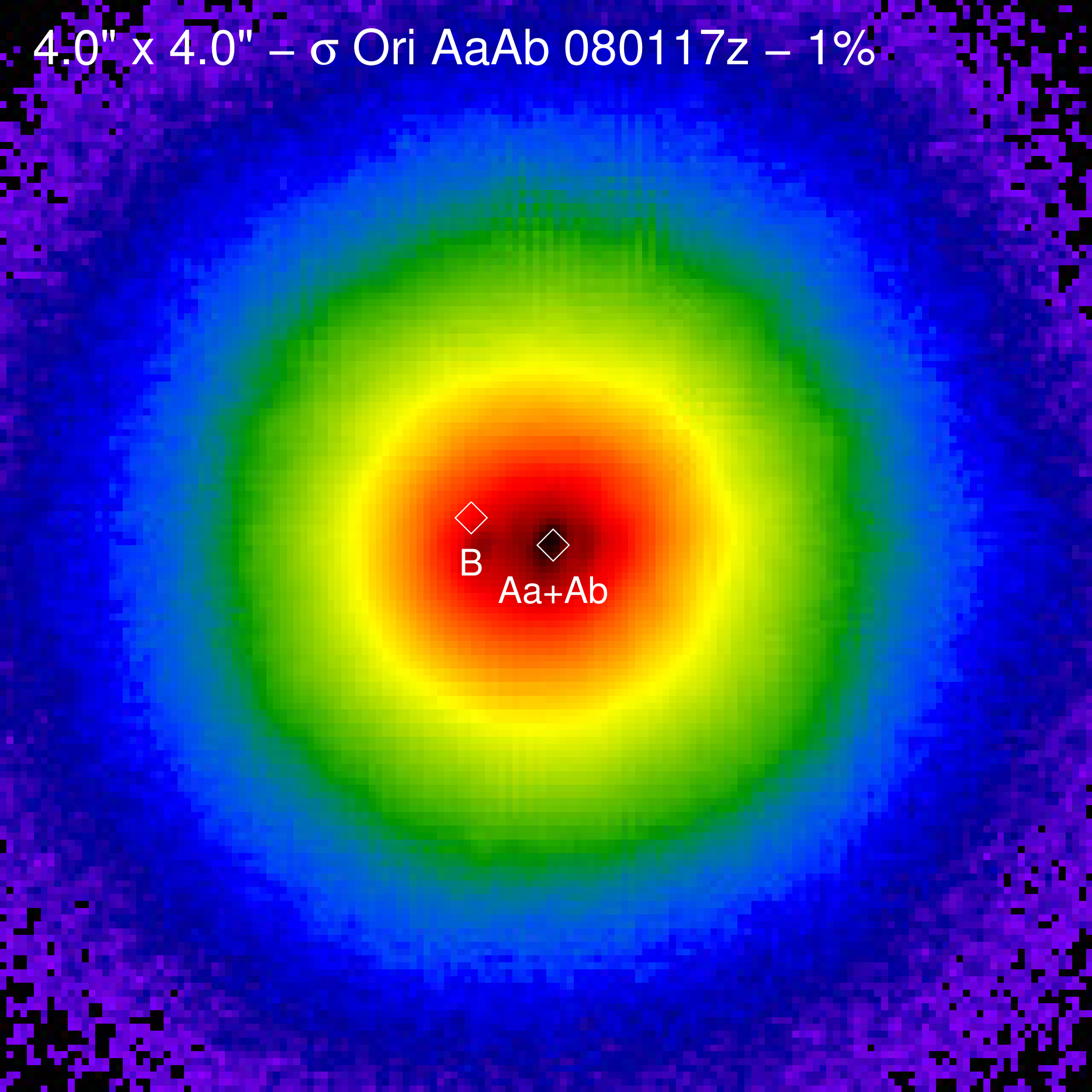} \
            \includegraphics*[width=0.240\linewidth]{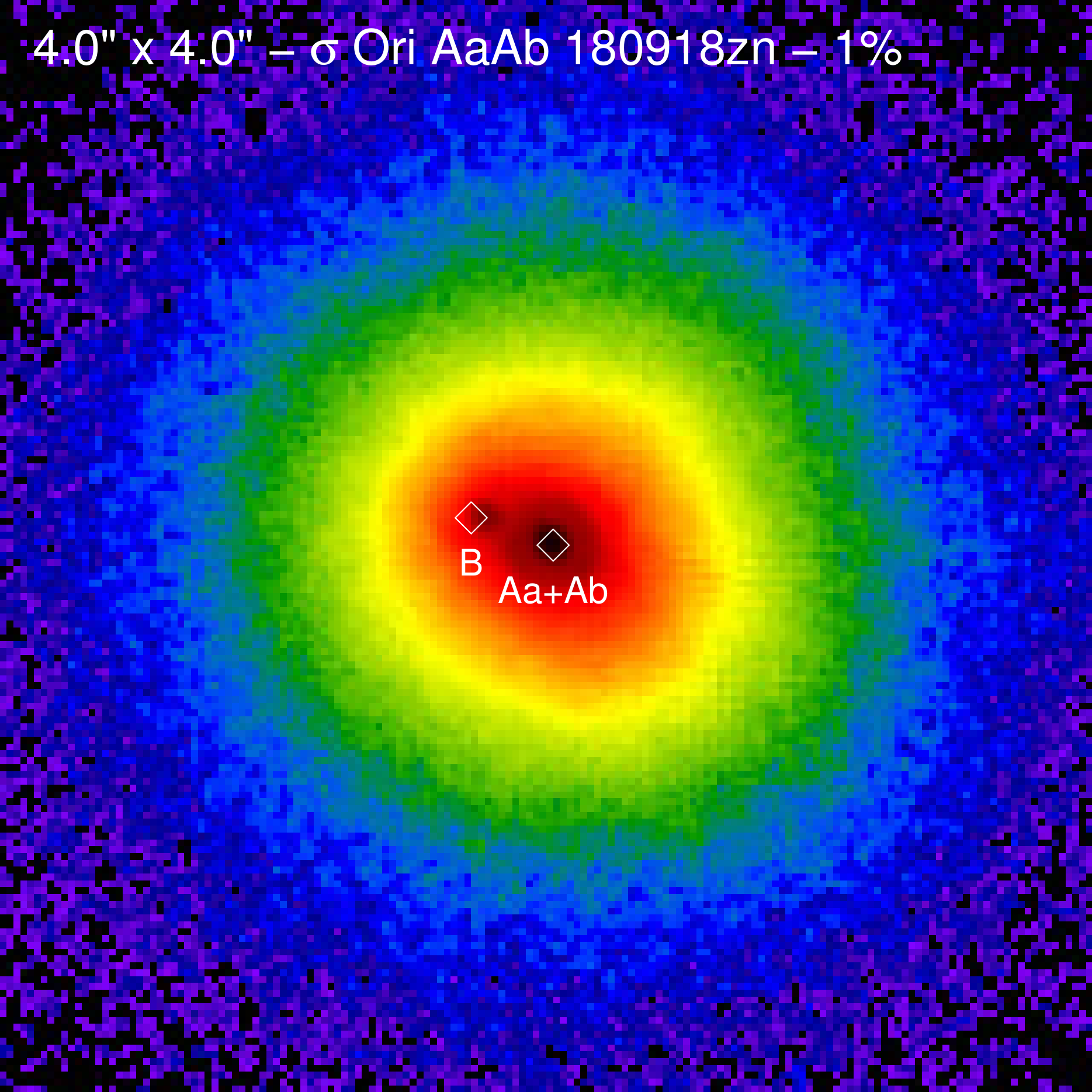} \
            \includegraphics*[width=0.240\linewidth]{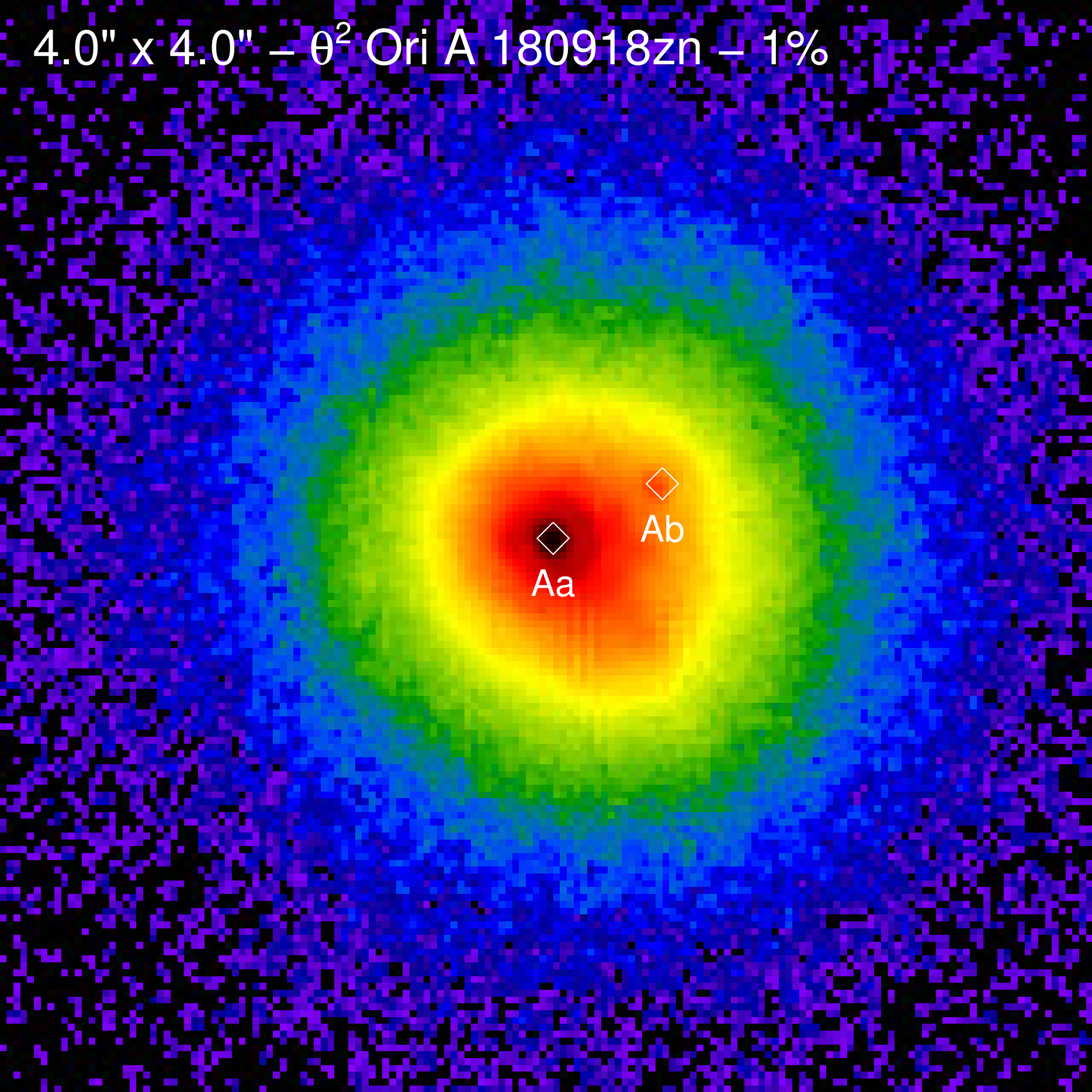}}
 \caption{Fragments of AstraLux fields for 13 of the targets in this paper (two of them with two epochs). 
          The field size, target, evening date (YYMMDD), band, and lucky image frame selection fraction
          are given in each case ($z$ stands for SDSS $z$, $zn$ is a narrow-band filter with a similar central wavelength). The intensity scale is logarithmic to show
          both bright and faint sources. Diamonds are used to mark the last position of the known components in the WDS catalog (if no entry exists there, then an A is
          used to mark the brightest detected component). Each frame is $4\arcsec\times 4\arcsec$ (160$\times$160 pixels) with North up and East left.}
\label{AstraLux1}
\end{figure*}

\begin{figure*}
\centerline{\includegraphics*[width=0.240\linewidth]{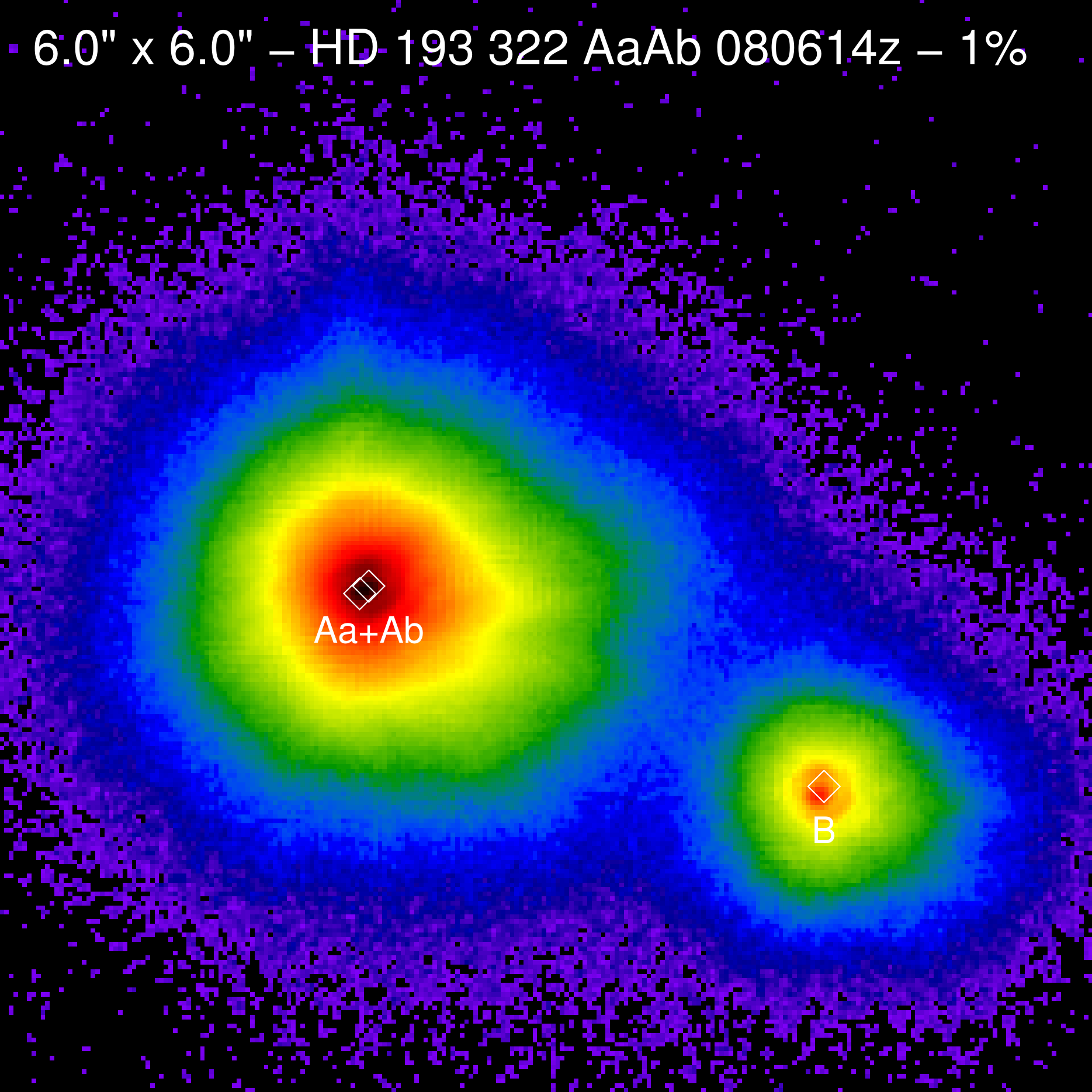} \
            \includegraphics*[width=0.240\linewidth]{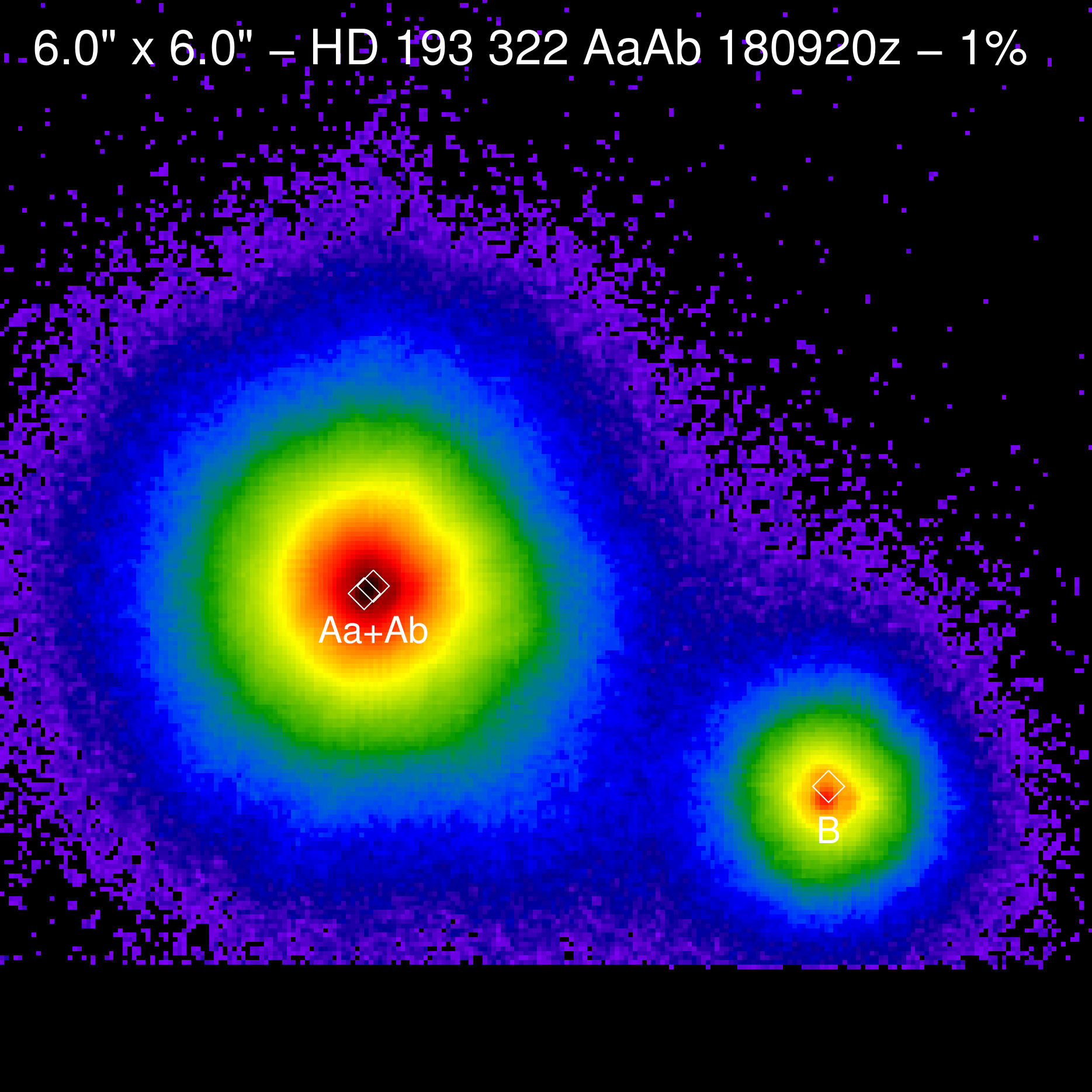} \
            \includegraphics*[width=0.240\linewidth]{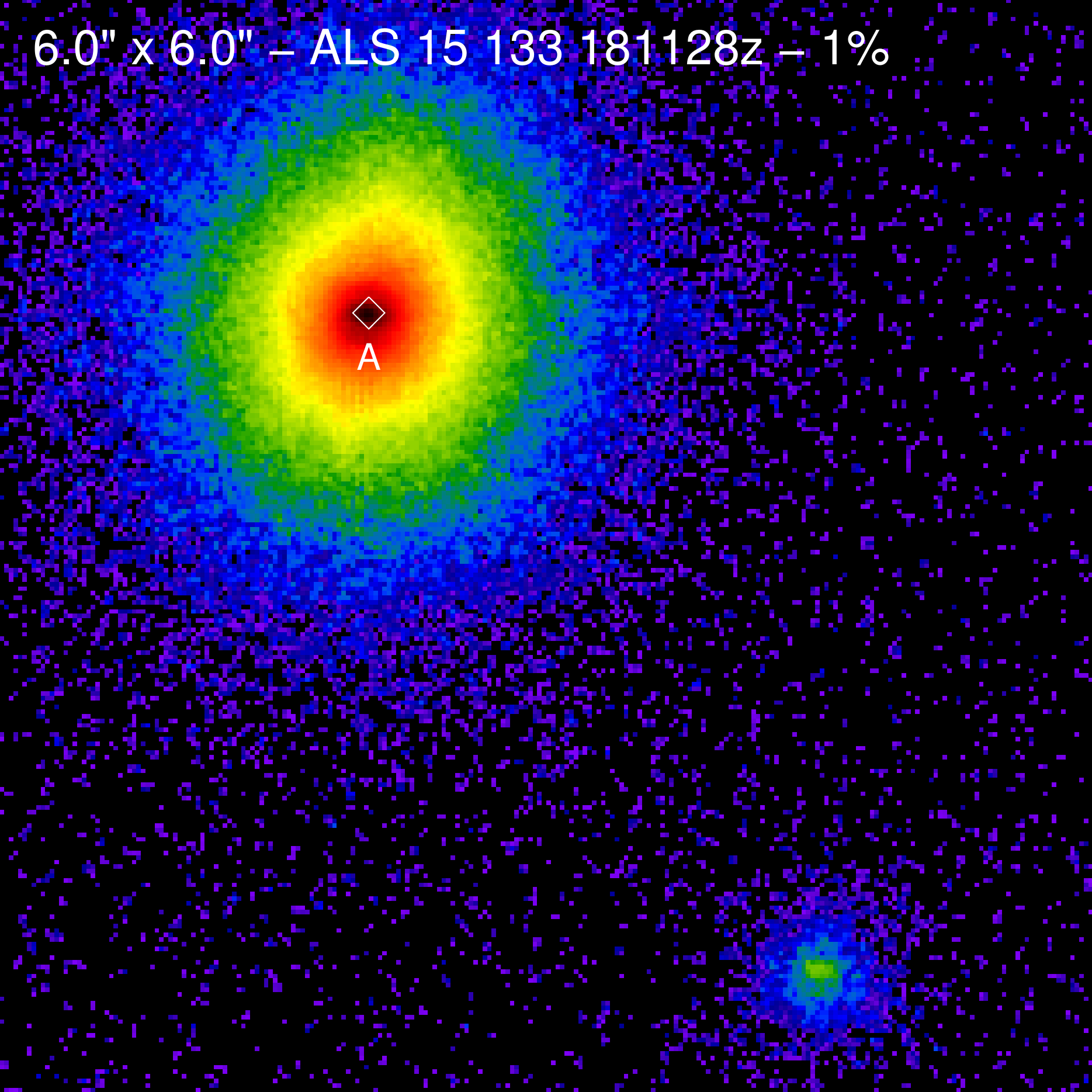} \
            \includegraphics*[width=0.240\linewidth]{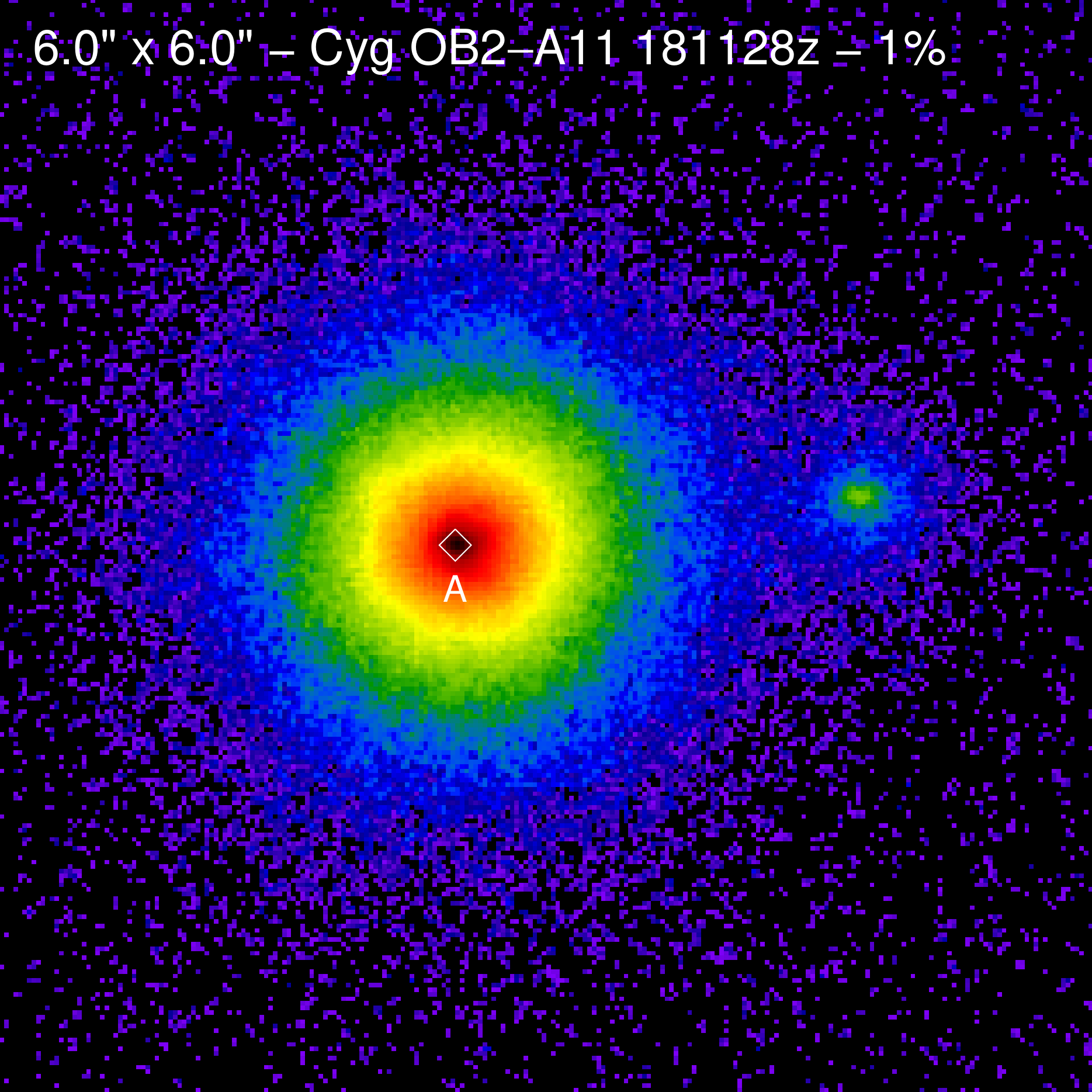}}
\vspace{1mm}
\centerline{\includegraphics*[width=0.240\linewidth]{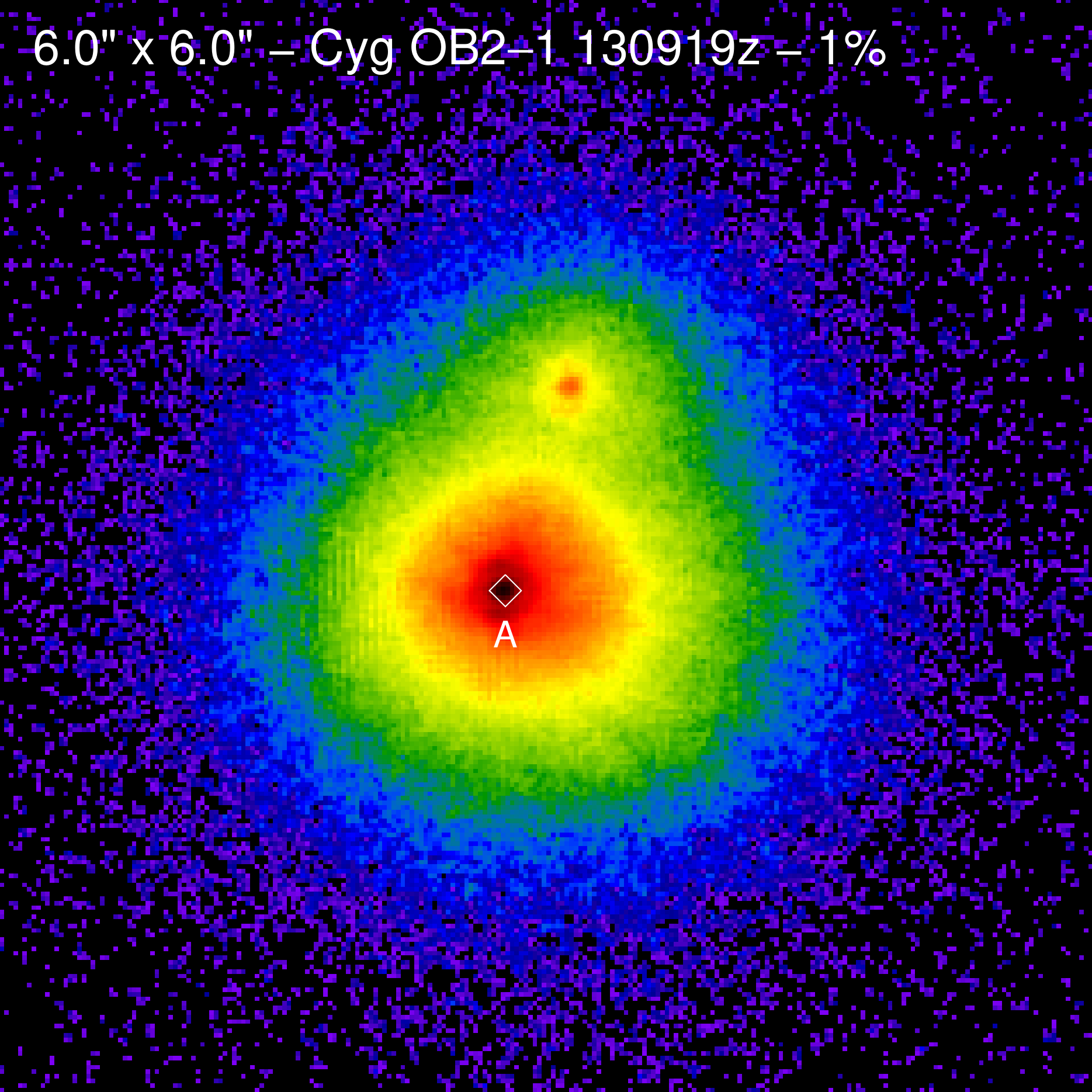} \
            \includegraphics*[width=0.240\linewidth]{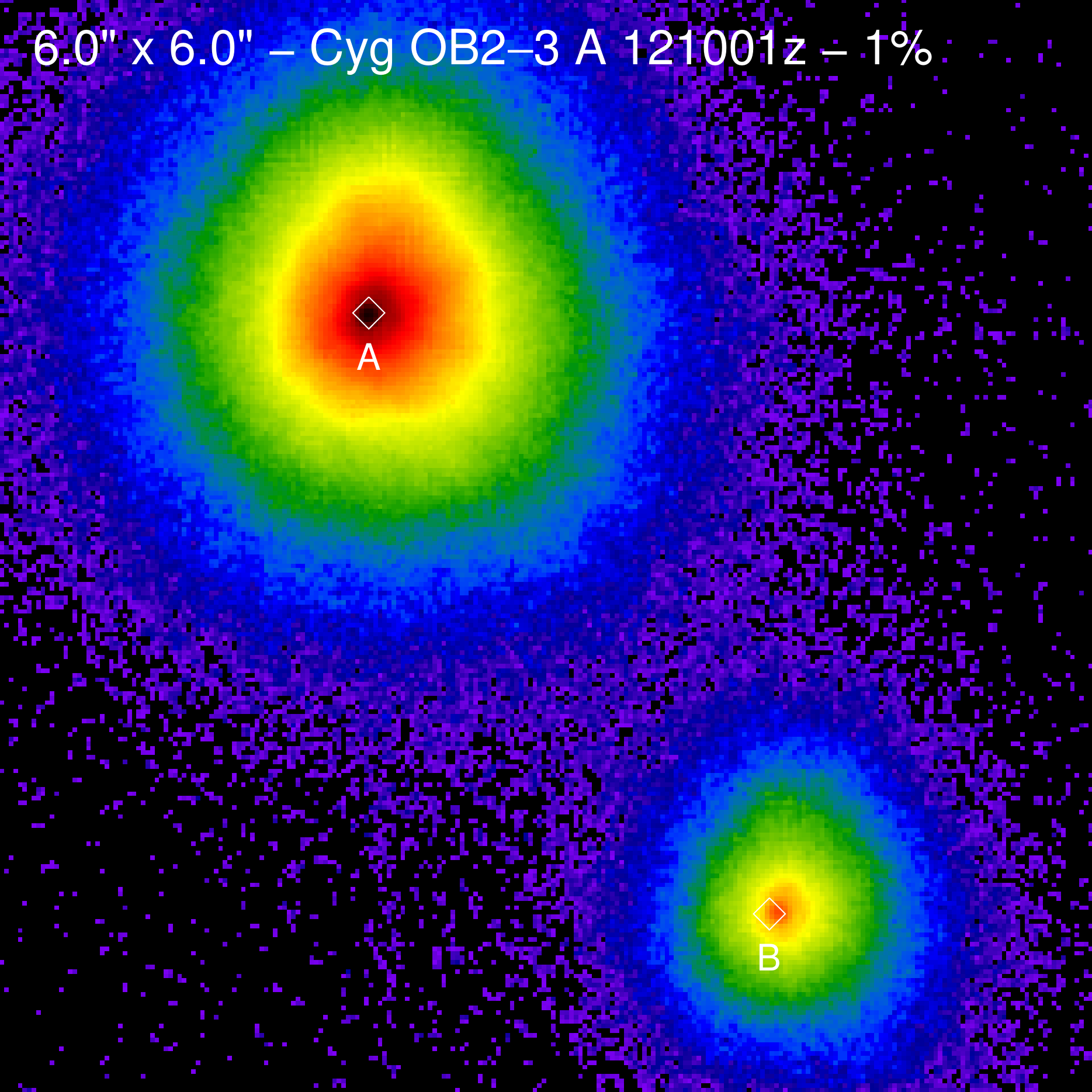} \
            \includegraphics*[width=0.240\linewidth]{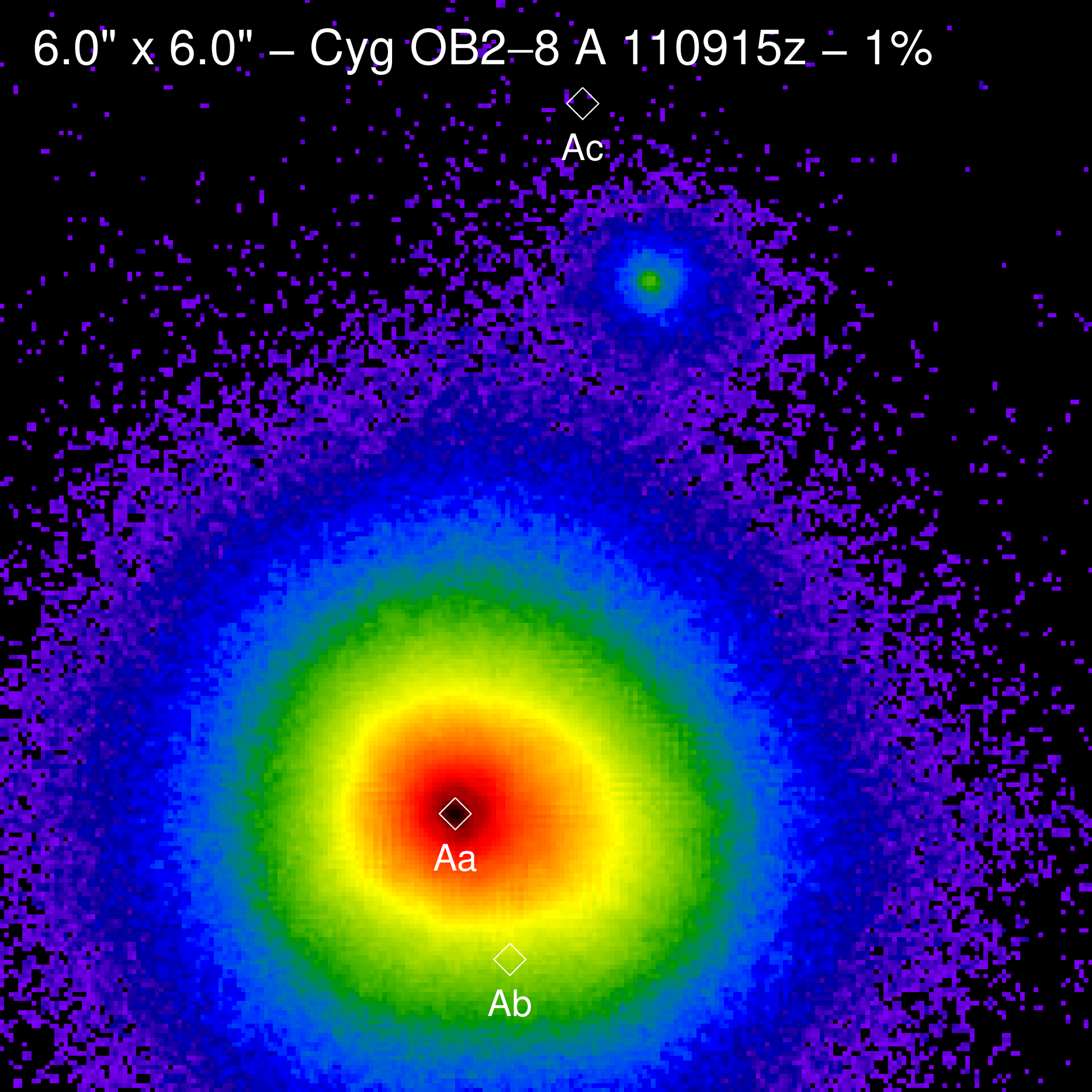} \
            \includegraphics*[width=0.240\linewidth]{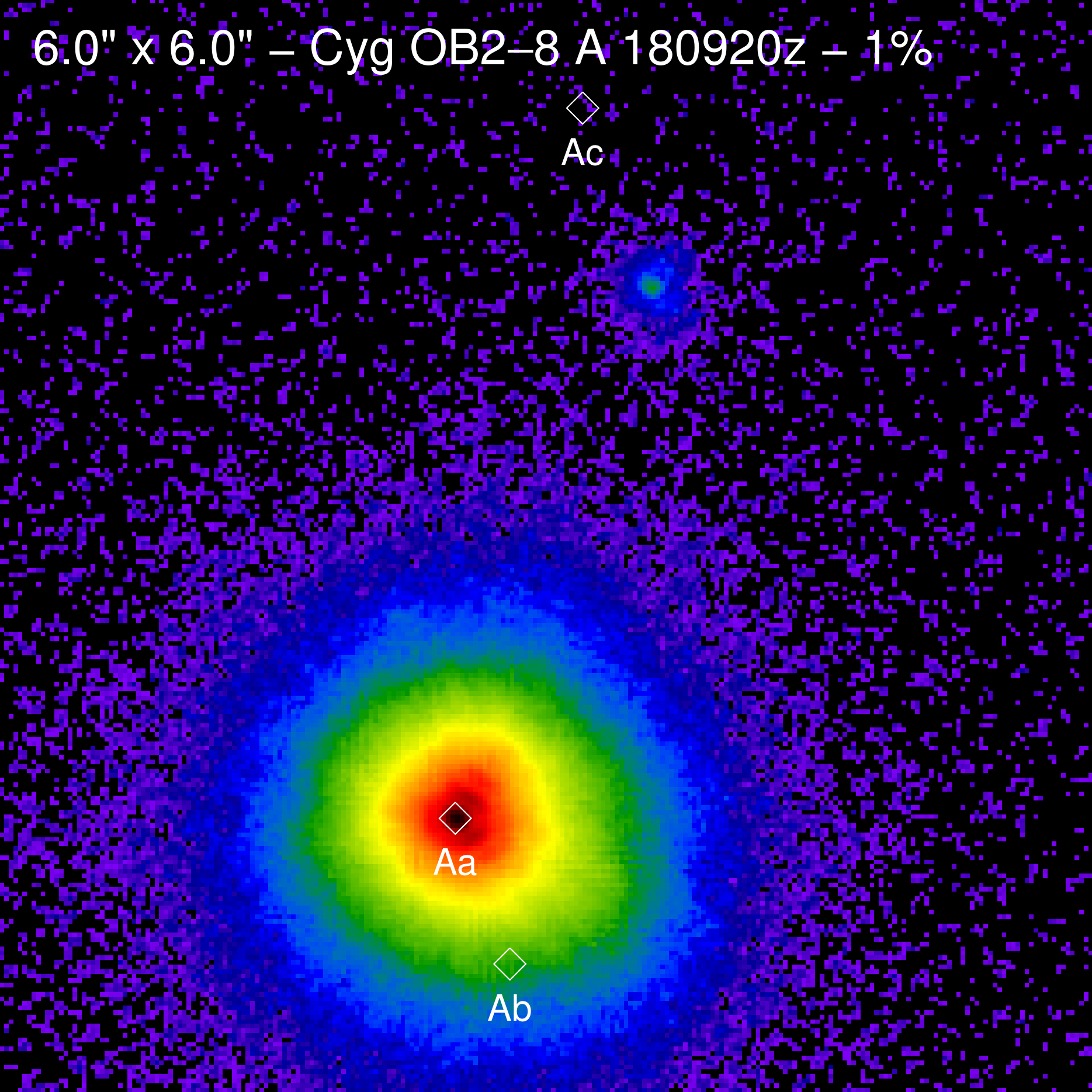}}  
\vspace{1mm}
\centerline{\includegraphics*[width=0.240\linewidth]{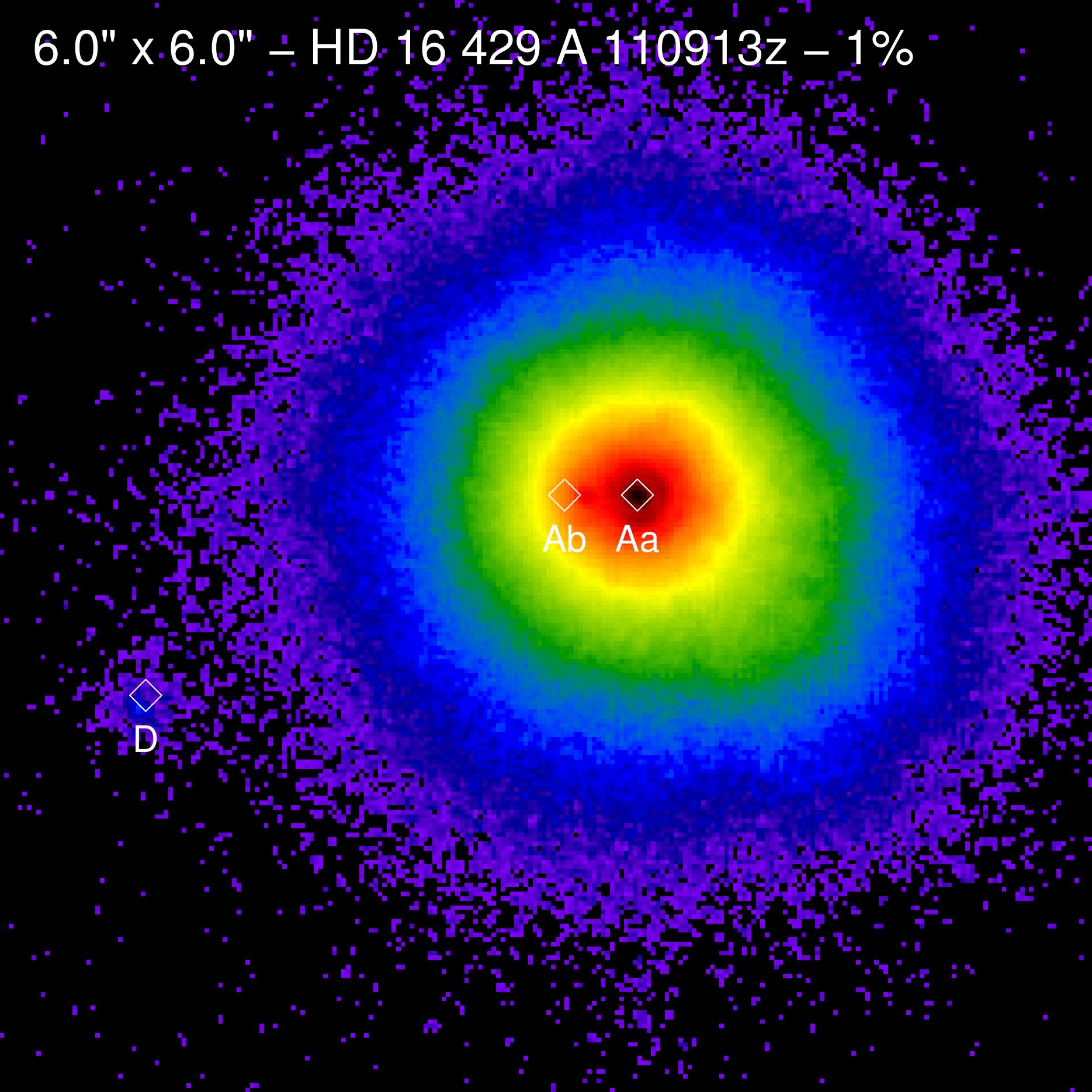} \
            \includegraphics*[width=0.240\linewidth]{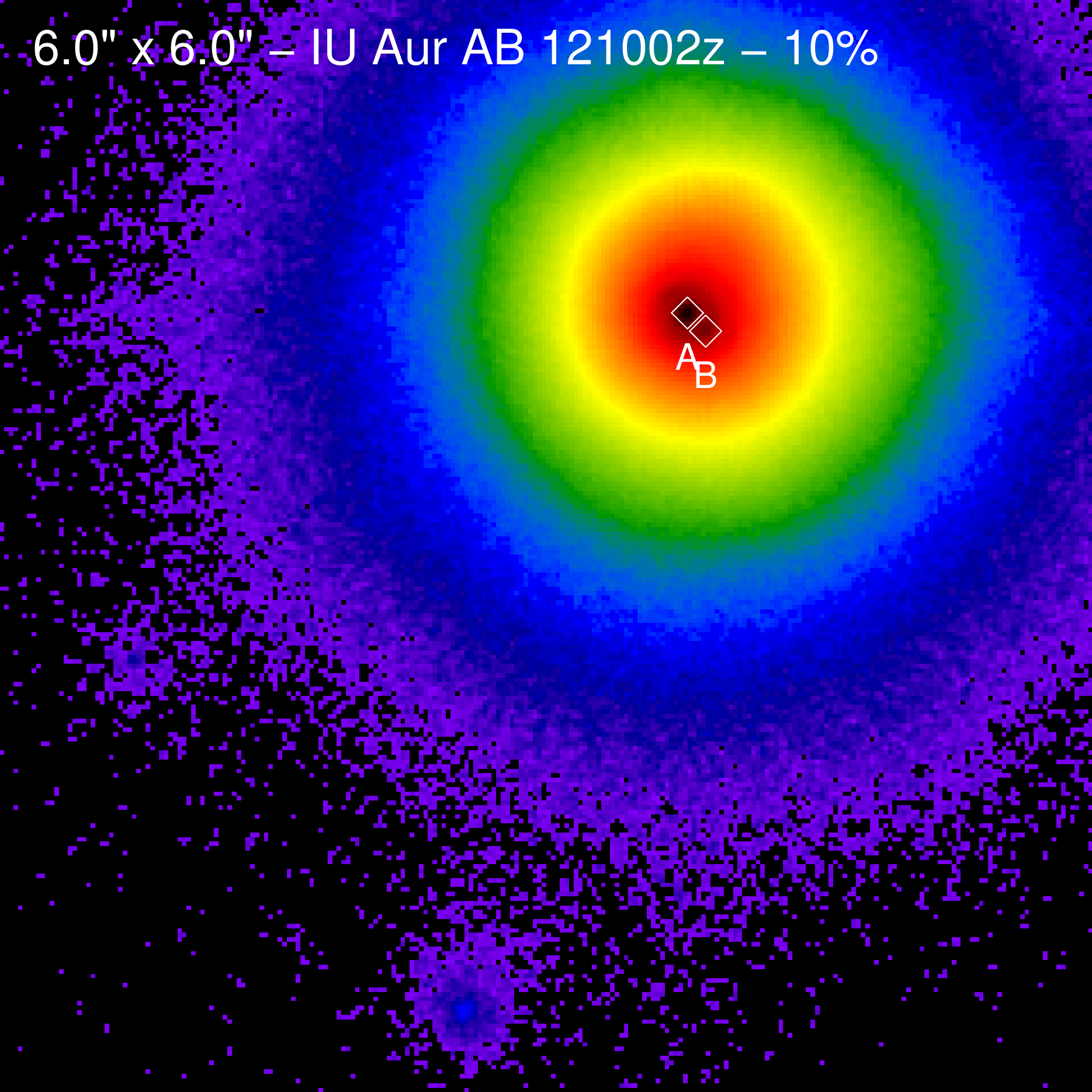} \
            \includegraphics*[width=0.240\linewidth]{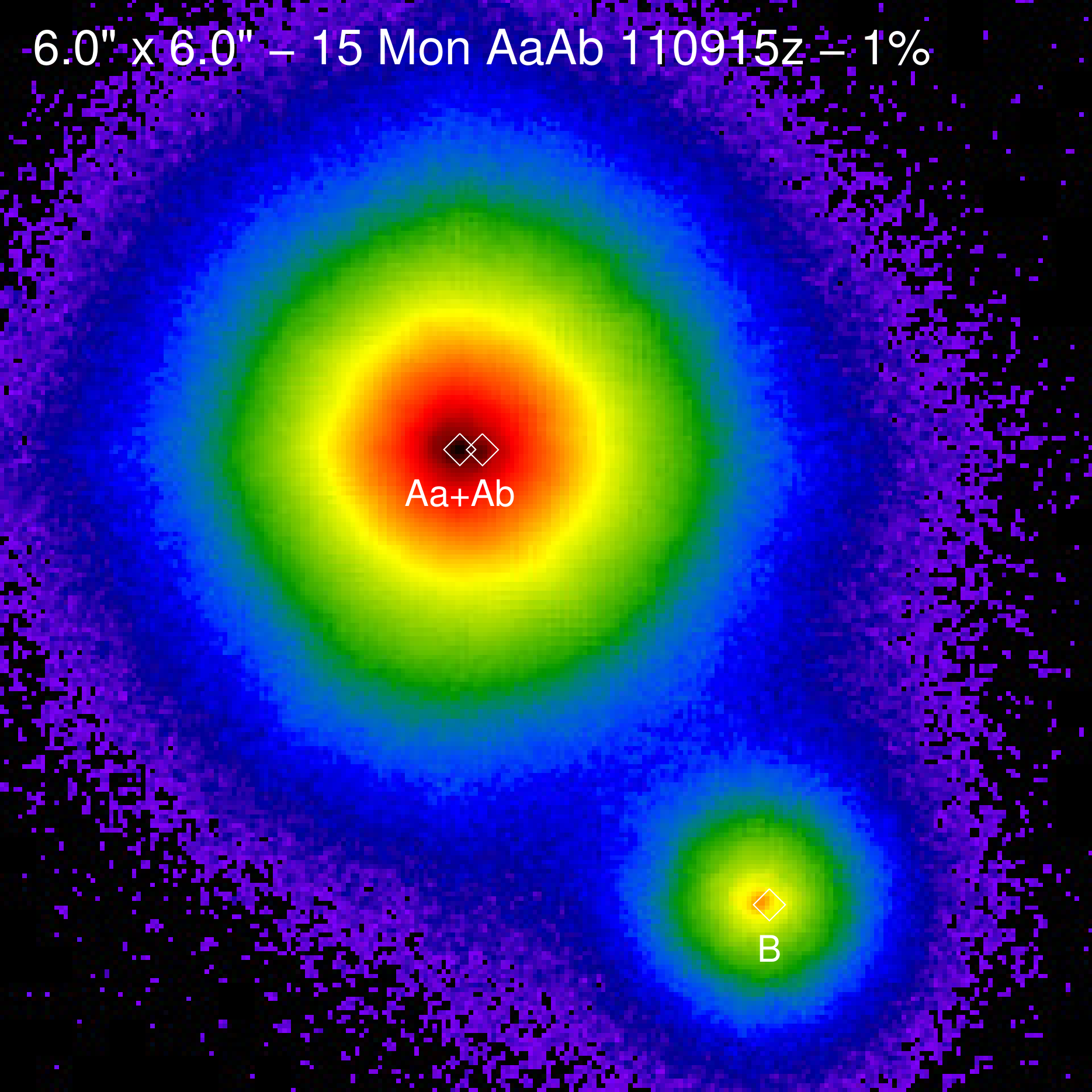}}
\vspace{1mm}
\centerline{\includegraphics*[width=0.240\linewidth]{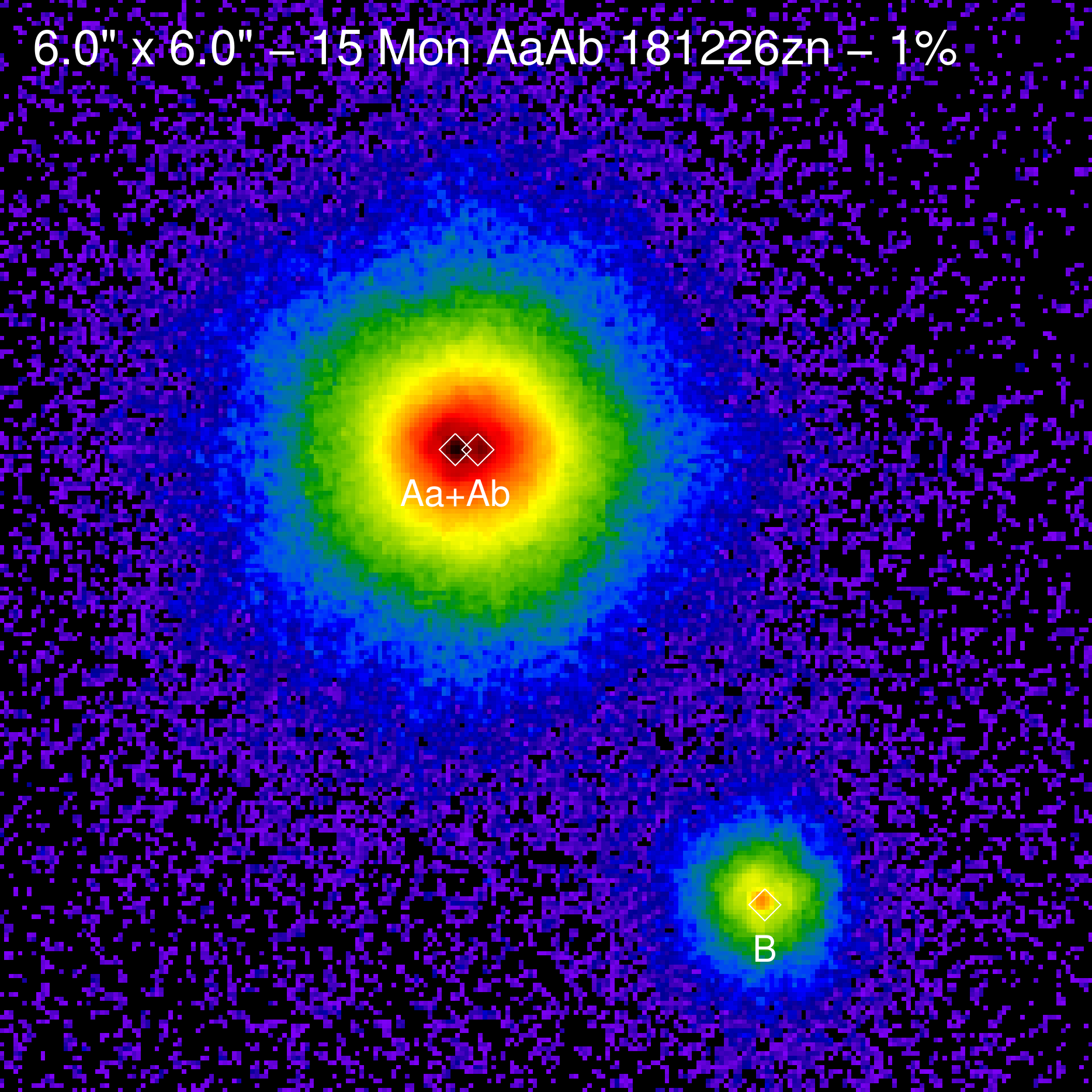} \
            \includegraphics*[width=0.240\linewidth]{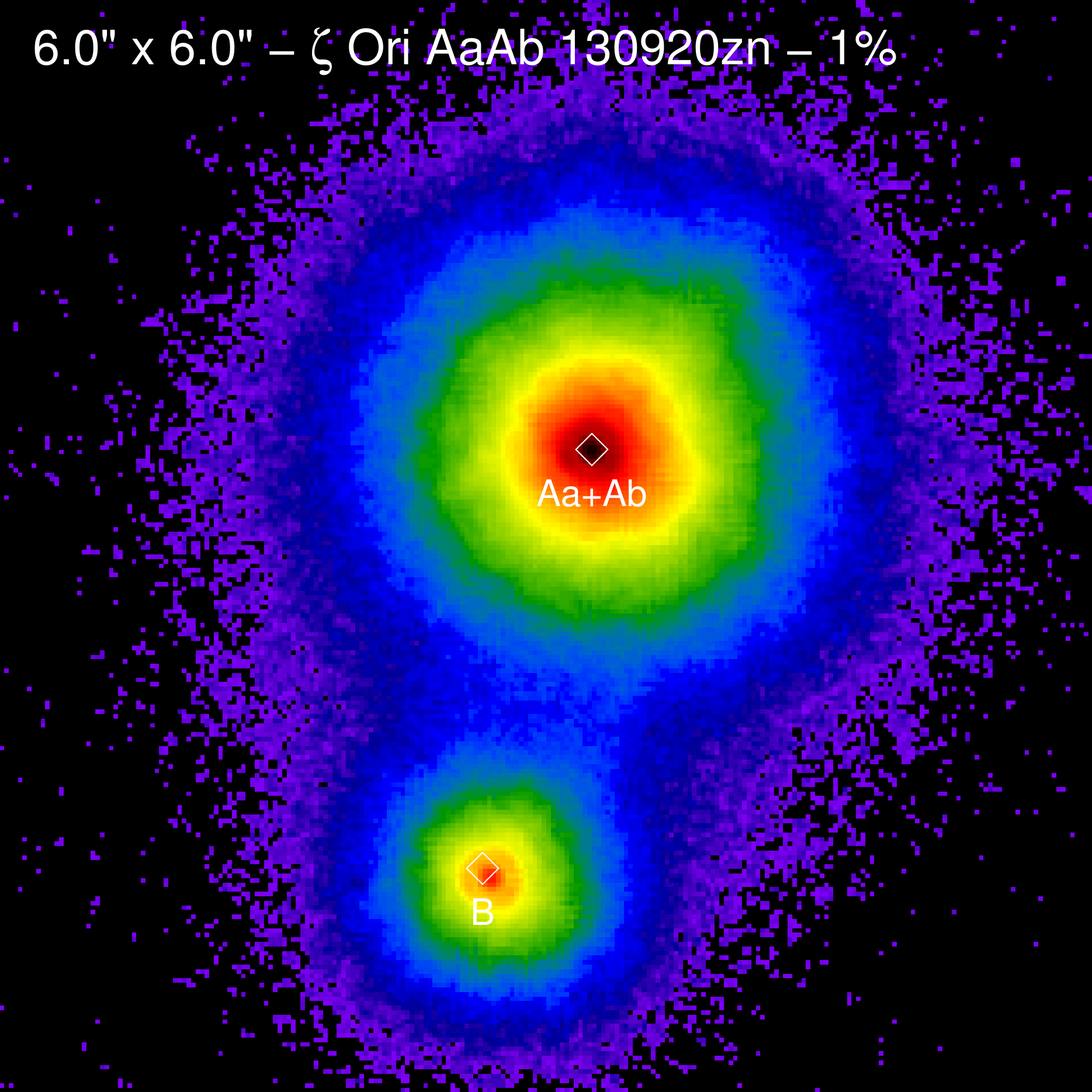} \
            \includegraphics*[width=0.240\linewidth]{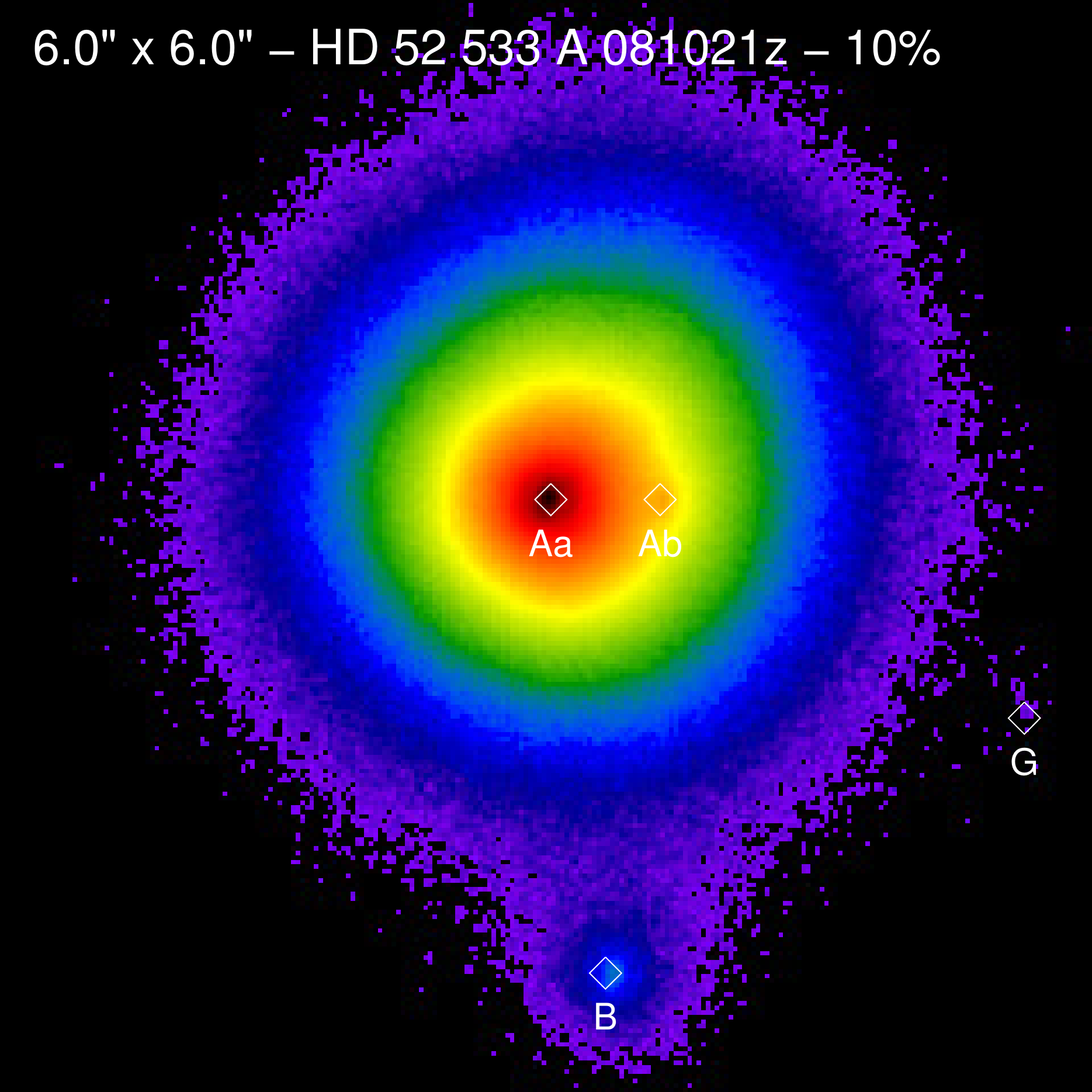}}
 \caption{Same as Fig.~\ref{AstraLux1} for eleven additional targets (three with two epochs) with fields of view of $6\arcsec\times 6\arcsec$ (240$\times$240 pixels).}
\label{AstraLux2}
\end{figure*}

\begin{figure*}
\begin{minipage}[t]{0.49\textwidth}
$\!\,$ \linebreak
\includegraphics*[width=\linewidth]{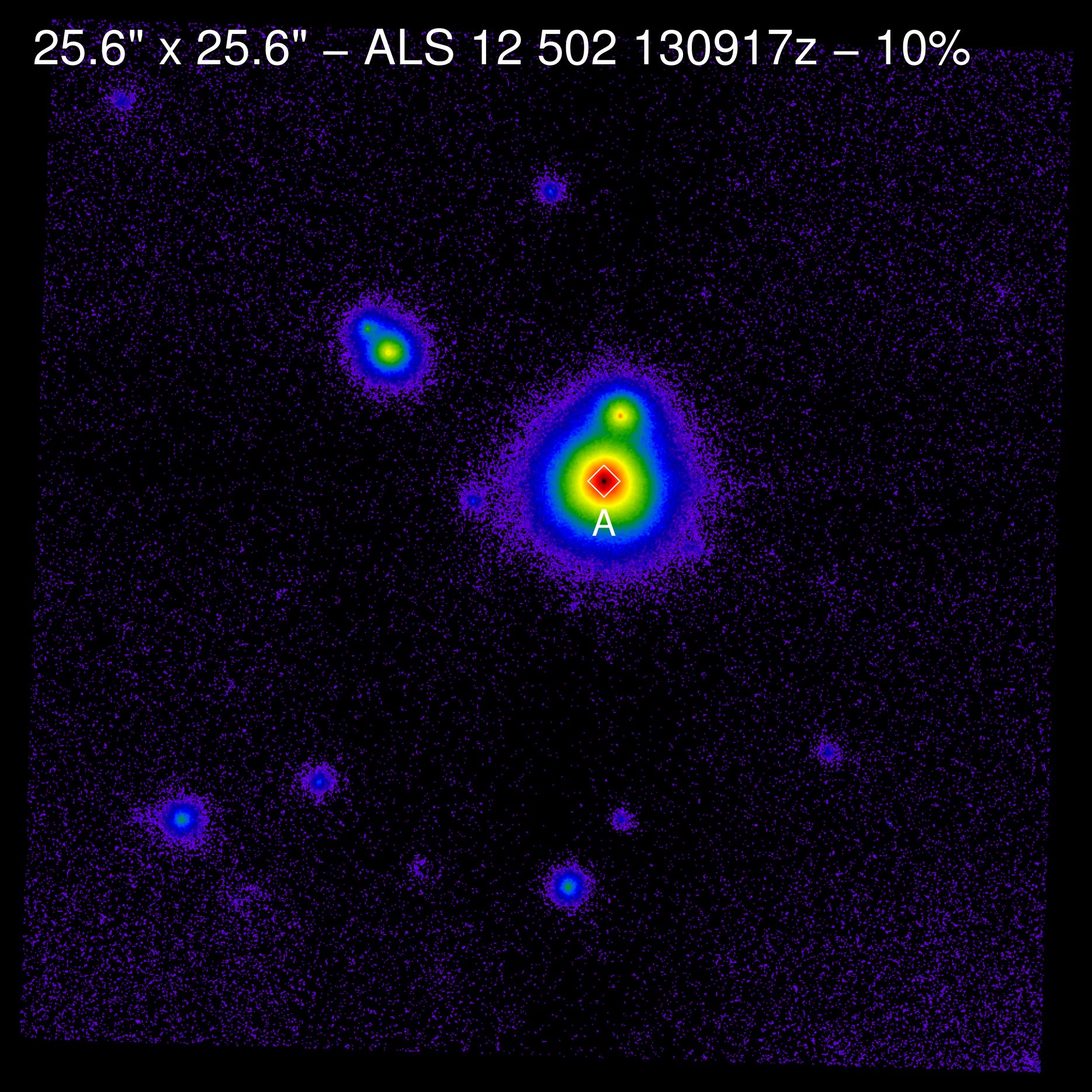}
\end{minipage} 
\begin{minipage}[t]{0.49\textwidth}
$\!\,$ \linebreak
\includegraphics*[width=\linewidth]{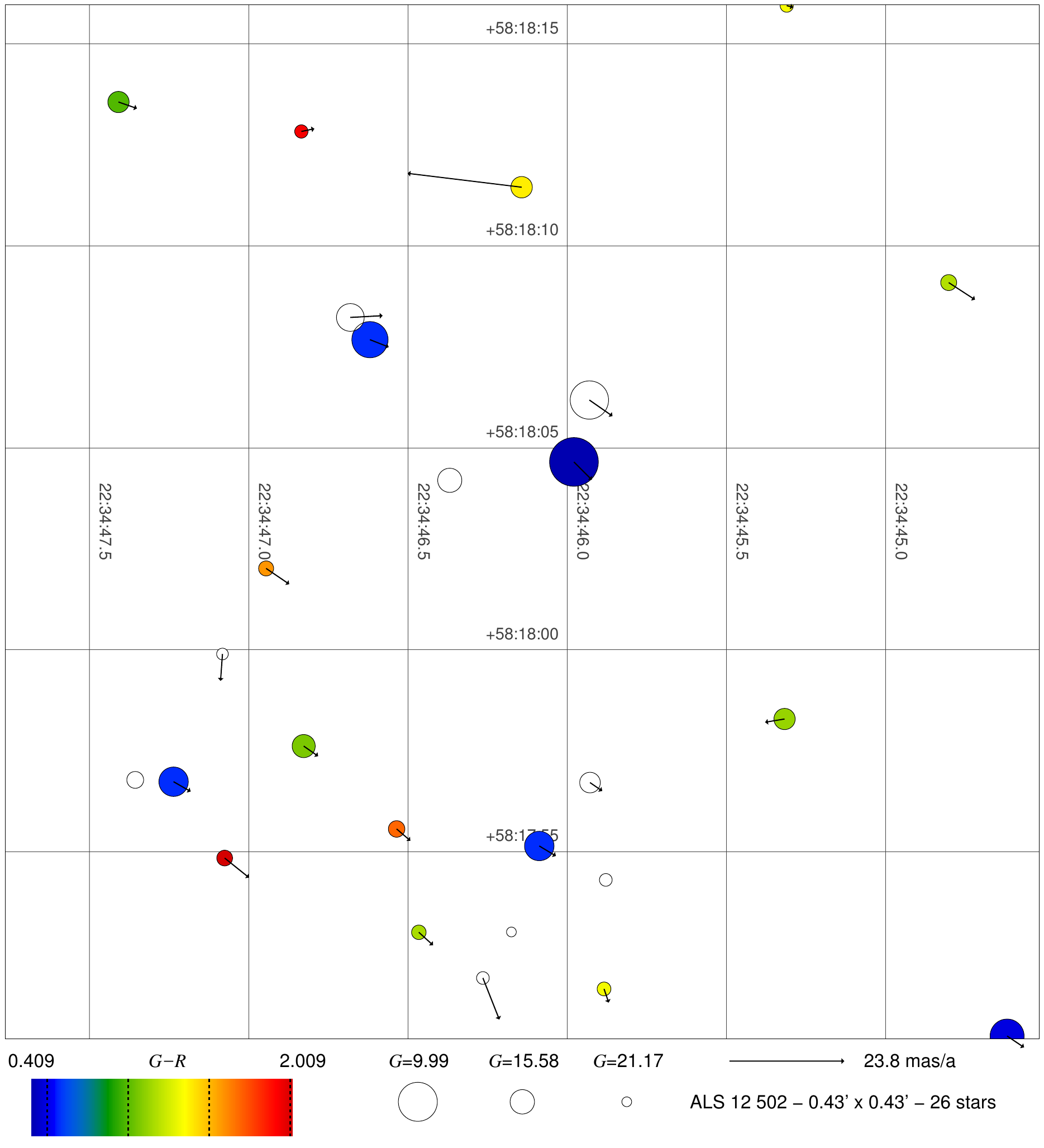}
\end{minipage}
\begin{minipage}[t]{0.49\textwidth}
$\!\,$ \linebreak
\includegraphics*[width=\linewidth]{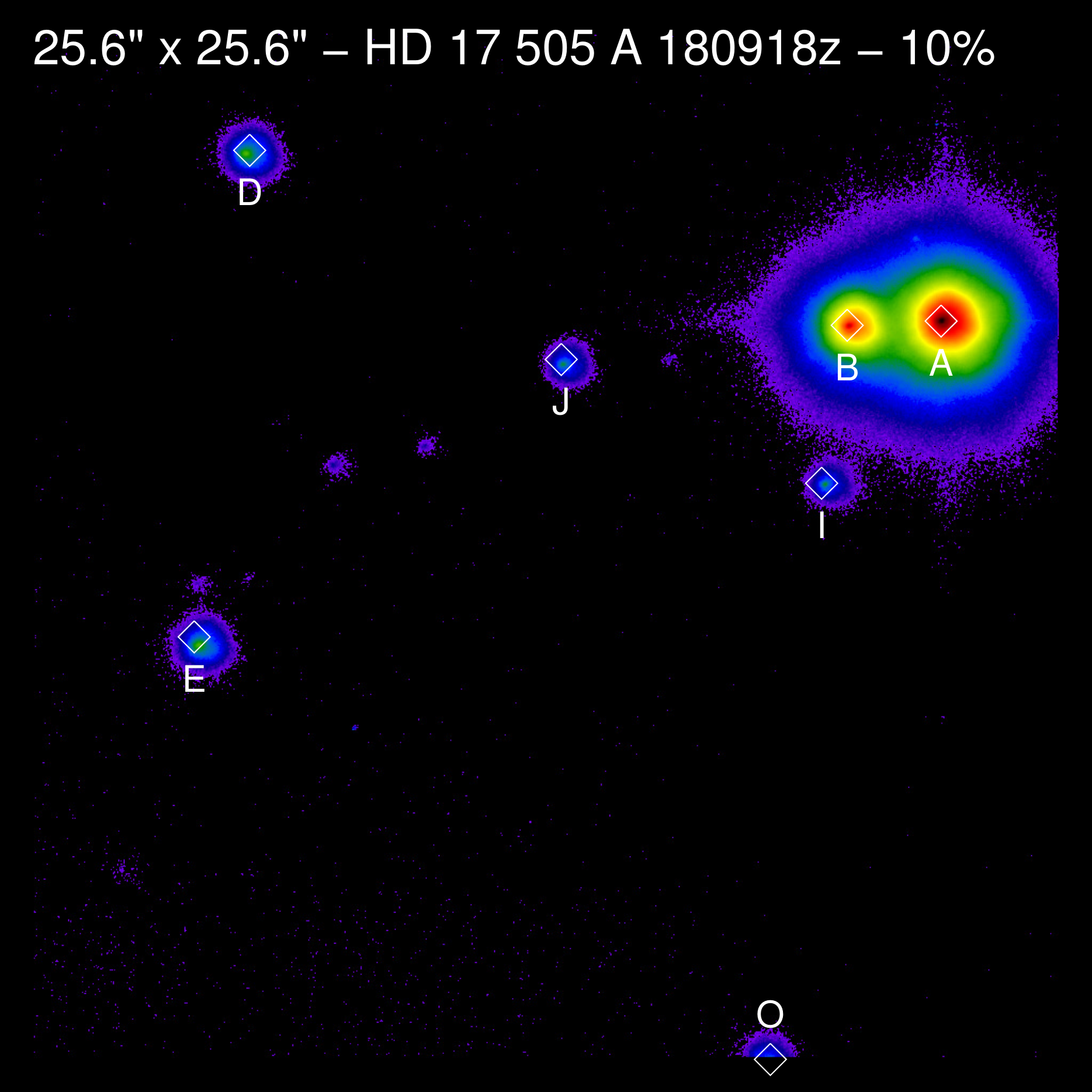}
\end{minipage} $\;\;\;$
\begin{minipage}[t]{0.49\textwidth}
$\!\,$ \linebreak
\includegraphics*[width=\linewidth]{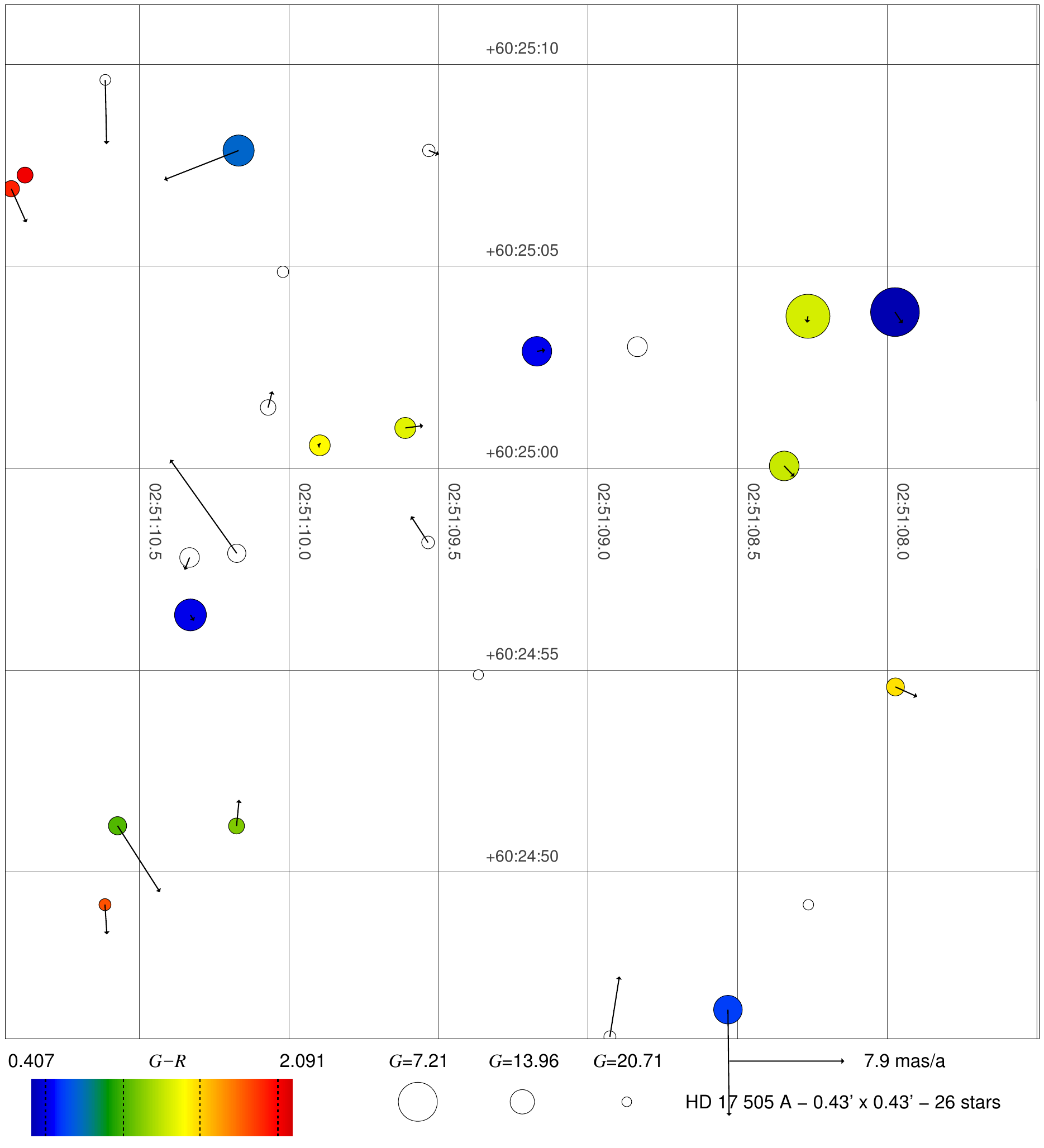}
\end{minipage}
\caption{(left column) Same as Fig.~\ref{AstraLux1} for two additional targets with the full AstraLux field of view. (right column) Charts of the same fields built 
         from Gaia~DR2 sources with magnitude coded as size, $G-G_{\rm RP}$ as color, and arrow size for total proper motion. Empty circles indicate no or bad-quality
         color information.}
\label{AstraLux3}
\end{figure*}

Our main source of information for visual multiplicity is the WDS catalog. In addition, we present new lucky images obtained with AstraLux at the 2.2~m Calar Alto
Telescope, a technique that allows the obtention of images close to the diffraction limit in the $z$ and $i$ bands. The reader is referred to \citet{Maiz10a}, where 
we present a first set of observations with that setup. Since that paper was published, we have obtained further images of many of the systems there. Here we 
present some images (Figs.~\ref{AstraLux1},~\ref{AstraLux2},~and~\ref{AstraLux3}) and results (Table~\ref{AstraLuxdata}) which we will expand in future papers. 
The values for the separations ($\rho$), position angles ($\theta$), and magnitude differences ($\Delta m$) were obtained using a custom-made PSF fitting code 
written in IDL, an optimized version of the one used in \citet{SimDetal15a}. Some of the results presented here are a reanalysis of the ones in \citet{Maiz10a} 
with the new code and represent a significant improvement due to the use of more accurate PSFs for fitting and of Gaia DR2 positions and proper motions for the 
calibration of the astrometric fields.

\begin{table*} 
\centerline{
\begin{tabular}{lcr@{.}lcrcccc}
\hline
Pair              & Even. date & \multicolumn{2}{c}{HJD$-2.4\cdot 10^6$} & $\rho$              & \multicolumn{1}{c}{$\theta$} & $\Delta i$    & $\Delta z$    & $\Delta zn$   & $\Delta Y$    \\
                  & (YYMMDD)   & \multicolumn{2}{c}{(d)}                 & ($^{\prime\prime}$) & \multicolumn{1}{c}{(deg)}    & (mag)         & (mag)         & (mag)         & (mag)         \\
\hline
HD 190\,967 A,B   & 180920     & $\;\;\;$58\,382&3                       & 1.124$\pm$0.006     & 145.14$\pm$0.21              & 5.20$\pm$0.01 & 4.80$\pm$0.01 & \ldots        & \ldots        \\ 
HD 191\,201 A,B   & 110915     & $\;\;\;$55\,819&9                       & 0.984$\pm$0.001     &  83.32$\pm$0.01              & \ldots        & 1.78$\pm$0.03 & \ldots        & \ldots        \\ 
                  & 121001     & $\;\;\;$56\,201&9                       & 0.983$\pm$0.001     &  83.36$\pm$0.03              & \ldots        & 1.81$\pm$0.01 & \ldots        & \ldots        \\ 
                  & 130916     & $\;\;\;$56\,551&9                       & 0.980$\pm$0.001     &  83.51$\pm$0.01              & 1.78$\pm$0.02 & \ldots        & \ldots        & \ldots        \\ 
HD 193\,322 Aa,Ab & 080614     & $\;\;\;$54\,632&1                       & 0.065$\pm$0.001     & 113.45$\pm$2.21              & \ldots        & 0.16$\pm$0.01 & \ldots        & \ldots        \\ 
                  & 110913     & $\;\;\;$55\,817&9                       & 0.067$\pm$0.001     & 129.82$\pm$3.18              & \ldots        & 0.23$\pm$0.03 & \ldots        & \ldots        \\ 
                  & 121001     & $\;\;\;$56\,201&9                       & 0.074$\pm$0.003     & 134.55$\pm$1.89              & \ldots        & 0.14$\pm$0.01 & \ldots        & \ldots        \\ 
                  & 130915     & $\;\;\;$56\,550&9                       & 0.066$\pm$0.002     & 136.12$\pm$2.31              & 0.32$\pm$0.06 & 0.19$\pm$0.02 & \ldots        & \ldots        \\ 
                  & 180920     & $\;\;\;$58\,381&8                       & 0.067$\pm$0.001     & 150.55$\pm$1.00              & \ldots        & 0.22$\pm$0.03 & \ldots        & \ldots        \\ 
HD 193\,322 Aa,B  & 080614     & $\;\;\;$54\,632&1                       & 2.719$\pm$0.003     & 245.22$\pm$0.08              & \ldots        & 1.63$\pm$0.01 & \ldots        & \ldots        \\ 
                  & 110913     & $\;\;\;$55\,817&9                       & 2.728$\pm$0.003     & 245.01$\pm$0.08              & \ldots        & 1.65$\pm$0.01 & \ldots        & \ldots        \\ 
                  & 121001     & $\;\;\;$56\,201&9                       & 2.744$\pm$0.003     & 244.80$\pm$0.08              & \ldots        & 1.61$\pm$0.01 & \ldots        & \ldots        \\ 
                  & 130915     & $\;\;\;$56\,550&9                       & 2.735$\pm$0.003     & 245.00$\pm$0.08              & 1.72$\pm$0.02 & 1.64$\pm$0.02 & \ldots        & \ldots        \\ 
                  & 180920     & $\;\;\;$58\,381&8                       & 2.758$\pm$0.003     & 244.89$\pm$0.08              & \ldots        & 1.65$\pm$0.02 & \ldots        & \ldots        \\ 
HD 194\,649 A,B   & 121002     & $\;\;\;$56\,202&9                       & 0.399$\pm$0.004     & 212.12$\pm$0.04              & \ldots        & 0.91$\pm$0.05 & \ldots        & \ldots        \\ 
                  & 130916     & $\;\;\;$56\,552&0                       & 0.399$\pm$0.001     & 212.53$\pm$0.13              & 0.99$\pm$0.06 & \ldots        & \ldots        & \ldots        \\ 
                  & 181128     & $\;\;\;$58\,450&8                       & 0.400$\pm$0.002     & 213.11$\pm$0.27              & 0.96$\pm$0.01 & 0.93$\pm$0.02 & \ldots        & \ldots        \\ 
ALS 15\,133 A,B   & 181128     & $\;\;\;$58\,450&8                       & 4.367$\pm$0.001     & 214.60$\pm$0.02              & \ldots        & 4.53$\pm$0.03 & \ldots        & \ldots        \\ 
Cyg OB2-A11 A,B   & 181128     & $\;\;\;$58\,450&8                       & 2.222$\pm$0.001     & 276.76$\pm$0.02              & \ldots        & 4.53$\pm$0.02 & \ldots        & \ldots        \\ 
Cyg OB2-5 A,B     & 071113     & $\;\;\;$54\,418&3                       & 0.931$\pm$0.008     &  54.92$\pm$0.12              & 2.75$\pm$0.07 & 2.96$\pm$0.08 & \ldots        & \ldots        \\ 
                  & 181128     & $\;\;\;$58\,451&3                       & 0.932$\pm$0.008     &  55.28$\pm$0.03              & \ldots        & \ldots        & 3.00$\pm$0.03 & 3.06$\pm$0.04 \\ 
Cyg OB2-22 A,Ba   & 110915     & $\;\;\;$55\,819&9                       & 1.530$\pm$0.001     & 146.16$\pm$0.02              & \ldots        & 0.63$\pm$0.02 & \ldots        & \ldots        \\ 
                  & 130919     & $\;\;\;$56\,554&9                       & 1.526$\pm$0.002     & 146.20$\pm$0.02              & \ldots        & 0.62$\pm$0.07 & \ldots        & \ldots        \\ 
                  & 180919     & $\;\;\;$58\,381&0                       & 1.525$\pm$0.001     & 146.07$\pm$0.05              & \ldots        & 0.66$\pm$0.01 & \ldots        & \ldots        \\ 
                  & 181228     & $\;\;\;$58\,450&8                       & 1.524$\pm$0.003     & 145.97$\pm$0.06              & \ldots        & 0.66$\pm$0.01 & \ldots        & \ldots        \\ 
Cyg OB2-22 Ba,Bb  & 110915     & $\;\;\;$55\,819&9                       & 0.217$\pm$0.003     & 180.28$\pm$1.00              & \ldots        & 2.50$\pm$0.14 & \ldots        & \ldots        \\ 
                  & 130919     & $\;\;\;$56\,554&9                       & 0.206$\pm$0.019     & 178.96$\pm$1.94              & \ldots        & 2.53$\pm$0.38 & \ldots        & \ldots        \\ 
                  & 180919     & $\;\;\;$58\,381&0                       & 0.191$\pm$0.024     & 174.15$\pm$5.44              & \ldots        & 2.16$\pm$0.13 & \ldots        & \ldots        \\ 
                  & 181228     & $\;\;\;$58\,450&8                       & 0.215$\pm$0.009     & 181.73$\pm$3.18              & \ldots        & 2.33$\pm$0.08 & \ldots        & \ldots        \\ 
Cyg OB2-1 A,B     & 130919     & $\;\;\;$56\,554&9                       & 1.174$\pm$0.001     & 341.90$\pm$0.05              & \ldots        & 2.66$\pm$0.01 & \ldots        & \ldots        \\ 
Cyg OB2-8 Aa,Ac   & 110915     & $\;\;\;$55\,820&3                       & 3.110$\pm$0.050     & 339.85$\pm$0.10              & \ldots        & 5.65$\pm$0.10 & \ldots        & \ldots        \\ 
                  & 180920     & $\;\;\;$58\,832&3                       & 3.106$\pm$0.050     & 339.89$\pm$0.10              & \ldots        & 5.83$\pm$0.10 & \ldots        & \ldots        \\ 
HD 206\,267 Aa,Ab & 110913     & $\;\;\;$55\,818&4                       & 0.091$\pm$0.006     & 220.64$\pm$6.27              & \ldots        & 1.63$\pm$0.30 & \ldots        & \ldots        \\ 
                  & 130916     & $\;\;\;$56\,552&6                       & 0.112$\pm$0.007     & 203.58$\pm$1.59              & 1.95$\pm$0.30 & \ldots        & \ldots        & \ldots        \\ 
                  & 181227     & $\;\;\;$58\,450&3                       & 0.105$\pm$0.007     & 196.18$\pm$9.00              & \ldots        & \ldots        & 2.02$\pm$0.06 & \ldots        \\ 
HD 206\,267 Aa,B  & 110913     & $\;\;\;$55\,818&4                       & 1.803$\pm$0.010     & 319.30$\pm$0.20              & \ldots        & 5.72$\pm$0.13 & \ldots        & \ldots        \\ 
                  & 130916     & $\;\;\;$56\,552&6                       & 1.778$\pm$0.010     & 319.74$\pm$0.20              & 5.73$\pm$0.25 & \ldots        & \ldots        & \ldots        \\ 
                  & 181227     & $\;\;\;$58\,450&3                       & 1.787$\pm$0.010     & 319.43$\pm$0.20              & \ldots        & \ldots        & 5.55$\pm$0.07 & \ldots        \\ 
ALS 12\,502 A,B   & 130917     & $\;\;\;$56\,553&484                     & 1.578$\pm$0.002     & 345.97$\pm$0.03              & 2.98$\pm$0.02 & 2.96$\pm$0.01 & \ldots        & \ldots        \\ 
                  & 181128     & $\;\;\;$58\,451&359                     & 1.587$\pm$0.004     & 345.88$\pm$0.13              & 3.16$\pm$0.02 & 3.12$\pm$0.03 & \ldots        & \ldots        \\ 
\hline
\end{tabular}
}
 \caption{Measurements for visual pairs using our AstraLux lucky images. The evening date, Heliocentric Julian Date (HJD), separation ($\rho$), position angle ($\theta$), and magnitude difference is given in each case.
         Four different filters were used: SDSS $i$ and $z$, $zn$ (a narrow filter with a central wavelength similar to that of $z$), and $Y$.}
\label{AstraLuxdata}
\end{table*}

\addtocounter{table}{-1}

\begin{table*} 
\centerline{
\begin{tabular}{lcr@{.}lcrcccc}
\hline
Pair              & Even. date & \multicolumn{2}{c}{HJD$-2.4\cdot 10^6$} & $\rho$              & \multicolumn{1}{c}{$\theta$} & $\Delta i$    & $\Delta z$    & $\Delta zn$   & $\Delta Y$    \\
                  & (YYMMDD)   & \multicolumn{2}{c}{(d)}                 & ($^{\prime\prime}$) & \multicolumn{1}{c}{(deg)}    & (mag)         & (mag)         & (mag)         & (mag)         \\
\hline
DN Cas A,B        & 181128     & $\;\;\;$58\,451&4                       & 1.073$\pm$0.011     & 128.21$\pm$0.17              & 4.96$\pm$0.07 & 4.68$\pm$0.16 & \ldots        & \ldots        \\ 
HD 16\,429 Aa,Ab  & 080118     & $\;\;\;$54\,483&9                       & 0.290$\pm$0.001     &  90.80$\pm$0.12              & \ldots        & 2.16$\pm$0.07 & \ldots        & \ldots        \\ 
                  & 110913     & $\;\;\;$55\,818&1                       & 0.282$\pm$0.004     &  90.62$\pm$0.52              & \ldots        & 2.19$\pm$0.07 & \ldots        & \ldots        \\ 
                  & 130918     & $\;\;\;$56\,554&2                       & 0.277$\pm$0.002     &  91.35$\pm$0.28              & 2.33$\pm$0.23 & \ldots        & \ldots        & \ldots        \\ 
                  & 181127     & $\;\;\;$58\,449&9                       & 0.275$\pm$0.006     &  91.17$\pm$0.30              & \ldots        & \ldots        & 2.26$\pm$0.07 & 2.29$\pm$0.07 \\ 
HD 16\,429 Aa,B   & 080118     & $\;\;\;$54\,483&9                       & 6.794$\pm$0.002     & 189.84$\pm$0.03              & \ldots        & 2.16$\pm$0.07 & \ldots        & \ldots        \\ 
                  & 110913     & $\;\;\;$55\,818&1                       & 6.817$\pm$0.001     & 189.83$\pm$0.03              & \ldots        & 2.19$\pm$0.07 & \ldots        & \ldots        \\ 
                  & 130918     & $\;\;\;$56\,554&2                       & 6.841$\pm$0.002     & 189.76$\pm$0.03              & 2.33$\pm$0.23 & \ldots        & \ldots        & \ldots        \\ 
                  & 181127     & $\;\;\;$58\,449&9                       & 6.859$\pm$0.001     & 189.84$\pm$0.03              & \ldots        & \ldots        & 2.26$\pm$0.07 & 2.29$\pm$0.07 \\ 
HD 16\,429 Aa,D   & 080118     & $\;\;\;$54\,483&9                       & 2.976$\pm$0.001     & 112.92$\pm$0.04              & \ldots        & 7.51$\pm$0.07 & \ldots        & \ldots        \\ 
                  & 110913     & $\;\;\;$55\,818&1                       & 2.973$\pm$0.001     & 113.02$\pm$0.04              & \ldots        & 7.42$\pm$0.07 & \ldots        & \ldots        \\ 
                  & 130918     & $\;\;\;$56\,554&2                       & 2.968$\pm$0.001     & 113.01$\pm$0.07              & 7.62$\pm$0.07 & \ldots        & \ldots        & \ldots        \\ 
                  & 181127     & $\;\;\;$58\,449&9                       & 2.964$\pm$0.021     & 113.03$\pm$0.56              & \ldots        & \ldots        & 7.54$\pm$0.07 & 7.78$\pm$0.07 \\ 
HD 17\,505 A,B    & 110915     & $\;\;\;$55\,820&6                       & 2.161$\pm$0.001     &  93.03$\pm$0.04              & \ldots        & 1.76$\pm$0.01 & \ldots        & \ldots        \\ 
                  & 180919     & $\;\;\;$58\,381&6                       & 2.163$\pm$0.002     &  93.09$\pm$0.03              & \ldots        & 1.75$\pm$0.03 & \ldots        & \ldots        \\ 
MY Cam A,B        & 130920     & $\;\;\;$56\,556&6                       & 0.725$\pm$0.004     & 141.93$\pm$0.06              & \ldots        & 2.77$\pm$0.03 & \ldots        & \ldots        \\ 
                  & 180918     & $\;\;\;$58\,380&6                       & 0.738$\pm$0.006     & 141.69$\pm$0.14              & 3.03$\pm$0.11 & 2.83$\pm$0.03 & \ldots        & \ldots        \\ 
IU Aur A,B        & 110915     & $\;\;\;$55\,820&646                     & 0.143$\pm$0.002     & 226.63$\pm$4.00              & \ldots        & 1.74$\pm$0.10 & \ldots        & \ldots        \\ 
                  & 121002     & $\;\;\;$56\,203&596                     & 0.144$\pm$0.001     & 228.61$\pm$4.00              & \ldots        & 1.40$\pm$0.10 & \ldots        & \ldots        \\ 
                  & 181127     & $\;\;\;$58\,450&391                     & 0.141$\pm$0.005     & 230.14$\pm$4.00              & \ldots        & 2.06$\pm$0.10 & \ldots        & \ldots        \\ 
                  & 181128     & $\;\;\;$58\,451&403                     & 0.137$\pm$0.001     & 231.47$\pm$4.00              & \ldots        & 1.84$\pm$0.10 & \ldots        & \ldots        \\ 
                  & 181226     & $\;\;\;$58\,479&446                     & 0.128$\pm$0.002     & 211.56$\pm$4.00              & \ldots        & 1.77$\pm$0.10 & \ldots        & \ldots        \\ 
                  & 181227     & $\;\;\;$58\,480&485                     & 0.131$\pm$0.001     & 221.41$\pm$4.00              & \ldots        & 1.82$\pm$0.10 & \ldots        & \ldots        \\ 
IU Aur A,C        & 181127     & $\;\;\;$58\,450&391                     & 4.012$\pm$0.005     & 162.36$\pm$0.08              & \ldots        & 7.85$\pm$0.10 & \ldots        & \ldots        \\ 
IU Aur A,D        & 181127     & $\;\;\;$58\,450&391                     & 3.592$\pm$0.010     & 122.21$\pm$0.12              & \ldots        & 8.87$\pm$0.10 & \ldots        & \ldots        \\ 
15 Mon Aa,Ab      & 080117     & $\;\;\;$54\,482&9                       & 0.109$\pm$0.004     & 252.26$\pm$1.29              & \ldots        & 1.49$\pm$0.13 & \ldots        & \ldots        \\ 
                  & 121002     & $\;\;\;$56\,203&1                       & 0.118$\pm$0.004     & 260.30$\pm$3.80              & \ldots        & 1.45$\pm$0.17 & 1.49$\pm$0.01 & \ldots        \\ 
                  & 130920     & $\;\;\;$56\,556&2                       & 0.127$\pm$0.004     & 259.51$\pm$2.62              & \ldots        & 1.26$\pm$0.20 & \ldots        & \ldots        \\ 
                  & 180918     & $\;\;\;$58\,380&2                       & 0.138$\pm$0.004     & 268.10$\pm$1.00              & \ldots        & \ldots        & 1.67$\pm$0.10 & \ldots        \\ 
                  & 181128     & $\;\;\;$58\,451&0                       & 0.135$\pm$0.004     & 268.79$\pm$1.00              & \ldots        & \ldots        & 1.48$\pm$0.06 & \ldots        \\ 
15 Mon Aa,B       & 080117     & $\;\;\;$54\,482&9                       & 2.977$\pm$0.003     & 213.58$\pm$0.02              & \ldots        & 3.04$\pm$0.02 & \ldots        & \ldots        \\ 
                  & 121002     & $\;\;\;$56\,203&1                       & 2.981$\pm$0.003     & 213.75$\pm$0.02              & \ldots        & 3.04$\pm$0.02 & 3.02$\pm$0.03 & \ldots        \\ 
                  & 130920     & $\;\;\;$56\,556&2                       & 2.985$\pm$0.003     & 213.88$\pm$0.05              & \ldots        & 2.97$\pm$0.10 & \ldots        & \ldots        \\ 
                  & 180918     & $\;\;\;$58\,380&2                       & 3.001$\pm$0.007     & 214.02$\pm$0.02              & \ldots        & \ldots        & 3.15$\pm$0.10 & \ldots        \\ 
                  & 181128     & $\;\;\;$58\,451&0                       & 2.991$\pm$0.003     & 213.94$\pm$0.04              & \ldots        & \ldots        & 2.99$\pm$0.02 & \ldots        \\ 
HD 52\,533 Aa,Ab  & 081021     & $\;\;\;$54\,761&2                       & 0.634$\pm$0.015     & 269.03$\pm$1.34              & \ldots        & 3.48$\pm$0.12 & \ldots        & \ldots        \\ 
HD 52\,533 Aa,B   & 081021     & $\;\;\;$54\,761&2                       & 2.640$\pm$0.006     & 187.83$\pm$0.19              & \ldots        & 5.68$\pm$0.29 & \ldots        & \ldots        \\ 
HD 52\,533 Aa,G   & 081021     & $\;\;\;$54\,761&2                       & 2.883$\pm$0.018     & 246.86$\pm$0.41              & \ldots        & 8.18$\pm$0.43 & \ldots        & \ldots        \\ 
\hline
\end{tabular}
}
\caption{(continued.)}
\end{table*}

\subsection{Spectroscopic/eclipsing multiplicity nomenclature}

$\,\!$\indent The previous subsections indicate that studying multiplicity in massive stars is complicated by the different possible orbital configurations and what is
detected through imaging/interferometry (visual systems), long-slit spectroscopy (GOSSS), and high-resolution spectroscopy (LiLiMaRlin and others). To clarify the status
of each of the spectroscopic systems in this paper, we have devised a spectroscopic binarity status (SBS) nomenclature that expands on the traditional notation and a 
series of diagrams (Fig.~\ref{sbstatus}) to accompany it. We use this nomenclature to classify each system based on the characteristics of the components that fall 
within the high-resolution spectroscopic aperture.

\begin{itemize}
 \item We start with the traditional notation for systems with two components: {\bf SB1} for single-lined spectroscopic binaries and {\bf SB2} for double-lined ones. 
       An {\bf E} is used for eclipsing binaries or added to SB1/SB2 if the system is also spectroscopic.
 \item If the two components of the pair are spatially separated with imaging/interferometry, we add an {\bf a} (astrometric or visual) to the SBS as in the SB2a case 
       in Fig.~\ref{sbstatus}.
 \item If a third component in an orbit around the inner pair falls within the high-resolution spectroscopic aperture, we add it after a plus sign. There are three
       possibilities: {\bf C} (constant) indicates that the third component is detected in the spectrum but is too distant to induce significant radial velocity variations in 
       the inner pair (or viceversa)\footnote{We include here the cases where the motion of the outer orbit can be detected but only through astrometry or light-time 
       effects.}, {\bf S} (spectroscopic) is used when such radial velocity variations are detected but the third component appears to be single itself, and {\bf SB1/SB2} are 
       used if a short-period orbit is detected for the third component itself (implying a fourth component). The nomenclature can get more complicated if three 
       spectroscopic orbits are detected but that does not happen for the sample here, at least at this point.
 \item If the third component is spatially separated with imaging/interferometry, we add an {\bf a} to the SBS as we did for the inner-orbit components.
 \item Finally, if the third component is spatially separated in GOSSS data, we add an {\bf s} to the SBS.
\end{itemize}

We identify 17 different SBS configurations in the 92 targets in our sample, as shown in the diagrams and listings in Fig.~\ref{sbstatus}. See the caption there for 
the meanings of the different symbols and lines used. When discussing each individual system below, we first indicate its current SBS.

\section{Individual systems}

$\,\!$\indent In this section we discuss the spectral classification, multiplicity, and special characteristics of each system.
The stars are sorted by Galactic longitude and grouped in regions of the sky. Each system has one paragraph that starts with the most common names used in the 
literature and its spectroscopic/eclipsing multiplicity status. If the status changes as a result of this paper, we indicate it in parenthesis. 

\subsection{Sagittarius-Sagitta}

\paragraph{HD~164\,438 = BD~$-$19~4800 = ALS~4567.}                                                
\textbf{SB1.}
\citet{Mayeetal17} identified this system as a SB1, which was classified as O9.2~IV in GOSSS~II.
HD~164\,438 has no WDS entry. M98 detected a possible companion with a 0\farcs05 separation indicating that confirmation was needed but \citet{Sanaetal14} 
did not detect it. It also appears single in our AstraLux data, hence the SB1 typing.

\paragraph{HD~167\,771 = BD~$-$18~4886 = ALS~4874.}                                                
\textbf{SB2.}
\citet{MorrCont78} classified this system as O7~III((f))~+~O9~III and was detected as a SB2 in GOSSS~I and in GOSSS~II we classified it as 
O7~III((f))~+~O8~III. A LiLiMaRlin epoch yields the same spectral classification.
The WDS catalog lists two visual components (Aa~and~Ab) with a small $\Delta m$ and a separation of 
0\farcs1. However, we have not included the components in the name because of the uncertain nature of that visual binary: there is only 
a single detection from 1923 and the modern attempts by \citet{Turnetal08} and \citet{Maiz10a} did not detect the alleged Ab companion 
(but the second paper detected the distant dim B companion).   

\paragraph{BD~$-$16~4826 = ALS~4944.}                                                              
\textbf{SB2 (previously SB1).}
The most recent multi-epoch study for this target by \citet{Willetal13} gave a SB1 orbital solution for this member of the 
NGC~6618 cluster. They only had a small number of epochs and covered a short period of time (less than an orbital period), making its 
orbital parameters only preliminary. The object was analyzed in GOSSS~III, where we gave it a spectral type of O5.5~V((f))z and 
noted that the lines were asymmetric, indicating the 
possibility of assigning a SB2 character with better data. In some of the LiLiMaRlin data we see double lines and we use one of the
epochs to give the first ever SB2 spectral classification for this object as O5~V((f))z~+~O9/B0~V. The object has no entry in the WDS 
catalog. 
 
\paragraph{HD~170\,097~A = V2349~Sgr = ALS~5061 = BD~$-$16~4888~A.}                                
\textbf{SB2E (previously E).}
This object was known to be an eclipsing binary \citep{Dvor04} but had only received previous spectral classifications as a B star.
It had not been previously observed by GOSSS and the spectrogram shown on Fig.~\ref{GOSSS} yields a SB2 spectral classification of
O9.5~V~+~B1:~V, with the uncertainty in the secondary classification caused by the small radial velocity difference and large $\Delta m$
between the two spectroscopic components. HD~170\,097~A also appears as SB2 in some LiLiMaRlin epochs and we derive the same spectral
types from them. Unpublished OWN data suggests the possible existence of a third spectroscopic component. There is a bright visual 
companion 17\farcs2 away (HD~170\,097~B) with its own ALS number (ALS~5060) but the system has no entry in the WDS catalog. 

\paragraph{QR~Ser = HD~168\,183 = BD~$-$14~4991 = ALS~4916.}                                       
\textbf{SB2E.}
This object was classified by \citet{Sanaetal09} as O9.5~III~+~B, with the secondary having an uncertain spectral subtype (likely mid-B)
based on a weak detection of \HeI{5876}. It has not previously appeared in a GOSSS paper. We cannot detect the
weak B-type companion in our spectra and we assign it a composite spectral type of O9.7~III. QR~Ser has no entry in the WDS catalog. 

\paragraph{V479~Sct = ALS~5039.}                                                                   
\textbf{SB1.}
\citet{McSwetal04} determined the SB1 orbit of this $\gamma$-ray binary, which is also a runaway star \citep{Maizetal18b}. We classified
it as ON6~V((f))z in GOSSS~III. V479~Sct has no WDS entry.

\paragraph{HD~168\,075 = BD~$-$13 4925 = ALS~4907.}                                                
\textbf{SB2.}
This spectroscopic binary has a long period of 43.627~d and has been previously classified as O6.5~V((f))~+~B0-1~V 
\citep{Sanaetal09,Barbetal10}. In GOSSS-III we 
obtained the same spectral type for the primary but we were unable to detect the secondary. The WDS catalog lists several components but 
caution should be used when assigning names, as the WDS entry 18186$-$1348 also includes another O-type system (HD~168\,076) in M16 at a 
distance of 26\arcsec\ away with its own BD and ALS entries. WDS~18186$-$1348~A corresponds to HD~168\,076~A (with a bright B component 
unresolved in most observations 0\farcs1 away) while WDS~18186$-$1348~H corresponds to HD~168\,075~A. The WDS catalog lists an unresolved 
companion to HD~168\,075~A (Ab or WDS~18186$-$1348~Hb) but with a $\Delta m$ larger than 3 mag \citep{Sanaetal14}, making it too faint to 
contribute to the combined spectrum. There are also three additional faint components within 6\arcsec\ of HD~168\,075~A. 

\paragraph{HD~168\,137~AaAb = BD~$-$13~4932 AaAb = ALS~4915~AaAb.}                                 
\textbf{SB2a.}
\citet{Sanaetal09} gave a spectral classification of O7~V~+~O8~V for this SB2 system. In GOSSS-III we could not detect the two components 
and derived a combined spectral type of O8~Vz. Using LiLiMaRlin data we derive new spectral types of O7.5~Vz~+~O8.5~V.
The WDS catalog lists a bright Ab component at a short distance (hence the designation 
AaAb for the system) which is likely the same component detected by \citet{LeBoetal17} with VLTI. It is likely that this visual Ab 
component is also the spectroscopic component detected by \citet{Sanaetal09}, given the extreme eccentricity of 0.9 of the preliminary 
orbit. 

\paragraph{BD~$-$13~4923 = ALS~4905.}                                                              
\textbf{SB2.}
\citet{Sanaetal09} gave a spectral classification of O4~V((f))~+~O7.5~V for this object. BD~$-$13~4923 had not been previously included in 
GOSSS and here we present a spectrogram that yields the same spectral type as those authors. We have also found a LiLiMaRlin epoch appropriate for 
spectral classification with MGB and found again the same spectral type. The WDS catalog lists a third component 0\farcs7 away with a 
$\Delta m$ larger than 3 magnitudes which we also see in AstraLux images.

\paragraph{MY~Ser~AaAb = HD~167\,971~AaAb = BD~$-$12~4980~AaAb = ALS~4894~AaAb.}                   
\textbf{SB2E+Sa.}
\citet{Leitetal87} derived spectral types of O8~If for the primary star and O5-8 for each of the secondary and tertiary components of 
this SB3 system. The latter two are an eclipsing binary. This complex system was also studied by \citet{Ibanetal13}, who found periods
of 3.321\,616~d and 21.2~a for the inner and outer orbits, respectively, and a large eccentricity of 0.53 for the outer orbit. They also derive
spectral types of O9.5-B0~III–I~+~O7.5~III~+~O9.5~III. In GOSSS-I we separated two of the kinematic components into O8~Iaf(n) and O4/5. With new 
GOSSS spectra and a reanalysis of the old spectrograms we are now able to resolve the three components as O8~Iaf~+~O4/5~If~+~O4/5~V-III. 
With LiLiMaRlin data we obtain a similar spectral classification of O8~Iabf~+~O4.5~If~+~O4:~V-III. The spectral type of the primary is relatively 
well established but those of the secondary and tertiary are more uncertain. The secondary has strong \HeII{4686} emission, hence the supergiant 
luminosity class (even though it is not clear if the emission arises on a typical supergiant wind or is the result of the interaction between the 
two close components). Our classifications are incompatible with the previous one of \citet{Ibanetal13}, as both the secondary and tertiary
have \HeII{4542} stronger than \HeI{4471} and, hence, they have to be earlier (not later) than O7. On the other hand, the earlier (and more imprecise)
classifications of \citet{Leitetal87} are closer to ours. \citet{Sanaetal14} detect two astrometric components (Aa and Ab) with a small magnitude 
difference and a separation of 17~mas. These are likely to be the spectroscopic primary on the one hand and the combination of secondary and 
tertiary spectroscopic components on the other hand, as the $\Delta m$ derived from the spectral fitting is also close to zero. The latter work of 
\citet{LeBoetal17} elaborate on this idea and present an astrometric outer orbit based on a yet incomplete arc (note that the previous outer orbit by 
\citet{Ibanetal13} was based on eclipse timings). Our LiLiMaRlin spectra span 9~a (from 2006 to 2015) and show small radial velocities differences, hence 
we type this system as SB2E+Sa instead of SB2E+Ca. The WDS catalog also lists a large number of dim companions, some of them possibly bound but the 
rest likely members of the NGC~6604 cluster, of which the MY~Ser system is the brightest member. 

\paragraph{HD~168\,112~AB = BD~$-$12~4988 = ALS~4912~AB.}                                          
\textbf{SB2a.}
The first orbital solution for this system is given in the OWN orbit paper. HD~168\,112~AB was classified as a single star in GOSSS~I~and~III and
had no previous separate spectral classifications but a new GOSSS spectrogram yields the two spectral types O5~IV(f)~+~O6:~IV:. 
Alternatively, with LiLiMaRlin data we obtain O4.5 III(f)~+~O5.5~IV((f)). \citet{Sanaetal14} 
detected two components (A and B) with a small $\Delta m$ and a separation of just over 3~mas. Given the long spectroscopic period of the 
system it is likely that the A and B visual components are also the spectroscopic ones. \citet{Sanaetal14} also detected two more distant 
and faint C and D companions (note that the current version of the WDS catalog follows a different component convention for this system). 
    
\paragraph{HD 166\,734 = V411~Ser = BD~$-$10~4625 = ALS~9405.}                                     
\textbf{SB2E.}
\citet{Walb73a} classified this eccentric system as O7~Ib(f)~+~O8-9~I. In GOSSS~I we derived a combined spectral type of O7.5~Iabf and 
here we present a new observation where the two components are resolved and we obtain a classification of O7.5~Iaf~+~O9~Iab. With 
LiLiMaRlin data we obtain spectral types of O7.5 Iaf~+~O8.5~Ib(f). This target 
has no entry in the WDS catalog and no significant companion is detected in our AstraLux images or by \citet{Sanaetal14}. 
    
\paragraph{HD~175\,514 = V1182~Aql = BD~$+$09~3928 = ALS~10\,048.}                                 
\textbf{SB2E+C.}
\cite{Mayeetal05} studied this three-object system and gave spectral types of O5.5, O9.5, and O9, noting that the first two stars
form an eclipsing system with a period of 1.621\,861~d. The outer orbit has a period of at least 50~a.
In GOSSS~III we obtained a spectral classification of O7~V(n)((f))z~+~B, 
likely because we are seeing a blend of the two earliest stars (hence the O7 type, intermediate between O5.5 and O9.5) plus the third later-type
star. In one of the LiLiMaRlin epochs we caught the system close to quadrature and derive spectral classifications of 
O5.5 V((f))~+~B0.5:~V+~O7.5~IV((f)), where the two first types correspond to the eclipsing binary (extreme radial velocities) and the last one to the
third object (central radial velocity). 
This target has no entry in the WDS catalog and no significant companion is detected in our AstraLux images. 

\paragraph{9~Sge = HD~188\,001 = QZ~Sge = BD~$+$18~4276 = ALS~10\,596.}                            
\textbf{SB1?.}
\citet{UndeMatt95} calculated a SB1 orbit for this runaway star \citep{Maizetal18b}. The orbit has not been confirmed and it is possible
that the radial velocity variations are caused by pulsations. In GOSSS-I we classified it as O7.5~Iabf. 9~Sge has no
entry in the WDS~catalog and our AstraLux images show no companions.

\subsection{Cygnus}

\paragraph{Cyg~X-1 = V1357~Cyg = HDE~226\,868 = BD~$+$34~3815 = ALS~10\,678.}                      
\textbf{SB1.}
This is arguably the most famous O-type SB1 system, as it is consists of an O supergiant orbiting the nearest known black hole. 
\cite{Orosetal11} is the last paper to give an orbital solution for it and also provides accurate values for the distance, masses, and 
inclination. GOSSS~I gives a spectral classification of O9.7~Iabp~var. Cyg~X-1-a has no entry in the WDS catalog and our  
AstraLux images do not show any significant close companion.  

\paragraph{HD~190\,967 = V448~Cyg = BD~$+$34~3871 = ALS~10\,828.}                                  
\textbf{SB2E.}
\citet{Harretal97} give a spectral type of O9.5~V~+~B1~Ib-II for this binary but the origin of that classification is uncertain, as
the work they cite (\citealt{Morgetal55} though they incorrectly say that paper was published in 1965) only gives the classification for the
B-type component. This system is quite peculiar, as it consists of an evolved B supergiant mass donor (the brighter star) and an O main-sequence 
gainer surrounded by a thick accretion disk \citep{Djuretal09}. This target had not been previously observed with GOSSS and the new spectral 
classification is O9.7:~V~+~B1.5~Iab. With a LiLiMaRlin epoch we obtain O9.7:~V~+~B1.5~II, that is, a lower luminosity class for the B-type component.
The object has no entry in the WDS catalog. Our AstraLux images reveal a previously undetected B companion with a large $\Delta m$ 
(Table~\ref{AstraLuxdata} and Fig.~\ref{AstraLux1}). Note that the magnitude difference is significantly smaller in $z$ than in $i$, indicating that the 
companion is redder. A comparison with isochrones with ages of 3 and 10~Ma indicates that the color difference is too large for the magnitude difference, 
so if the system is coeval and bound it is likely that the B visual companion has not reached the ZAMS. Alternatively, the color difference may be distorted by the 
existence of the thick accretion disk around the O-type spectroscopic component in the A visual component. However, that additional extinction around
the A component would change its color in the opposite direction and our high-resolution spectroscopy reveals no emission lines in the $i$ band, so that alternative
explanation does not work.

\paragraph{HD~191\,201 A = BD~$+$35~3970~A = ALS~10\,843~A.}                                       
\textbf{SB2+Cas.}
This object is a hierarchical multiple system formed by an inner spectroscopic object (A) and an external B component at a distance of 
1\farcs0 and a $\Delta m$ of 1.8 magnitudes (between the sum of the two stars in A and B, Fig.~\ref{AstraLux1} and Table~\ref{AstraLuxdata}). 
The system was known to be an O-type SB2 as 
far back as  \citet{Plas26} but it was not until GOSSS~I that it was determined that the B component is also an O star with spectral 
type O9.7~III. The A component has a GOSSS classification of O9.5~III~+~B0~IV. As the system includes three objects of similar spectral 
type within a small aperture, special care has to be taken to ensure what light is entering the spectrograph. For example, \citet{ContAlsc71} classified 
this system as O9~III~+~O9~V and is quite possible that their spectral classification for the secondary included the B component as well as 
the lower luminosity star in A. In GOSSS we fit spatial
profiles to the long-slit data to obtain separate spectrograms and spectral classifications, something not possible with the echelle data,
and that spatial separation is indicated by the ``s'' in the SB2+Cas typing (Fig.~\ref{HD_191_201_HeI_4922}). 
Our LiLiMaRlin data do not show significant radial velocity differences 
for the external component in a 14~a time span (from 2004 to 2018) and the WDS catalog shows little astrometric change in the relative position
of the A-B system over 170~a (with many measurements in between but noisy ones), indicating that the period is measured in millennia. The AstraLux
measurements in Table~\ref{AstraLuxdata} are also consistent with a very long period.

\paragraph{HD~191\,612 = BD~$+$35~3995 = ALS~10\,885.}                                             
\textbf{SB2.}
This is one of the five previously known Of?p stars in the Galaxy (\citealt{Walbetal10a} but see below for $\theta^1$~Ori~CaCb). 
Its magnetic field is the main source of its variability, 
tied to the 538~d rotational period \citep{Donaetal06}. Superimposed on the strong line variations caused by the changes in the relative 
orientation of the magnetic field, \cite{Howaetal07} detected the signal of a double-lined spectroscopic binary with small radial velocity amplitudes 
and a period of 1542~d (later updated to 1548~d by \citealt{Wadeetal11}). Those authors give a spectral type of O6.5-8f?p~var~+~B0-2. Our similar
GOSSS-I classification is O6-8~f?p~var, as the combination of spectral resolution and S/N of the spectrogram there is not
sufficient to reveal the weak lines that originate in the secondary component. This object has no entry in the WDS 
catalog and no apparent companion is seen in our AstraLux images. 

\paragraph{HDE~228\,854 = V382~Cyg = BD~$+$35~4062 = ALS~11\,132.}                                 
\textbf{SB2E.}
\cite{Pear52} classified this short-period binary as O6.5~+~O7.5. The GOSSS~I spectral classification is
O6~IVn~var~+~O5~Vn~var, with variations in the spectral type likely caused by different parts of the star being exposed or eclipsed
during different orbital phases. The WDS catalog lists a companion at a distance of 11\farcs3 (with a single observation from 1896) but our 
AstraLux images show nothing at that position, so it is likely to be a spurious detection.  
    
\paragraph{HDE~228\,766 = BD~$+$36~3991 = ALS~11\,089.}                                            
\textbf{SB2.}
\citet{Walb73a} classified this system as O4~If~+~O8-9~In and in GOSSS~I we refined the classification of the secondary to obtain 
O4~If~+~O8:~II:. \citet{Rauwetal14} claimed that the GOSSS-I classification was incorrect and that the early-type component should be an 
early Of/WN star (or ``hot slash'', see GOSSS-II). However, that goes against the morphological criterion established by \citet{CrowWalb11} 
based on the H$\beta$ profile: if that line is in absorption, the object should be classified as an early Of star; if it has a P-Cygni profile,
as an early Of/WN star; and if it is in emission as a WN star. The H$\beta$ profile of HDE~228\,766 is variable but it never shows a clear
P-Cygni profile. Therefore, even though it is true that other emission lines are strong, the early type component should not be classified 
as an early Of/WN star. This object has no entry in the WDS catalog and no apparent companion is seen in our AstraLux images. 

\paragraph{HD~193\,443~AB = BD~$+$37~3879~AB = ALS~11\,137~AB.}                                    
\textbf{SB2+Ca.}
\cite{Mahyetal13} classified this system as O9~III/I~+~O9.5~V/III. In GOSSS~I~and~II we were only able to obtain a combined spectral 
classification of O9~III. The WDS catalog reveals that the situation is complex, as there is a visual companion at a distance of
0\farcs1-0\farcs2 with a $\Delta m$ of 0.3 magnitudes (hence the designation AB, Fig.~\ref{AstraLux1}) and an appreciable motion in the last 
century. This visual companion is also detected in our AstraLux images. Therefore, there is a third body in the system contributing to the
spectrum besides the two in the short-period binary, making HD~193\,443~AB a hierarchical system with three late-O stars (one of them may
be a B0 instead). 

\paragraph{BD~+36~4063 = ALS~11\,334.}                                                             
\textbf{SB1.}
\citet{Willetal09b} measured the SB1 orbit of this interacting system and argued that the companion is invisible due to a thick and opaque 
disk around it. In GOSSS~I we classified it as ON9.7~Ib. BD~+36~4063 has no WDS entry and appears single in our AstraLux data.

\paragraph{HD~193\,611 = V478~Cyg = BD~$+$37~3890 = ALS~11\,169.}                                  
\textbf{SB2E.}
\citet{Martetal17b} classified this system as O9.5~V~+~O9.5~V. HD~193\,611 had no previous GOSSS classification and here we derive 
O9.5~II~+~O9.5~III. With LiLiMaRlin we derive a similar classification of O9.5~III~+~O9.7~III. The dwarf classification of 
\citet{Martetal17b} is excluded on the basis of the low depth of \HeII{4686} for both components.
The WDS catalog lists a dim component (also seen in our AstraLux images) 3\farcs7 away that does not influence the classification.

\paragraph{HDE~228\,989 = BD~$+$38~4025 = ALS~11\,185.}                                             
\textbf{SB2E.}
\cite{Mahyetal13} determined an O8.5~+~O9.7 spectral classification for this object and \citet{Lauretal17} identified it as an eclipsing
binary. HDE~228\,989 was not previously included in GOSSS and here we derive a spectral classification of O9.5~V~+~B0.2~V. Using a LiLiMaRlin
epoch we derive a similar classification of O9.5~V~+~B0~V. Both classifications are of later type for the two components than the
\citet{Mahyetal13} ones. More specifically, the secondary must be a B star because \HeII{4542} is very weak or absent and \HeII{4686} is 
weak. Also, the \SiIII{4552}/\HeII{4542} for the primary does not allow for a spectral classification as early as O8.5. The system has no 
entry in the WDS catalog and our AstraLux observations show no significant companion. 

\begin{figure*}
\centerline{\begin{tabular}{llllll}
\includegraphics[width=0.15\textwidth]{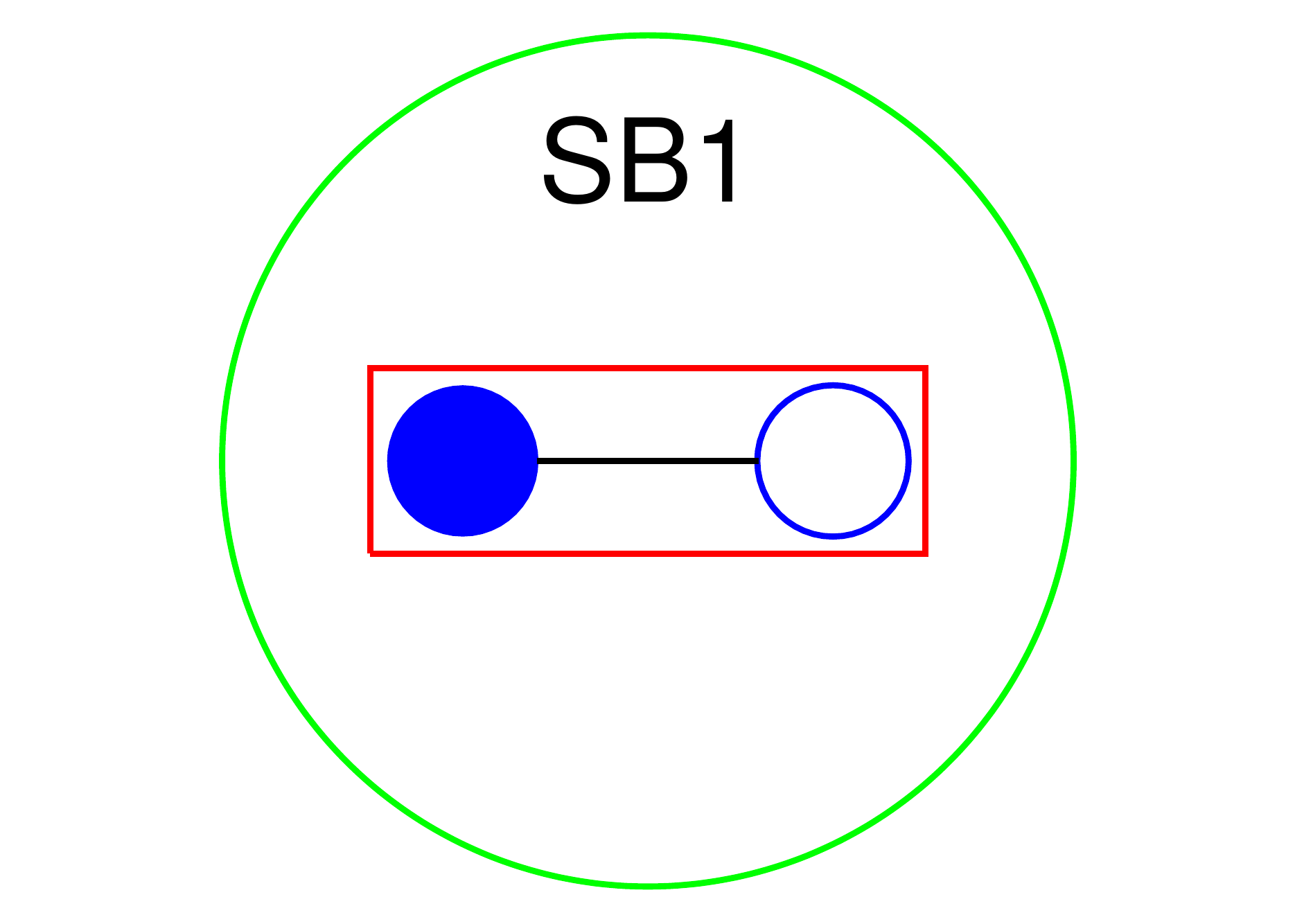}         & \includegraphics[width=0.15\textwidth]{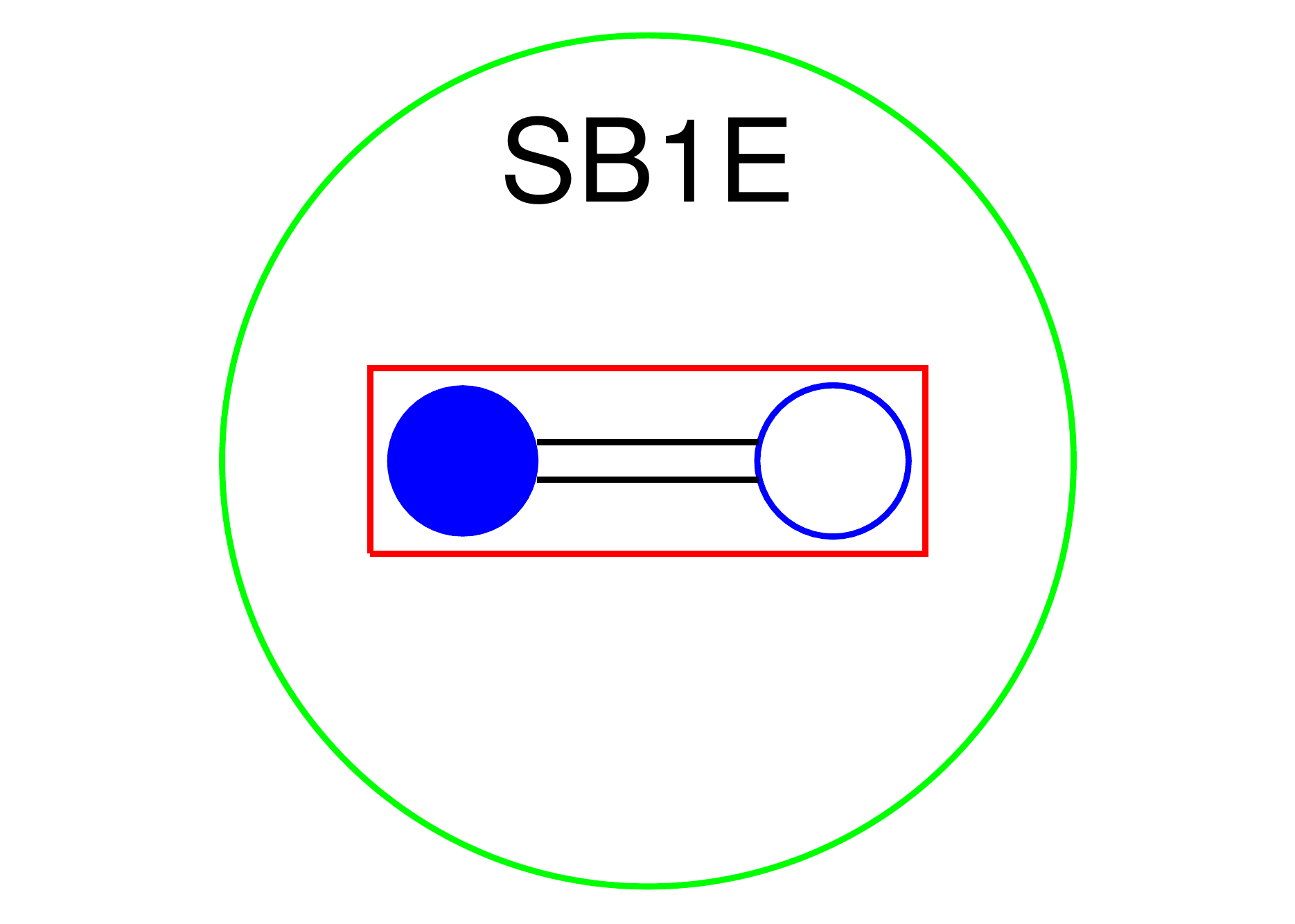}        & \includegraphics[width=0.15\textwidth]{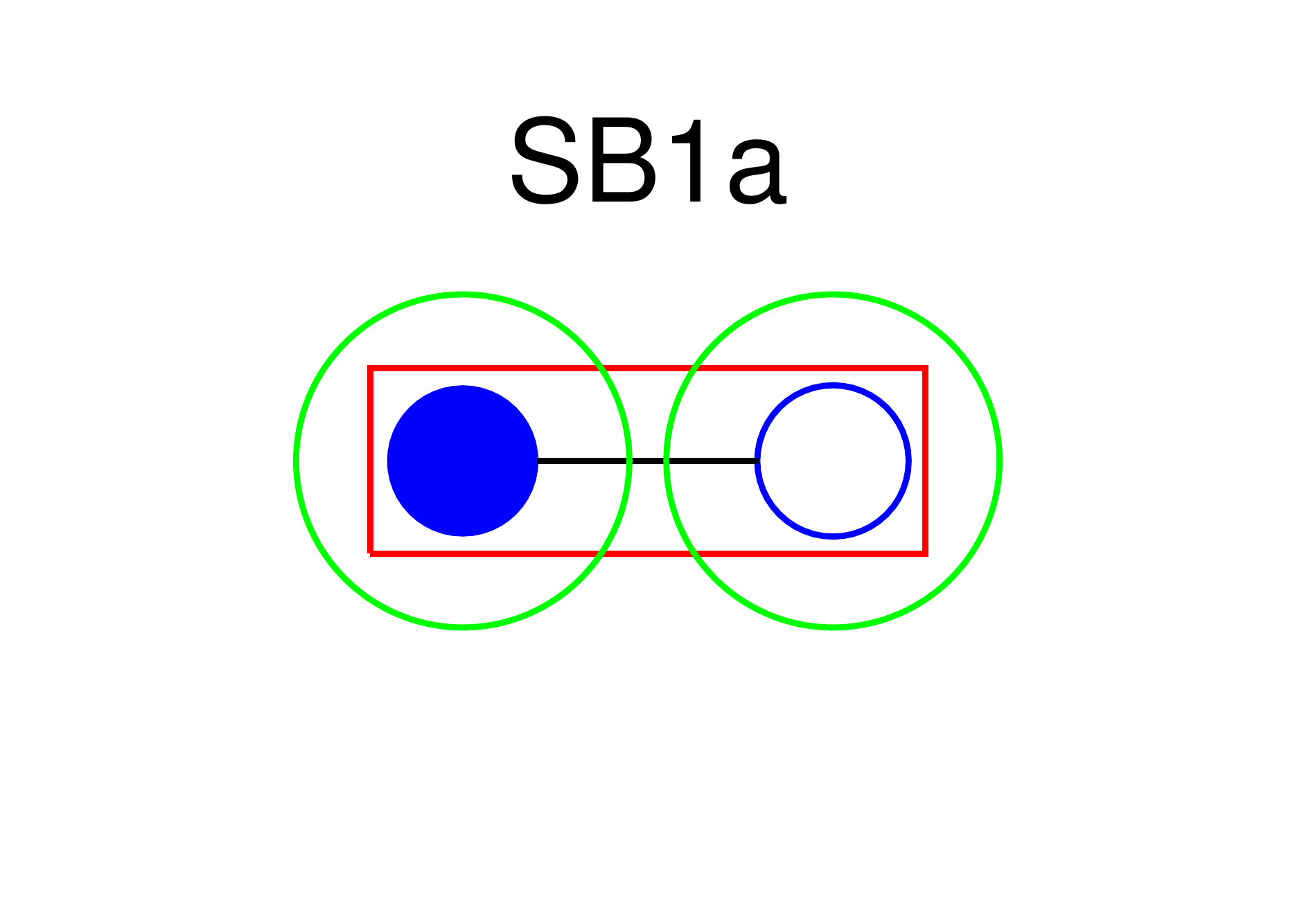}        & \includegraphics[width=0.15\textwidth]{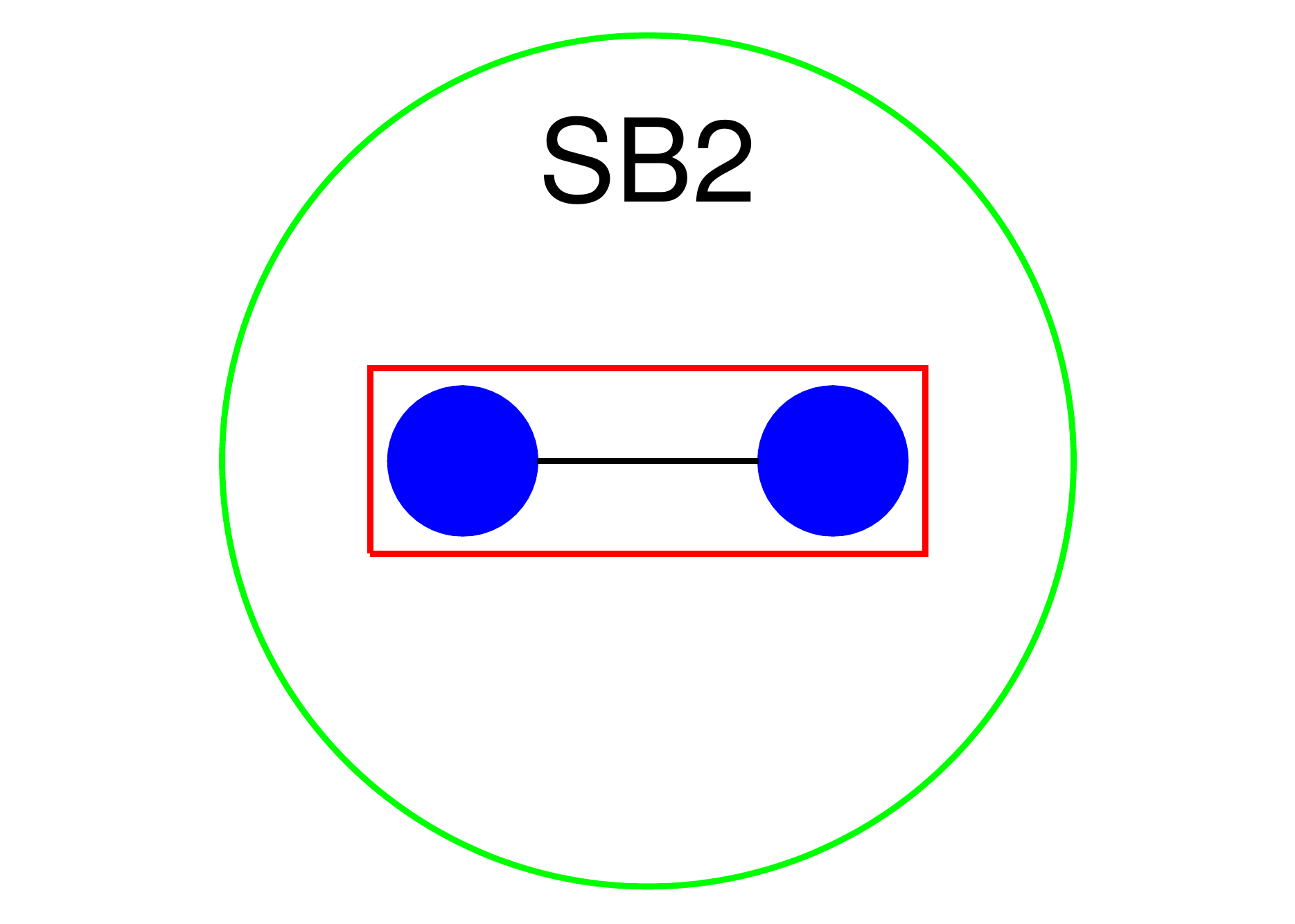}         & \includegraphics[width=0.15\textwidth]{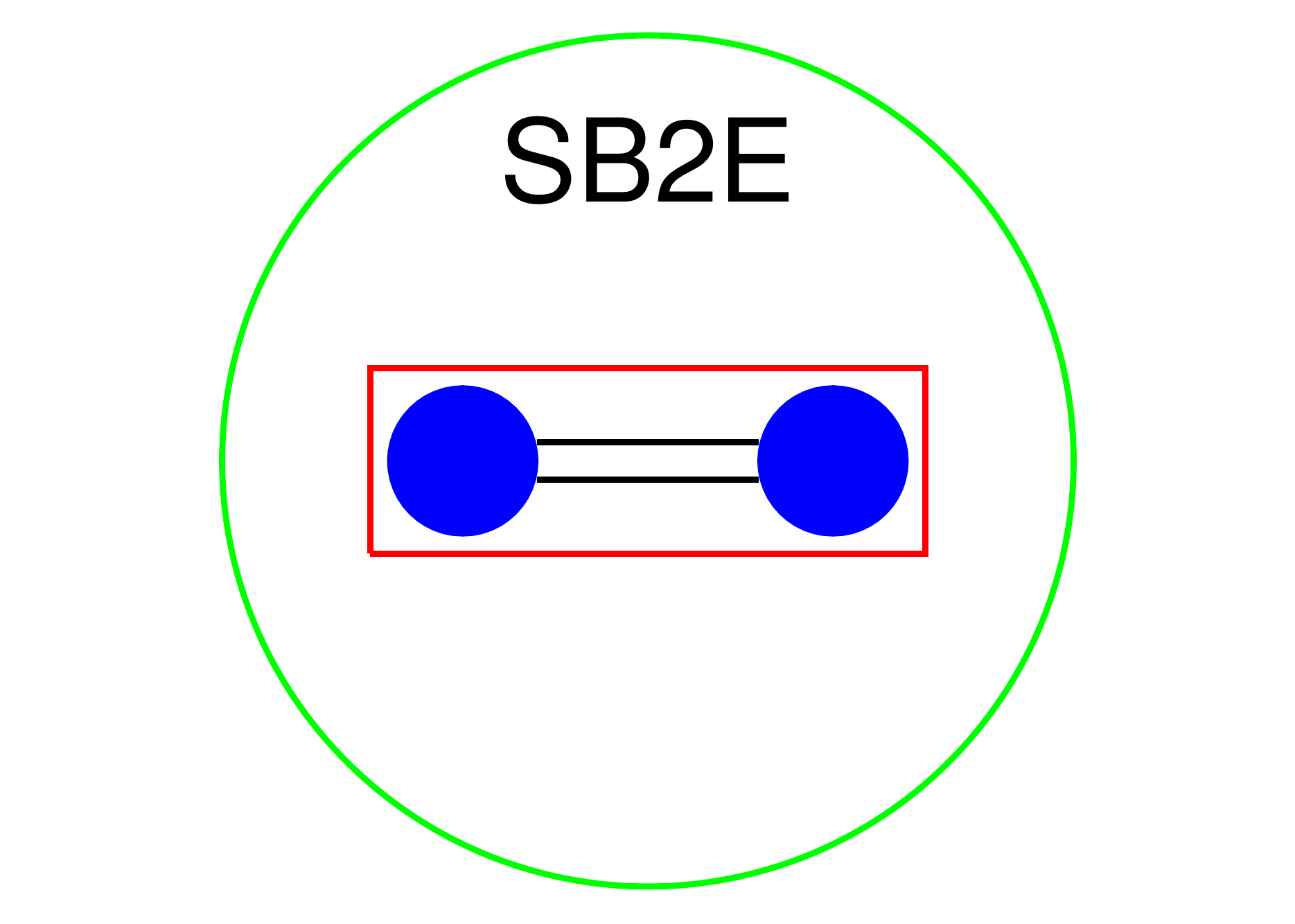}        & \includegraphics[width=0.15\textwidth]{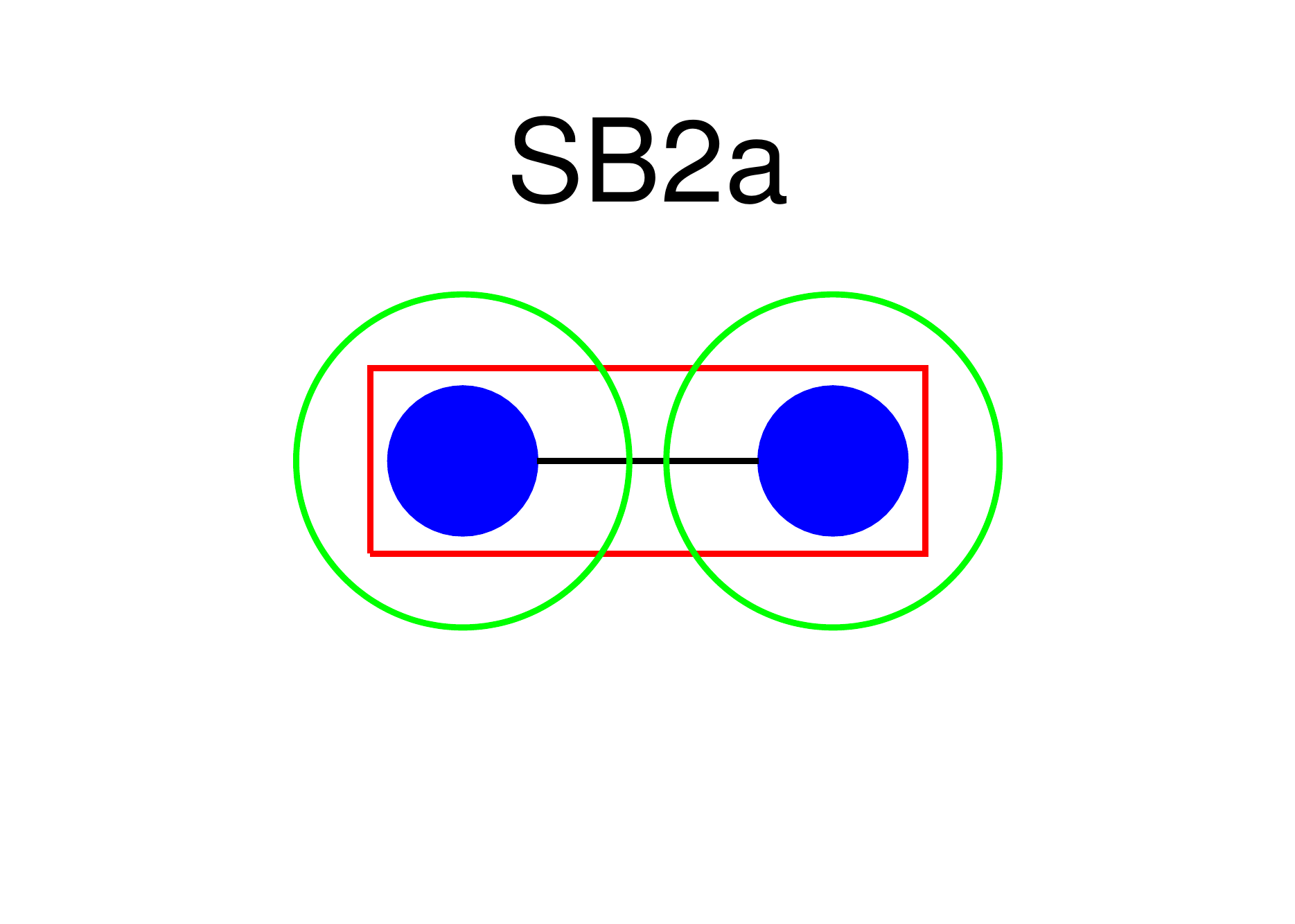}        \\
\hphantom{xxx}HD 164\,438                               & \hphantom{xxx}Cyg OB2-22 C                              & \hphantom{xxx}15 Mon AaAb                               & \hphantom{xxx}HD 167\,771                               & \hphantom{xxx}HD 170\,097 A                             & \hphantom{xx}HD 168\,137 AaAb                            \\
\hphantom{xxx}V479 Sct                                  & \hphantom{xxx}V747 Cep                                  &                                                         & \hphantom{xxx}BD $-$16 4826                             & \hphantom{xxx}QR Ser                                    & \hphantom{xx}HD 168\,112 AB                              \\
\hphantom{xxx}9 Sge                                     &                                                         &                                                         & \hphantom{xxx}HD 168\,075                               & \hphantom{xxx}HD 166\,734                               & \hphantom{xx}HD 54\,662 AB                               \\
\hphantom{xxx}Cyg X-1                                   &                                                         &                                                         & \hphantom{xxx}BD $-$13 4923                             & \hphantom{xxx}HD 190\,967                               &                                                         \\
\hphantom{xxx}BD $+$36 4063                             &                                                         &                                                         & \hphantom{xxx}HD 191\,612                               & \hphantom{xxx}HDE 228\,854                              &                                                         \\
\hphantom{xxx}HDE 229\,234                              &                                                         &                                                         & \hphantom{xxx}HDE 228\,766                              & \hphantom{xxx}HD 193\,611                               &                                                         \\
\hphantom{xxx}HD 192\,281                               &                                                         &                                                         & \hphantom{xxx}Cyg OB2-9                                 & \hphantom{xxx}HDE 228\,989                              &                                                         \\
\hphantom{xxx}ALS 15\,133                               &                                                         &                                                         & \hphantom{xxx}Cyg OB2-8 A                               & \hphantom{xxx}Y Cyg                                     &                                                         \\
\hphantom{xxx}Cyg OB2-A11                               &                                                         &                                                         & \hphantom{xxx}Cyg OB2-73                                & \hphantom{xxx}Cyg OB2-B17                               &                                                         \\
\hphantom{xxx}Cyg OB2-22 B                              &                                                         &                                                         & \hphantom{xxx}ALS 15\,114                               & \hphantom{xxx}Cyg OB2-3 A                               &                                                         \\
\hphantom{xxx}Cyg OB2-41                                &                                                         &                                                         & \hphantom{xxx}LS III $+$46 11                           & \hphantom{xxx}ALS 12\,688                               &                                                         \\
\hphantom{xxx}ALS 15\,148                               &                                                         &                                                         & \hphantom{xxx}HD 199\,579                               & \hphantom{xxx}AO Cas                                    &                                                         \\
\hphantom{xxx}ALS 15\,131                               &                                                         &                                                         & \hphantom{xxx}14 Cep                                    & \hphantom{xxx}CC Cas                                    &                                                         \\
\hphantom{xxx}Cyg OB2-20                                &                                                         &                                                         & \hphantom{xxx}DH Cep                                    &                                                         &                                                         \\
\hphantom{xxx}Cyg OB2-70                                &                                                         &                                                         & \hphantom{xxx}BD $+$60 497                              &                                                         &                                                         \\
\hphantom{xxx}Cyg OB2-15                                &                                                         &                                                         & \hphantom{xxx}HD 15\,558 A                              &                                                         &                                                         \\
\hphantom{xxx}ALS 15\,115                               &                                                         &                                                         & \hphantom{xxx}HD 37\,366                                &                                                         &                                                         \\
\hphantom{xxx}Cyg OB2-29                                &                                                         &                                                         & \hphantom{xxx}HD 47\,129                                &                                                         &                                                         \\
\hphantom{xxx}Cyg OB2-11                                &                                                         &                                                         & \hphantom{xxx}HD 48\,099                                &                                                         &                                                         \\
\hphantom{xxx}68 Cyg                                    &                                                         &                                                         & \hphantom{xxx}HD 46\,149                                &                                                         &                                                         \\
\hphantom{xxx}HD 108                                    &                                                         &                                                         & \hphantom{xxx}$\zeta$ Ori AaAb                          &                                                         &                                                         \\
\hphantom{xxx}HD 12\,323                                &                                                         &                                                         & \hphantom{xxx}$\theta^{2}$ Ori A                        &                                                         &                                                         \\
\hphantom{xxx}HD 15\,137                                &                                                         &                                                         & \hphantom{xxx}$\iota$ Ori                               &                                                         &                                                         \\
\hphantom{xxx}$\alpha$ Cam                              &                                                         &                                                         & \hphantom{xxx}HD 53\,975                                &                                                         &                                                         \\
\hphantom{xxx}HD 37\,737                                &                                                         &                                                         &                                                         &                                                         &                                                         \\
\hphantom{xxx}HD 46\,573                                &                                                         &                                                         &                                                         &                                                         &                                                         \\
\hphantom{xxx}HD 52\,533 A                              &                                                         &                                                         &                                                         &                                                         &                                                         \\
\\
\end{tabular}}

 \caption{Spectroscopic binary status (SBS) diagrams and systems that belong to each category. Blue circles are used to represent components: filled if they are 
         significantly detected in the spectrum, empty otherwise. Lines connecting the components represent orbits: dashed if they do not induce detected radial velocity
         variations in the observed lines, single-solid if they do, and double-solid if they do and also produce eclipses. Empty green circles identify the 
         observed astrometric components. Red rectangles identify the components included in the GOSSS long-slit spectra, note that if two red rectangles are shown in
         a diagram because GOSSS spatially separates components then the secondary spectrum is listed in parenthesis in the list below. Components outside the 
         high-resolution aperture or low-mass components that do not induce detected radial velocity variations in the observed lines are not included. This subfigure 
         shows systems with two components and the next subfigure those with three or four.}
\label{sbstatus}
\end{figure*}	

\addtocounter{figure}{-1}

\begin{figure*}
\centerline{\begin{tabular}{llllll}
\includegraphics[width=0.15\textwidth]{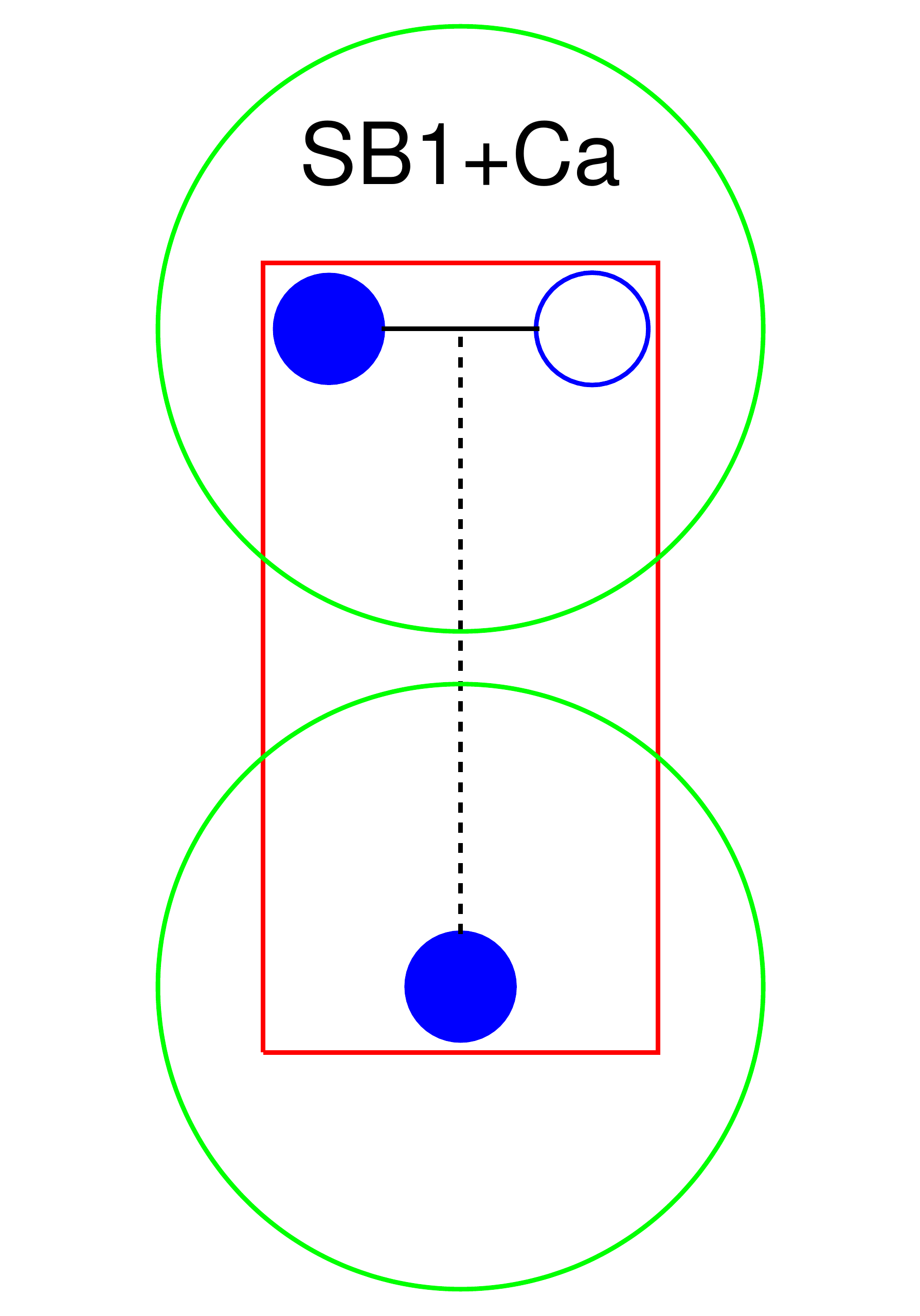}      & \includegraphics[width=0.15\textwidth]{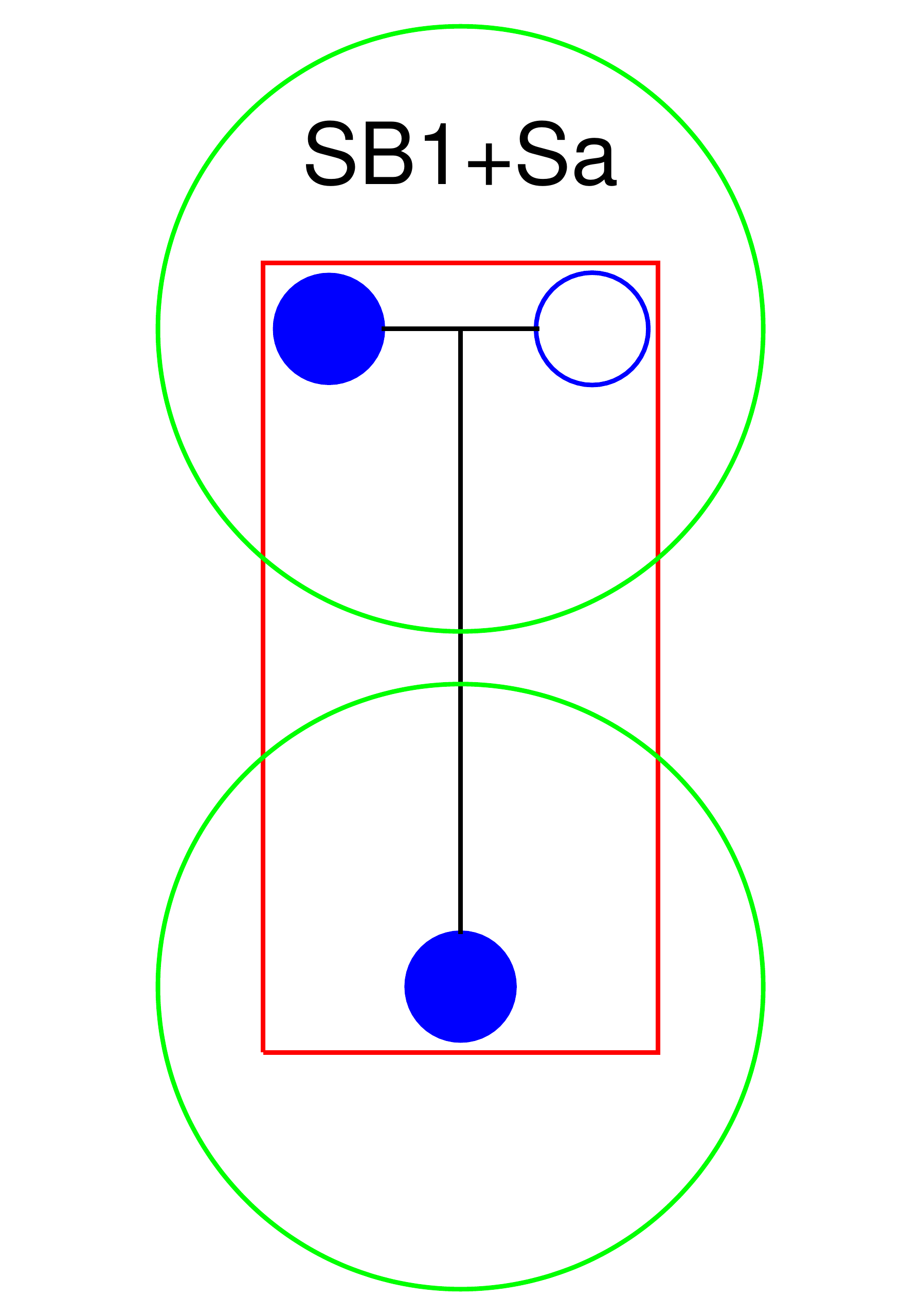}      & \includegraphics[width=0.15\textwidth]{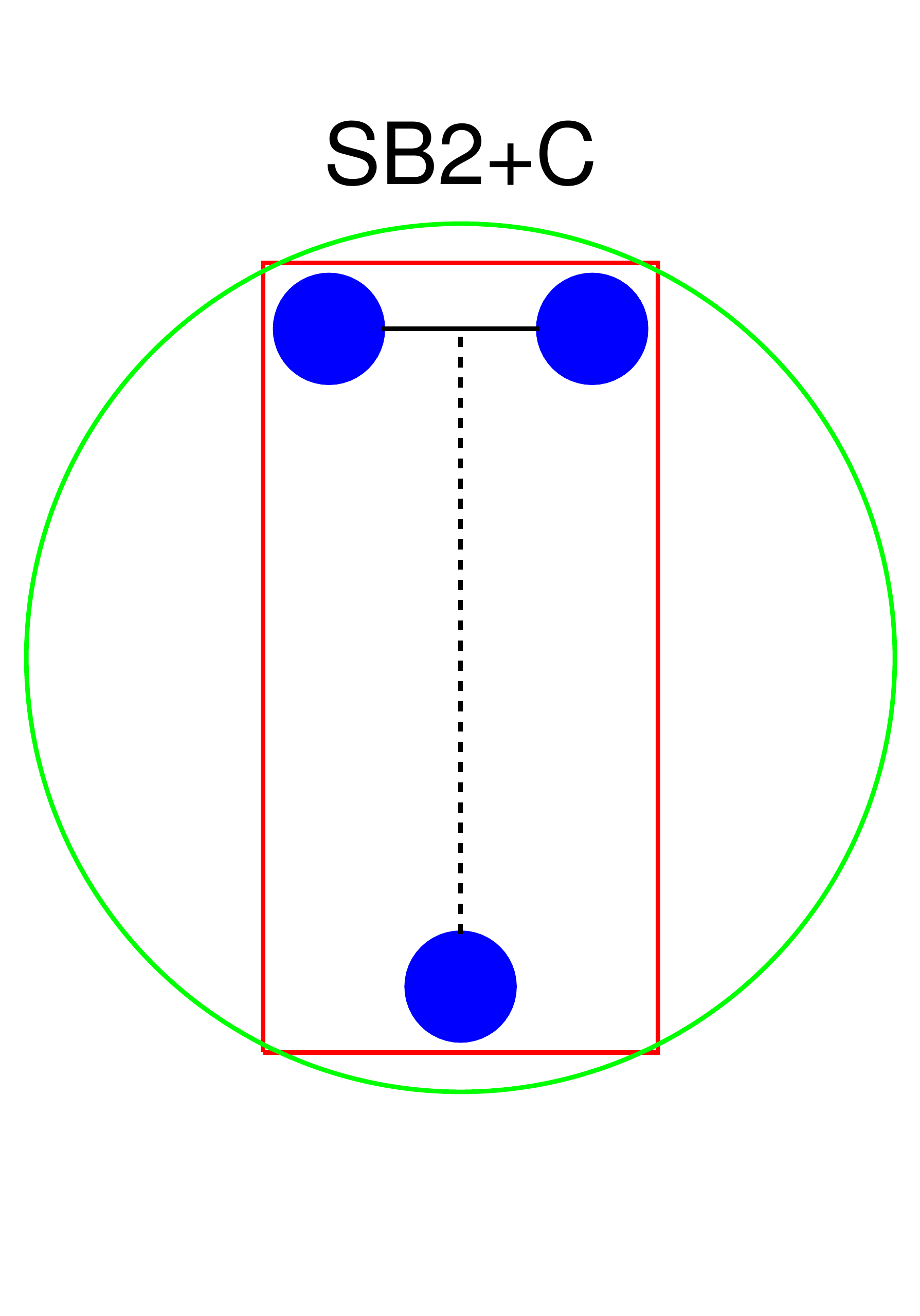}       & \includegraphics[width=0.15\textwidth]{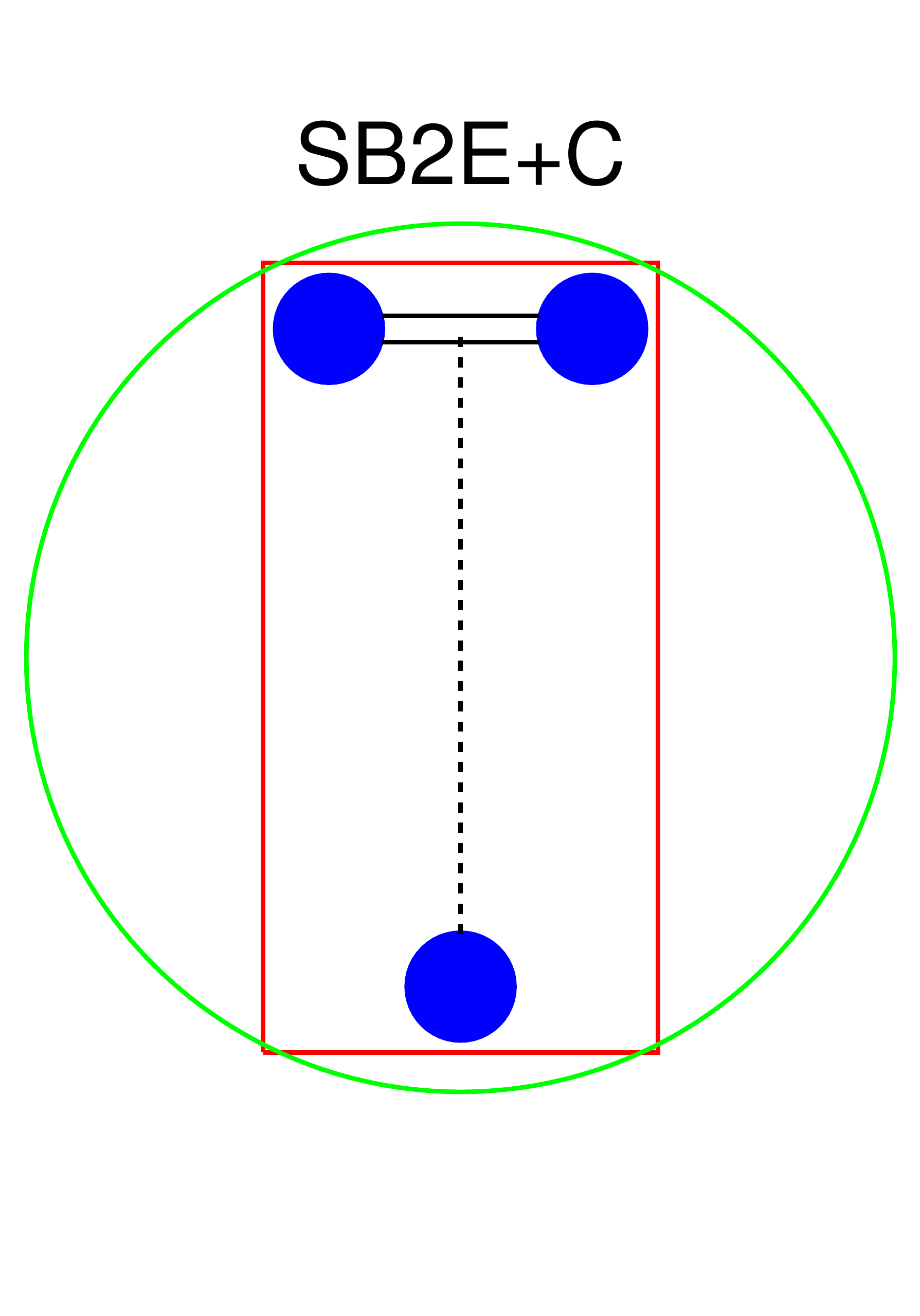}      & \includegraphics[width=0.15\textwidth]{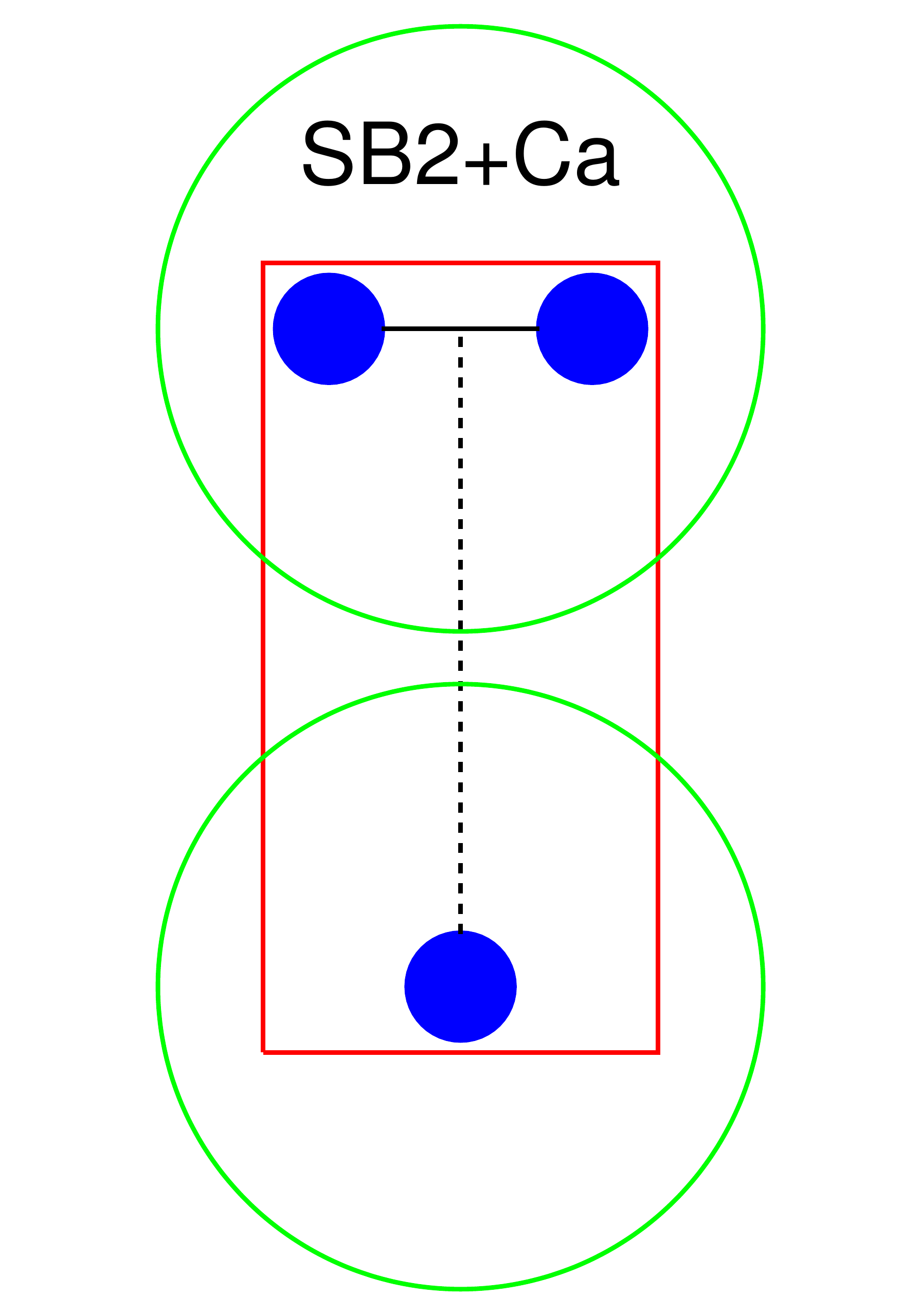}      & \includegraphics[width=0.15\textwidth]{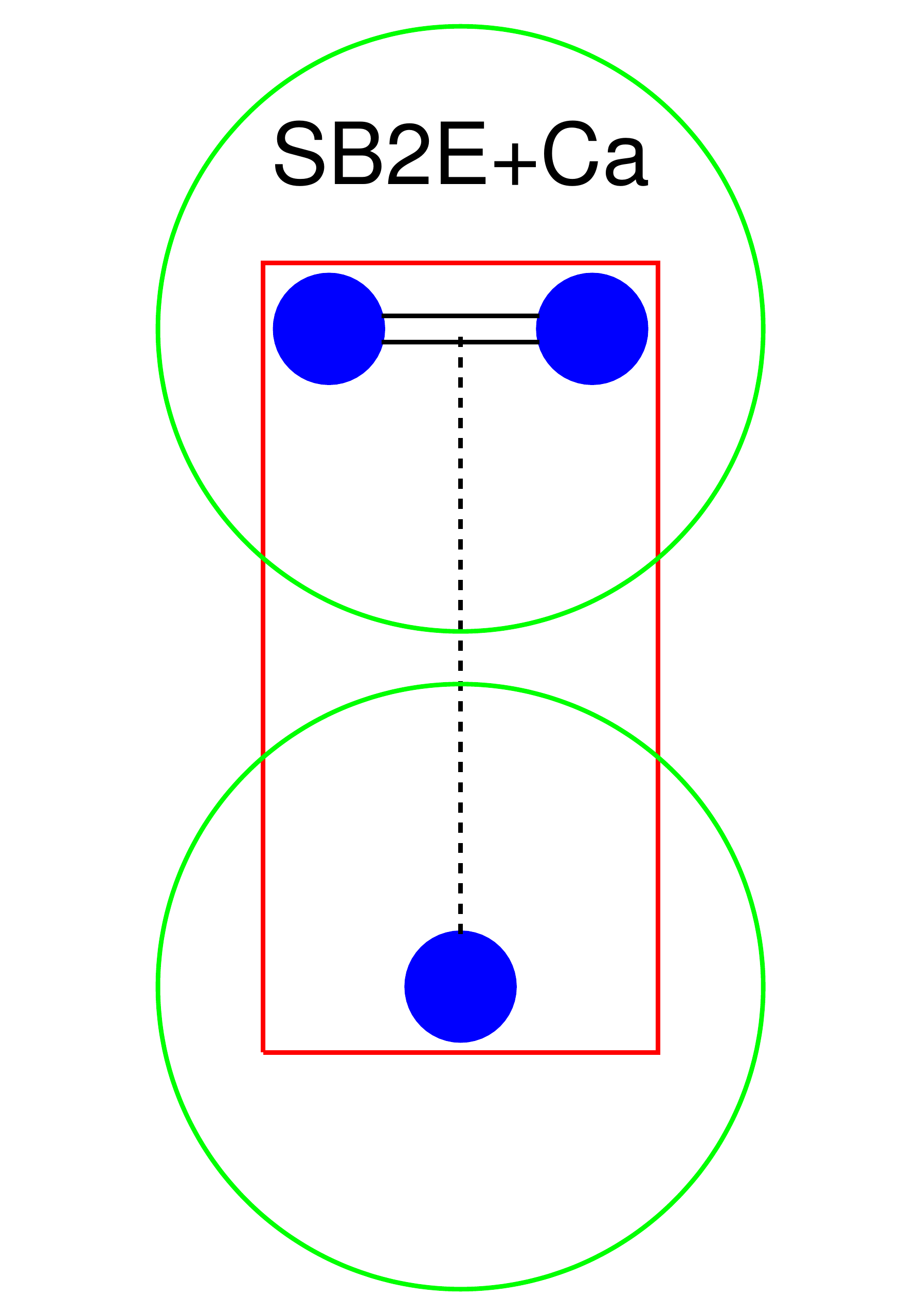}     \\
\hphantom{xxx}HDE 229\,232 AB                           & \hphantom{xxx}$\theta^{1}$ Ori CaCb                     & \hphantom{xxx}HD 17\,505 A                              & \hphantom{xxx}HD 175\,514                               & \hphantom{xxx}HD 193\,443 AB                            & \hphantom{xxx}Cyg OB2-27 AB                             \\
\hphantom{xxx}Cyg OB2-1                                 &                                                         &                                                         &                                                         & \hphantom{xxx}HD 194\,649 AB                            & \hphantom{xxx}ALS 12\,502                               \\
\hphantom{xxx}HD 16\,429 A                              &                                                         &                                                         &                                                         & \hphantom{xxx}HD 206\,267 AaAb                          & \hphantom{xxx}DN Cas                                    \\
\hphantom{xxx}HD 14\,633 AaAb                           &                                                         &                                                         &                                                         &                                                         & \hphantom{xxx}IU Aur AB                                 \\
\\
\includegraphics[width=0.15\textwidth]{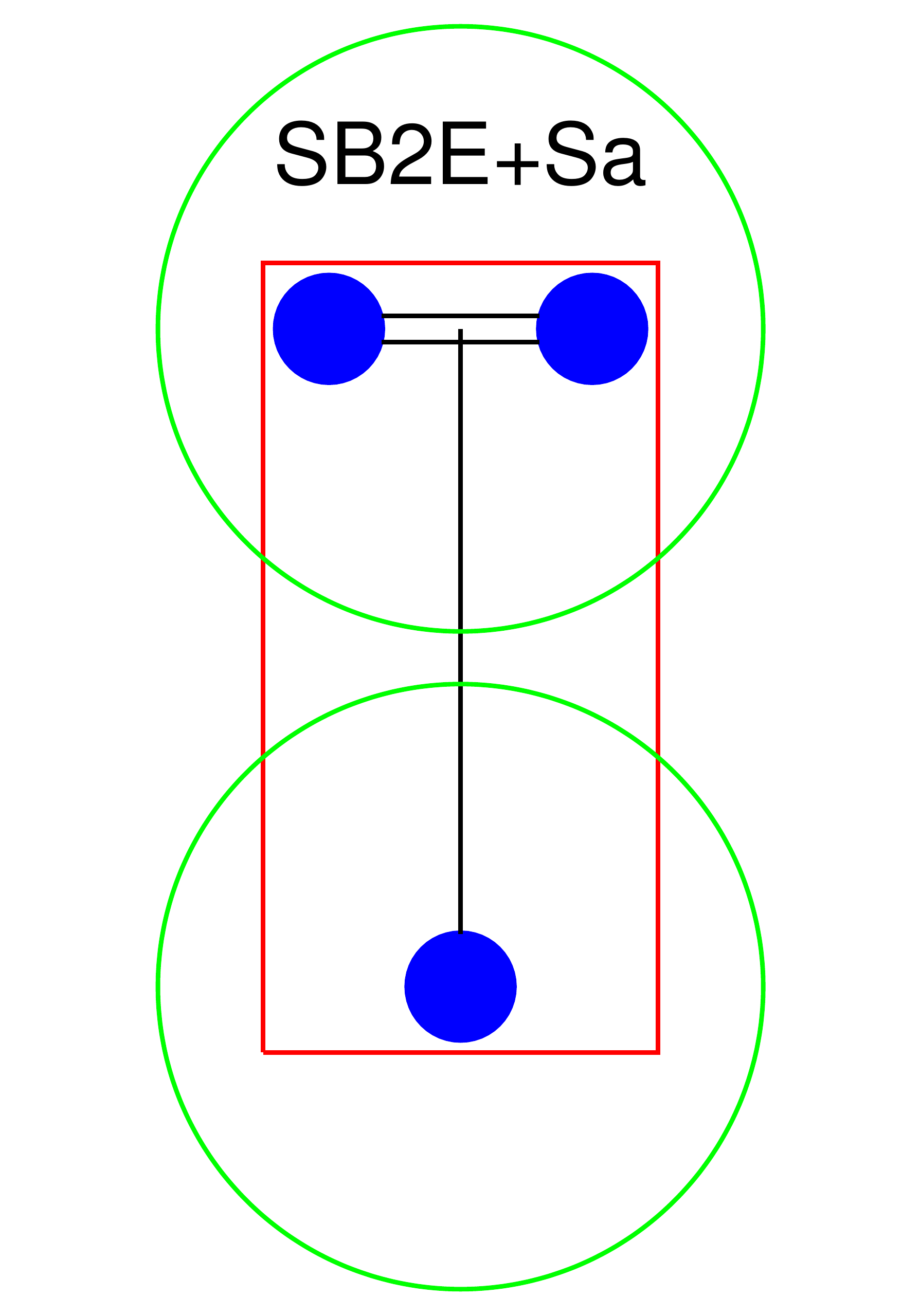}     & \includegraphics[width=0.15\textwidth]{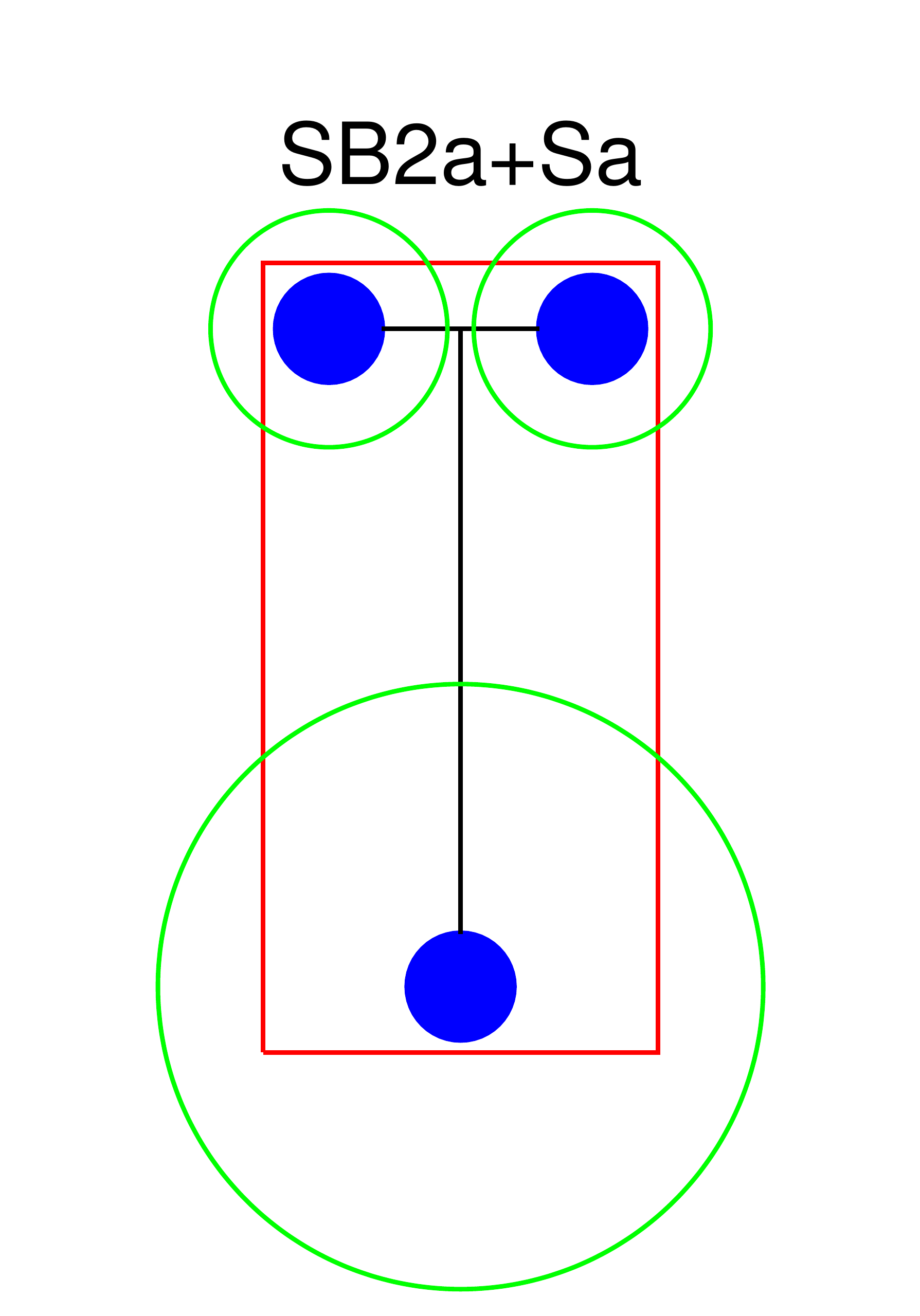}     & \includegraphics[width=0.15\textwidth]{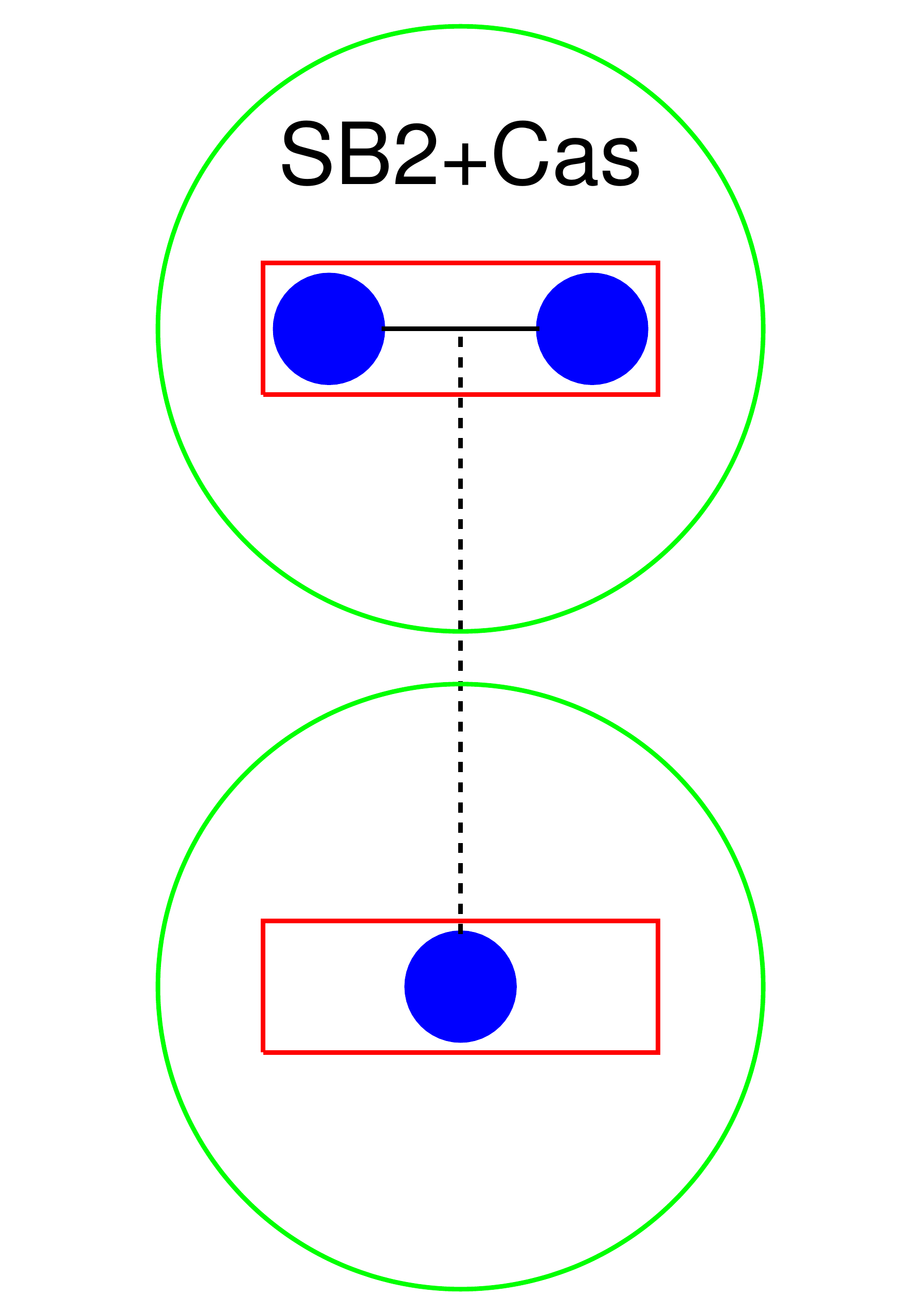}     & \includegraphics[width=0.15\textwidth]{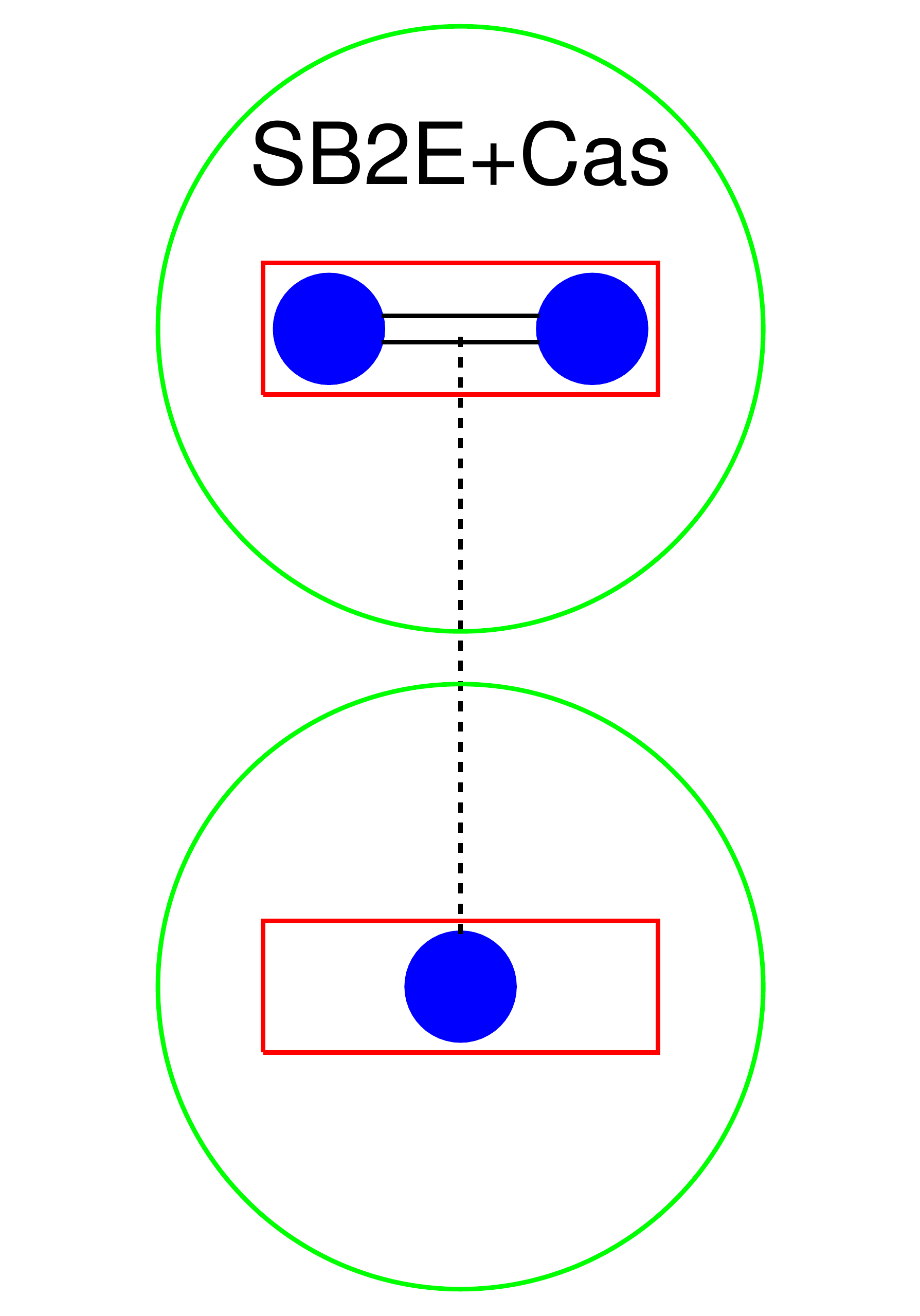}    & \includegraphics[width=0.15\textwidth]{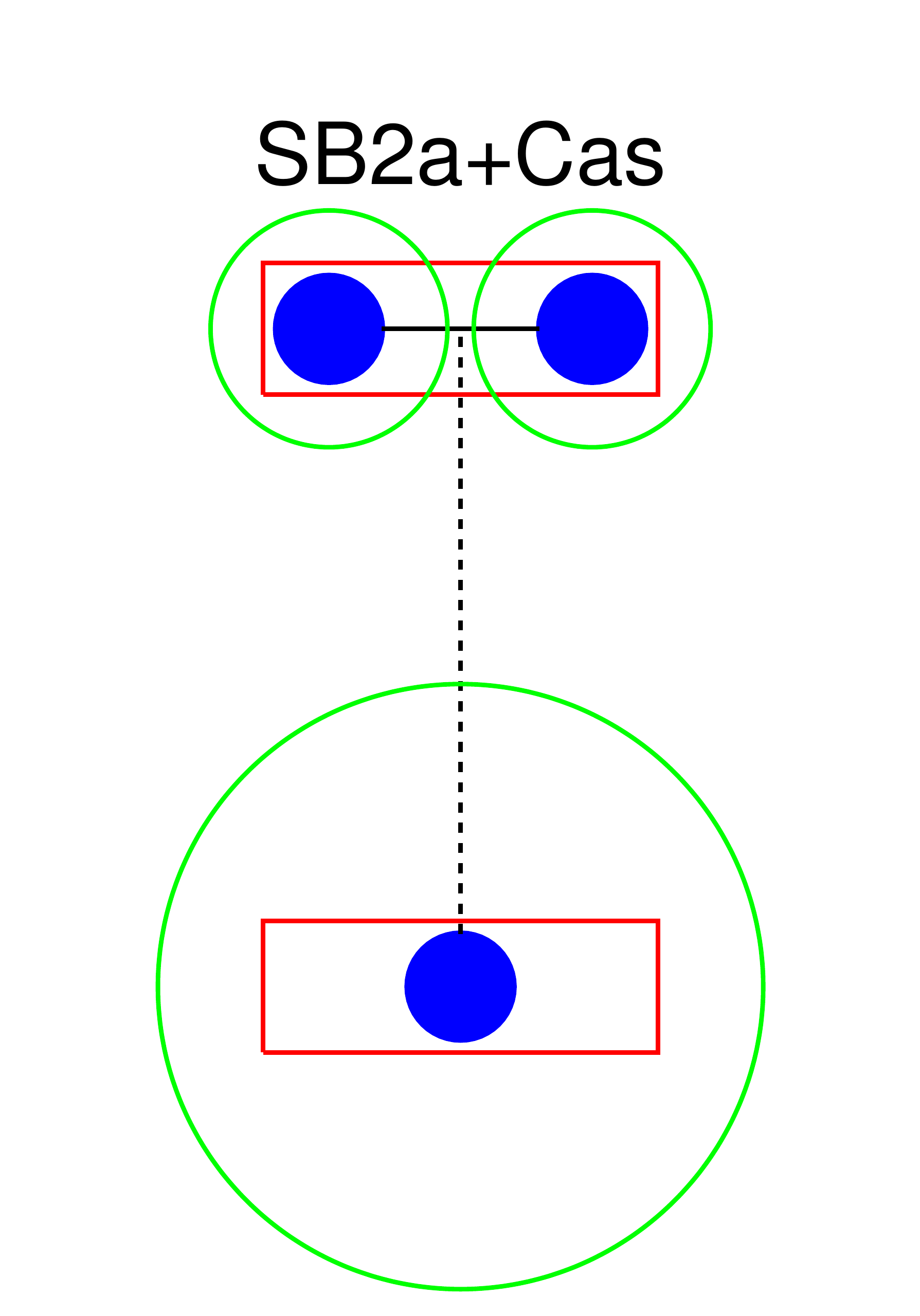}    & \includegraphics[width=0.15\textwidth]{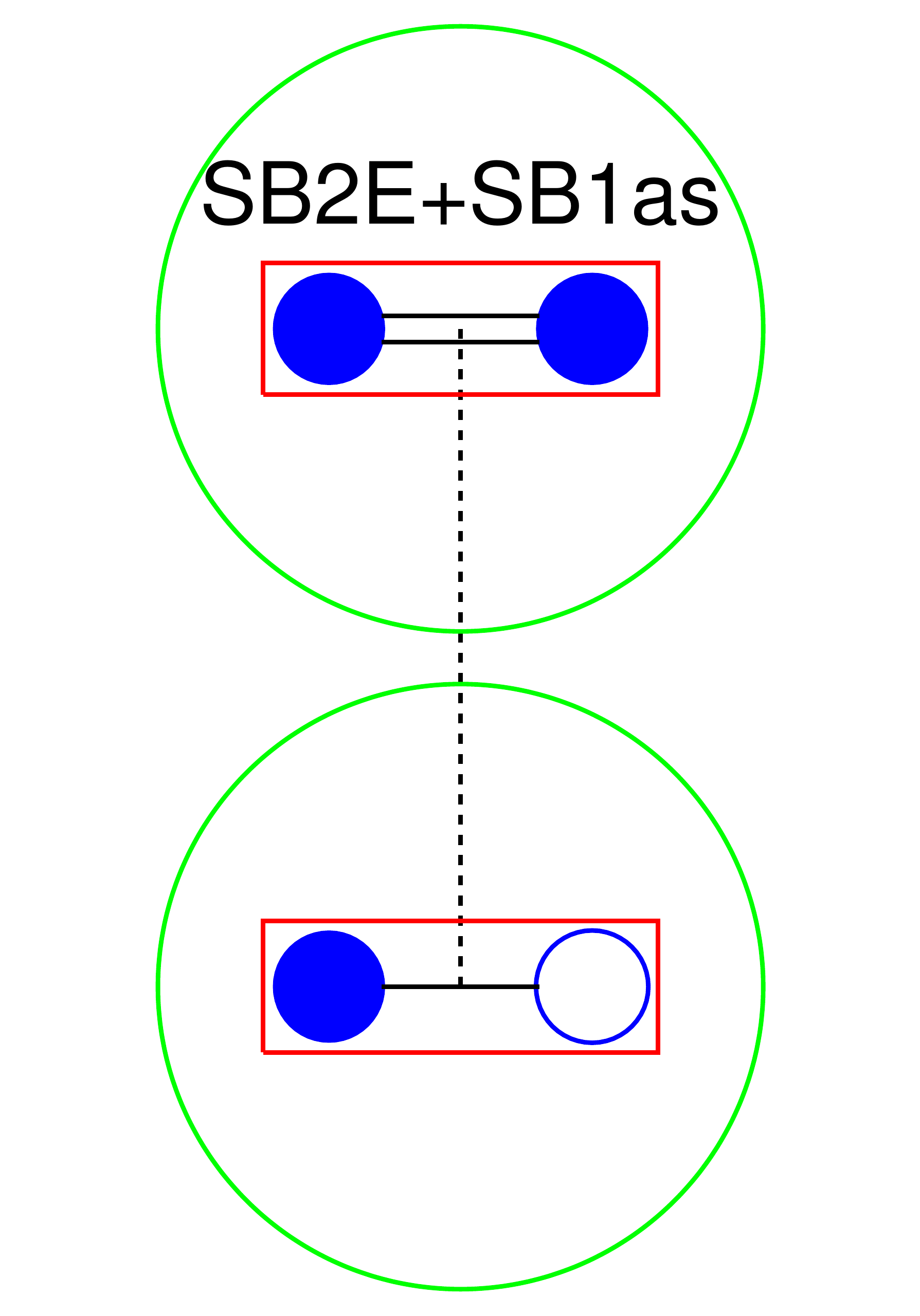}  \\
\hphantom{xxx}MY Ser AaAb                               & \hphantom{xxx}HD 193\,322 AaAb                          & \hphantom{xxx}HD 191\,201 A (+B)                        & \hphantom{xxx}Cyg OB2-5 A (+B)                          & \hphantom{xxx}$\sigma$ Ori AaAb (+B)                    & \hphantom{xxx}LY Aur A (+B)                             \\
                                                        &                                                         &                                                         & \hphantom{xxx}MY Cam A (+B)                             &                                                         &                                                         \\
                                                        &                                                         &                                                         & \hphantom{xxx}$\delta$ Ori Aa (+Ab)                     &                                                         &                                                         \\
\\
\end{tabular}}

\caption{(continued).}
\end{figure*}	

\begin{figure}
\centerline{\includegraphics*[width=\linewidth]{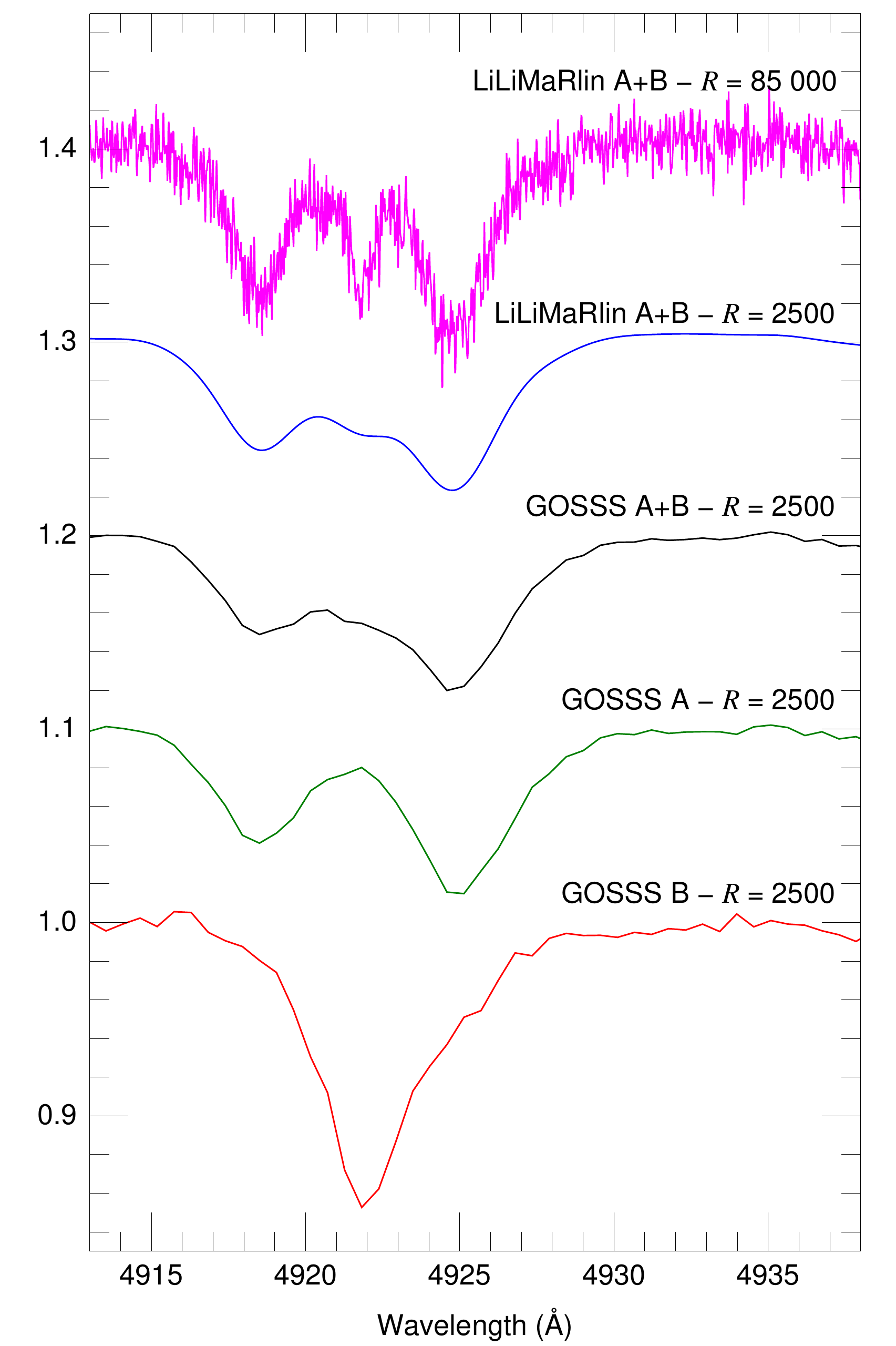}}
\caption{Comparison of HD~191\,201 spectra for the \HeI{4922} region that highlights the importance of combining spatially and spectrally
         resolved information to analyze multiple systems. The two bottom plots show the spatially resolved, rectified, and normalized  
         GOSSS spectra of HD~191\,201~A~and~B, respectively (all plots are offset by 0.1 continuum units with respect to the adjacent ones).
         The middle plot is a linear combination of the bottom two (using the measured $\Delta m$) to simulate the A+B spectrum, note how the
         B component leaves just a small imprint around $\lambda\approx$~4922~\AA\ due to its relative faintness. The top plot shows a LiLiMaRlin
         spectrum of A+B taken at a similar phase as the GOSSS spectra and the one below it is the same spectrum degraded to the GOSSS spectral
         resolution. Note the similarity between the two spectra of A+B at $R$~=~2500 but the different paths taken to obtain them.}
\label{HD_191_201_HeI_4922}
\end{figure}

\paragraph{HDE~229\,234 = BD~$+$38~4069 = ALS~11\,297.}                                             
\textbf{SB1.}
This system had not appeared in GOSSS before and here we obtain a single spectral classification of O9~III with no sign of the secondary. 
The object has no entry in the WDS catalog and appears single in our AstraLux images. 

\paragraph{HD~192\,281 = V2011~Cyg = BD~+39~4082 = ALS~10\,943.}                                   
\textbf{SB1.}
\citet{Bara93} identified this object as a SB1 but its nature has not been confirmed afterwards. In GOSSS~III we classified it as
O4.5~IV(n)(f). It is a runaway star \citep{Maizetal18b}. HD~192\,281 has no WDS entry and no significant companion is seen in our AstraLux data.

\paragraph{Y~Cyg = HD~198\,846 = BD~$+$34~4184 = ALS~11\,594.}                                     
\textbf{SB2E.}
This eccentric short-period binary has a rapid apsidal motion \citep{Harmetal14} and was classified by 
\citet{Burketal97} as O9~V~+~O9.5~V. In GOSSS~I we obtained a spectral classification of O9.5~IV~+~O9.5~IV, which we reproduce here using a 
LiLiMaRlin epoch. The object has no entry in the WDS~catalog and appears single in our AstraLux images. It is a runaway star \citep{Maizetal18b}.

\paragraph{HDE~229\,232~AB = BD~$+$38~4070~AB = ALS~11\,296~AB.}                                   
\textbf{SB1+Ca?.}
This system has one low-amplitude SB1 orbit published \citep{Willetal13}. In GOSSS~III we classified it as O4~V:n((f)) and noted the anomalous profiles.
The WDS catalog lists a bright nearby companion (hence the AB designation) but with only partial information. There is only one detection of B \citep{Aldoetal15}
which seems robust (S. Caballero-Nieves, private communication) and with a maximum separation of 40~mas but a large uncertainty in the magnitude difference.
No companion is seen in our AstraLux images but note that B is too 
close to be resolved with that technique. It is a runaway star \citep{Maizetal18b}. 
    
\paragraph{HD~193\,322~AaAb  = BD~+40~4103~AaAb = ALS~11\,113~AaAb.}                               
\textbf{SB2a+Sa.}
This hierarchical system is quite complex and its four inner components include two O stars and two B stars within 3\arcsec, with two late-type B stars
farther away (\citealt{tenBetal11}, from where some of the information in this paragraph comes from). An inner SB2 binary Ab1+Ab2 has a period of 312.4~d and
is composed of an O8.5~III star (Ab1) with a dimmer B2.5:~V: companion (Ab2, note that we do not design the system as AaAb1Ab2 because the contribution of Ab2 to the 
integrated spectrum is small). The Ab subsystem orbits around Aa, an O9~Vnn star, with a period of 35.2~a. In GOSSS-I we were unable to separate any of the
three combined components and we were only able to give a combined O9~IV(n) classification.
In \citet{Maiz10a} we measured (based on a single epoch) that Ab was brighter than Aa by 0.04~mag but the uncertainty was large enough to allow for 
the situation to be reversed. Our reanalysis of the original AstraLux epoch and of three new ones with improved PSFs yields that Aa is brighter than Ab
in the $z$ band by 0.15-0.20~mag (Table~\ref{AstraLuxdata} and Fig.~\ref{AstraLux2}), with a possible (but unconfirmed at this stage) small color term that would make 
the difference larger at shorter wavelengths. Note that Aa being slightly brighter than Ab1+Ab2 is incompatible with the first being of luminosity class V and Ab1
of luminosity class III (with both having similar spectral subtypes), leading us to think that the luminosity classes of the two O stars may need to be revised.
Our AstraLux observations cover a significant part of the 35.2~a orbit of Aa+Ab and indeed we detect a significant astrometric motion. HD~193\,322~B is located
2\farcs7 away with a $\Delta z$ of 1.6~mag, far enough not to contaminate significantly observations with high-resolution spectrographs and normal seeing. 
\citet{Robeetal10} classified it as B1.5~V and the GOSSS spectrum shown here concurs with that classification but adding a (n)p suffix, as there is substantial
broadening with respect to the standard and the He\,{\sc i}
lines are too deep, leading us to suspect a He enrichment that should be confirmed with further data. Finally, we note that we detect a significant 
outward motion ($\sim$39~mas in 10~years) for B with respect to Aa, something that is not seen in the historical WDS data or in the recent Gaia~DR2 proper motions. 
However, those other measurements refer to the A,B separation, not to Aa,B. The detected outward motion is actually compatible with the effect of the Aa,Ab orbit
in the position of Aa. This system and its cluster, Collinder~419, will be analyzed in a separate paper (Ma{\'\i}z Apell\'aniz 2019, in preparation). 

\paragraph{HD~194\,649~AB = BD~$+$39~4177~AB = ALS~11\,324.}                                       
\textbf{SB2+Ca.}
\citet{Mahyetal13} classified this system as O6~III(f)~+~O8~V. Despite its short period there is no mention of eclipses in the literature.
In GOSSS~III we gave it an unresolved classification of O6.5~V((f)). Here we present a new GOSSS spectrogram from which we are able to derive a separate 
spectral classification of O6~V((f))~+~O9.7:~V. In one of the LiLiMaRlin epochs we caught the system close to quadrature and derive a similar
separate spectral classification of O6~IV((f))~+~O9.5~V((f)). Both the GOSSS and LiLiMaRlin results
differ from the \cite{Mahyetal13} one in that the primary has a lower luminosity class (\HeII{4686} is too 
strong to be a giant) and in that the secondary is of later spectral type.
As a further complication, the WDS catalog lists a component 0\farcs4 away with a $\Delta m$ of 0.9~mag which is also seen in our 
AstraLux images (hence the AB designation). This implies that HD~194\,649~AB is another example of a hierarchical triple 
system where three stars contribute to the spectrum. Our AstraLux data shows that B has moved 1~degree with respect to A in 6 years in a clockwise direction, 
which is consistent with the existing WDS information and yields a period of 2-3~ka. 

\paragraph{Cyg~OB2-B17 = V1827~Cyg = [CPR2002]~B17.}                                               
\textbf{SB2E.}
This is the dimmest and one of the most extinguished targets in our sample. \citet{Stroetal10} found Cyg~OB2-B17 to be an eclipsing SB2 
composed by two supergiants with spectral types O7~Iaf~+~O9~Iaf. In GOSSS-III we derived an earlier classification for the primary, 
yielding O6~Iaf~+~O9:~Ia:. It does not have an entry in the WDS catalog and our AstraLux image does not reveal any 
significant companion.

\paragraph{Cyg~OB2-3~A = BD~$+$40~4212~A = ALS~11\,403~A = Schulte~3~A.}                           
\textbf{SB2E.}
\citet{Kimietal08} classified this system as O9~III~+~O6~IV: and \citet{Salaetal15} identified it as an eclipsing binary. 
Cyg~OB2-3~A was not previously included in GOSSS. Here we obtain a GOSSS spectral type of O8.5~Ib(f)~+~O6~III:. The WDS catalog gives a bright companion 
4\farcs0 away which also appears in our AstraLux images (Fig.~\ref{AstraLux2}).
We aligned the GOSSS slit to obtain simultaneous spectra of the two targets and we derived a spectral type 
of B0~IV for Cyg~OB2-3~B. Note that the B component is too far away to contribute to the LiLiMaRlin spectra. 

\paragraph{ALS~15\,133.}                                                                           
\textbf{SB1.}
This object was identified as a SB1 by \citet{Kobuetal12}, who classified it as O9~III. ALS~15\,133 had not appeared in any previous GOSSS papers
and here we classify it as O9.5~IV. It has no entry in the WDS and here we identify a B component 4\farcs4 away (Table~\ref{AstraLuxdata} and 
Fig.~\ref{AstraLux2}) that also appears in Gaia DR2. \citet{Salaetal15} identified it as having low-amplitude long-term photometric variability.

\paragraph{Cyg~OB2-A11 = ALS~21\,079 = [CPR2002]~A11.}                                             
\textbf{SB1.}
\citet{Neguetal08a} classified Cyg~OB2-A11 as O7~Ib-II and \citet{Kobuetal12} identified it as a SB1. In GOSSS~III we classified it as O7~Ib(f). This 
object has no entry in the WDS and here we detect a B companion 2\farcs2 away (Table~\ref{AstraLuxdata} and Fig.~\ref{AstraLux2}) that also appears 
in Gaia DR2.

\begin{figure}
\centerline{\includegraphics*[width=\linewidth]{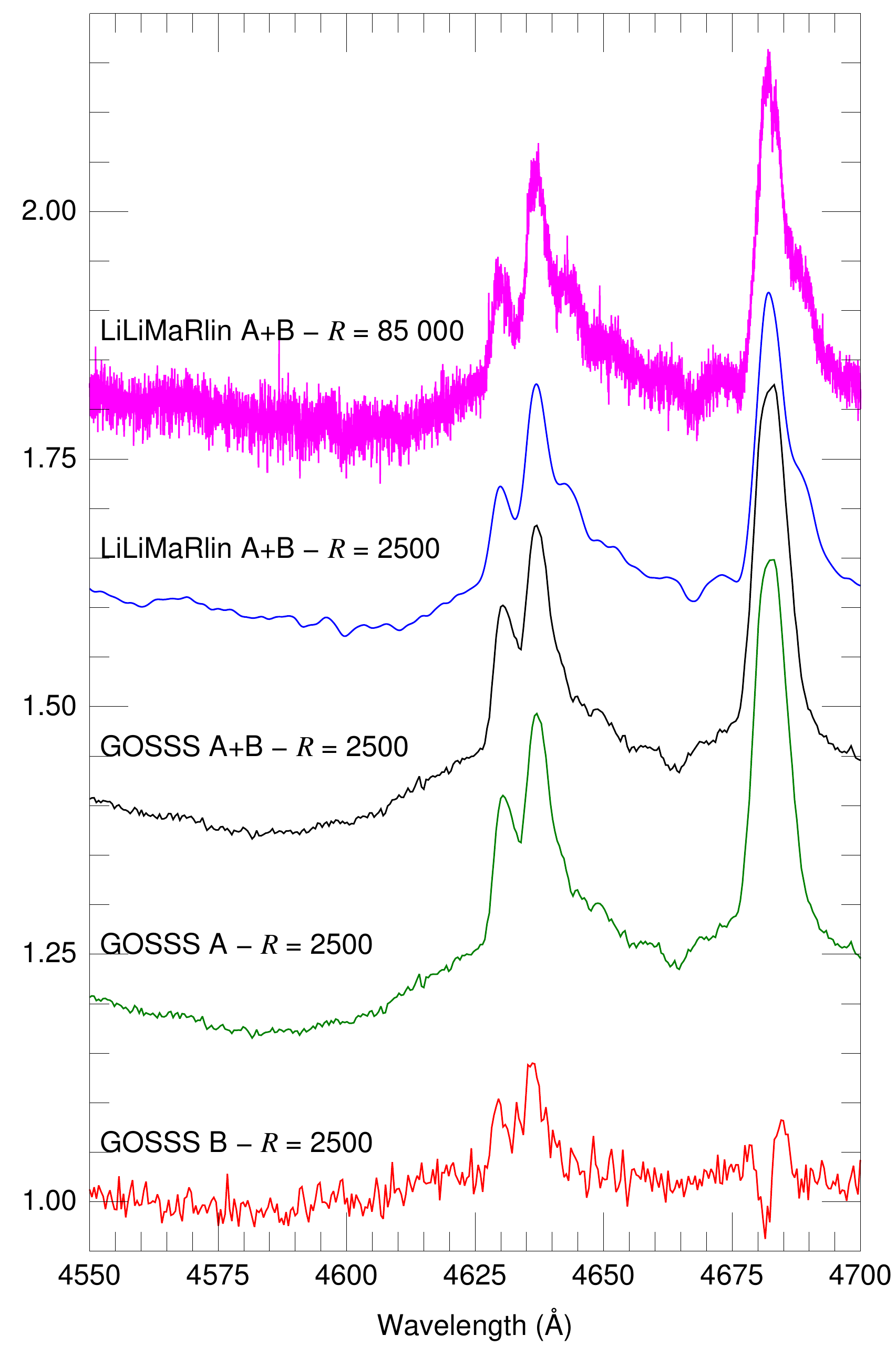}}
\caption{Plot equivalent to Fig.~\ref{HD_191_201_HeI_4922} for Cyg~OB2-5 and the 4550-4700~\AA\ region (all plots are offset by 0.2 continuum 
         units with respect to the adjacent ones). Note that in this case there is a
         small phase difference between the GOSSS and LiLiMaRlin epochs for the A orbit. As the magnitude difference between A and B is larger than for 
         HD~191\,201 here the signature of the B component is much harder to detect in the spatially unresolved spectra.}
\label{Cyg_OB2-5_HeII_4686}
\end{figure}

\paragraph{Cyg~OB2-5~A = V279~Cyg~A = BD~$+$40~4220~A = ALS~11\,408~A.}                            
\textbf{SB2E+Cas.}
In \citet{Rauwetal99} they classified this SB2 supergiant as O5.5-6.5~+~O6.5. In GOSSS~I we gave a composite spectral classification 
of O7~Iafpe. In the new spectrogram presented here we separate the two components to give new spectral types of O6.5:~Iafe~+~O7~Iafe. The
emission lines associated with both components are quite strong, likely as a result of the interaction between the two winds. Note that 
there is a third lower-luminosity supergiant in the system, Cyg~OB2-5~B, for which in GOSSS~III we gave a spectral type of O6.5~Iabfp. It 
is located at an angular distance of 0\farcs93 with a $\Delta z$ of 3 magnitudes \citep{Maiz10a}, so its light falls at least partially in the 
aperture of the LiLiMaRlin spectrographs. However, given the large magnitude difference it contributes little to the combined spectra
(Fig.~\ref{Cyg_OB2-5_HeII_4686}). Here we have obtained a new spectrogram for Cyg~OB2-5~B where the contamination noise due to the brighter 
A component is significantly lower. The classification is slightly revised to O7~Ib(f)p var? with the suffix indicating an anomalous 
\HeII{4686} profile and a possible variability (unclear whether it has a physical origin or is due to contamination from A). We have reanalyzed
our previous AstraLux observations from 2007 and added a new epoch from 2018 (Table~\ref{AstraLuxdata}). We detect no motion along the line that joins the stars
and only a slight one in position angle over a span of 11 years. This contrasts with a significant difference in separation (0\farcs6) and position
angle (11\degr) between the first (from 1967, \citealt{Herb67}) and last (from 2007, our previous data) epochs in the WDS, making us suspect that 
the 1967 measurement of the separation and position angle was incorrect. The consistency between our separation and position angle
and those measured by Hipparcos point in the same direction.
Another peculiarity of this visual pair is that the dimmer one (B) is bluer than the brighter one (A). As all stars
involved have similar effective temperatures there should be no photospheric color differences. The difference could be caused by small-scale extinction 
differences (not surprising in a region with complex extinction such as Cyg~OB2) or by wind-induced variability/emission lines in the A component.

\paragraph{Cyg~OB2-22~C = V2185~Cyg = ALS~15\,127.}                                                
\textbf{SB1E?.}
\citet{PiguKola98} identified this object as an eclipsing binary (see also \citealt{Salaetal15}) and \citet{Kimietal12} as a potential SB1
(but without an orbit). In GOSSS-I we classified it as O9.5~IIIn. It is a member of the Cyg~OB2-22 cluster (in the Cygnus OB2 association) 
that includes at least five other O-type visual components \citep{Maiz10a}. 

\paragraph{Cyg~OB2-22~B = ALS~19\,499~A.}                                                          
\textbf{SB1.}
This object is the second brightest star in the Cyg~OB2 cluster and is located 1\farcs5 away from the brightest star in the cluster, Cyg~OB2-22~A
(Fig.~1 in \citealt{Maiz10a} and Table~\ref{AstraLuxdata}), which was one of the first two O3 stars to be identified in the Northern hemisphere \citep{Walbetal02b}. 
In GOSSS~I we classified Cyg~OB2-22~B as O6~V((f)). \citet{Kobuetal14} identified it as a SB1 but note that paper provides an incorrect position angle between
A and B of 50~degrees: the correct value is 146~degrees (Table~\ref{AstraLuxdata} and Table~2 in \citealt{Maiz10a}), with no appreciable motion over a time span of
13 years. Those values actually refer to the A,Ba pair, as there is a dim companion, Bb, that is included in the spectroscopic aperture but is too dim to significantly 
influence the spectral classification. The Ba,Bb pair shows no appreciable motion over the same time span but in that case the uncertainties in the individual measurements
are larger (Table~\ref{AstraLuxdata}). All three components (plus others in the cluster) are included in the WDS.

\paragraph{Cyg~OB2-41 = ALS~15\,144.}                                                              
\textbf{SB1.}
Cyg~OB2-41 was classified as O9.7~III(n) in GOSSS~III; previous papers misidentified it as a B star. It was identified as a SB1 by \citet{Kobuetal14}.
This object has no WDS entry.

\paragraph{ALS~15\,148.}                                                                           
\textbf{SB1.}
\citet{Kobuetal14} identified this object as a SB1. It has not appeared in a GOSSS paper before and here we classify it as O6.5:~V, noting that the
spectrum has a S/N lower than most GOSSS spectra and needs to be repeated. \citet{Salaetal15} 
found a quasi-sinusoidal photometric variation with a period of 3.170~d, which corresponds to the orbital period of \citet{Kobuetal14}, so this 
object is likely an ellipsoidal variable. ALS~15\,148 has no WDS entry and appears to be single in our AstraLux data.

\paragraph{Cyg~OB2-9 = ALS~11\,422.}                                                               
\textbf{SB2.}
This target is a long-period highly eccentric SB2 system with strong non-thermal radio emission. \citet{Nazeetal12c} classified it as 
O5-5.5~I~+~O3-4~III. In GOSSS~I we derived a composite spectral type of O4.5~If, as the high eccentricity yields double lines only during
a small fraction of the orbit. With LiLiMaRlin we caught the system close to a periastron passage and derive a spectral classification of 
O4~If~+~O5.5~III(f), that is, the earlier-type component is the supergiant as opposed to the \citet{Nazeetal12c} result (note that our 
LiLiMaRlin spectrogram was obtained as part of NoMaDS, whose spectrograph does not include the 4710-4760~\AA\ region). In other words, the 
He, N, and C emission lines do not take place at the radial velocity of the stronger He\,{\sc i} absorption lines but at the radial velocity of the 
weaker ones. Cyg~OB2-9 has only a distant and weak companion that is likely unbound \citep{Maiz10a}. 
    
\paragraph{Cyg~OB2-1 = ALS~11\,401.}                                                               
\textbf{SB1+Ca.}
This object was identified as a SB1 by \citet{Kimietal08}. It has not appeared in a GOSSS paper before and here we classify it as O8~IV(n)((f)).
Cyg~OB2-1 has no entry in the WDS but with AstraLux (Table~\ref{AstraLuxdata} and Fig.~\ref{AstraLux2}) we detect a bright companion 1\farcs2 
away that also appears in Gaia~DR2.

\paragraph{ALS~15\,131.}                                                                           
\textbf{SB1.}
\citet{Kobuetal14} identified this target as a SB1. It has not appeared in a GOSSS paper before and here we classify it as O7.5~V((f)).
ALS~15\,131 has no WDS entry.

\paragraph{Cyg~OB2-20 = ALS~11\,404.}                                                              
\textbf{SB1.}
This target was identified as a SB1 by \citet{Kimietal09}. It has not appeared in a GOSSS paper before and here we classify it as O9.7~IV.
This object has no WDS entry and appears to be single in our AstraLux data.

\paragraph{Cyg~OB2-8~A = BD~$+$40~4227~A = ALS~11\,423~A.}                                         
\textbf{SB2.}
Cyg~OB2-8 is a cluster in the Cyg~OB2 association with at least four O-type visual components (see GOSSS~I+II). 
\citet{DeBeetal04}\footnote{Those authors published the only SB2 orbit available for this system but their paper does not give a value for $\omega$.}
classified Cyg~OB2-8~A as O6~+~O5.5 and in GOSSS-II we classified it as O6~Ib(fc)~+~O4.5:~III:(fc). Here, we use a LiLiMaRlin epoch to 
derive the same classification but without the uncertainties in the secondary i.e. O6~Ib(fc)~+~O4.5~III(fc).
The WDS catalog lists two dim nearby companions from \citet{Herb67}, Ab and Ac, but our AstraLux results (Table~\ref{AstraLuxdata} and 
Fig.~\ref{AstraLux2}) do not agree with the data there. Ab is not seen so it is either dimmer than
announced, has moved significantly (the single measurement in the WDS catalog is from 1961), or is a spurious detection. Ac is detected 
but it appears to have moved significantly since 1961, from a distance of 4\farcs0 and a position angle of 350\degr\ to a distance of 
3\farcs1 and a position angle of 340\degr\ in 2018, with no significant motion in the seven previous years. Therefore, as with the Cyg~OB2-5~A,B
pair, we suspect an error in the separation and position angle of the first epoch.

\paragraph{Cyg~OB2-70 = ALS~15\,119.}                                                              
\textbf{SB1.}
GOSSS~III classified this object as an O star (O9.5~IV(n)) for the first time. \citet{Kobuetal14} identified it as a SB1.
Cyg~OB2-70 has no WDS entry.

\paragraph{Cyg~OB2-15 = ALS~15\,102.}                                                              
\textbf{SB1.}
This target was identified as a SB1 by \citet{Kimietal09}. It has not appeared in a GOSSS paper before and here we classify it as O8~III.
Cyg~OB2-15 has no WDS entry and no significant companion appears in our AstraLux data.

\paragraph{ALS~15\,115.}                                                                           
\textbf{SB1.}
\citet{Kobuetal14} identified this target as a SB1. In GOSSS~III we classified it as O8~V.
ALS~15\,115 has no WDS entry.

\paragraph{Cyg~OB2-27~AB = ALS~15\,118~AB.}                                                        
\textbf{SB2E+Ca.}
\citet{RiosDeGi04} identified this object as a photometrically variable SB2. \citet{Salaetal15} published a light curve that showed 
that it is an eclipsing variable with a period of 1.46917$\pm$0.00002~d. In GOSSS~III we classified it as O9.7~V(n)~+~O9.7~V:(n), that is, with
the two stars more similar between them than the previous spectral classification of O9.5~V~+~B0~V of \citet{Kobuetal12}.
The WDS lists a close companion with a small $\Delta m$ \citep{CabNetal14}, hence the AB designation, that is not resolved 
in our AstraLux data.

\paragraph{Cyg~OB2-73.}                                                                            
\textbf{SB2.}
\citet{Kimietal09} identified Cyg~OB2-73 as an O+O SB2 system and gave it the spectral classification O8~III~+~O8~III. In GOSSS~III we give a 
classification of O8~Vz~+~O8~Vz, as the \HeII{4686} line is too deep to be a giant star in either case.
This object has no WDS entry and appears to be single in our AstraLux data.

\paragraph{ALS~15\,114.}                                                                           
\textbf{SB2.}
This object was identified as a SB2 by \citet{Kimietal12}, who classified it as O7~V~+~O9~V. It was included in GOSSS~III but at 
that time we could not resolve the two components as the spectrum was obtained at an unfavorable epoch. A new GOSSS spectrum obtained 
since then allows us to derive a new classification of O7~V((f))~+~O7~IV((f)), where the secondary is significantly earlier than the
previous classification. \citet{Salaetal15} identified a quasi-sinusoidal photometric variation with a 1.4316$\pm$0.0006~d period, 
which corresponds to one half of the orbital period of \citet{Kimietal12}, making it a likely ellipsoidal variable.
This object has no WDS entry.

\paragraph{Cyg~OB2-29 = ALS~15\,110.}                                                              
\textbf{SB1.}
\citet{Kobuetal14} identified this object as a SB1. In GOSSS~III we classified it as O7.5~V(n)((f))z.
Cyg~OB2-29 has no WDS entry and no significant companion appears in our AstraLux data.

\paragraph{Cyg~OB2-11 = BD~+41~3807 = ALS~11\,438.}                                                
\textbf{SB1.}
This object was identified as a SB1 by \citet{Kobuetal12}. In GOSSS~I we classified it as O5.5~Ifc.
Cyg~OB2-11 has no WDS entry and appears to be single in our AstraLux data.

\paragraph{LS~III~$+$46~11 = ALS~11\,448.}                                                         
\textbf{SB2.}
This target is a very massive eccentric SB2 binary composed of two near-twin stars, both O3.5~If*, whose nature was discovered with GOSSS
observations \citep{Maizetal15a}. It is located at the center of the Berkeley~90 cluster and has several dim companions (see Fig.~1 in 
\citealt{Maizetal15c}). 

\paragraph{HD 199\,579 = BD~$+$44~3639 = ALS~11\,633.}                                             
\textbf{SB2.}
\citet{Willetal01} published a SB1 orbit for this O star and used a Doppler tomography algorithm to search for the weak signal of the 
secondary, suggesting a B1-2~V spectral type for it and a mass ratio of $4\pm 1$ for the system (effectively making it a SB2 orbit with a
large $K_2$ uncertainty). In GOSSS~I we gave a spectral type of O6.5~V((f))z for the primary. The WDS catalog
lists a very dim companion 3\farcs8 away. 

\paragraph{68~Cyg = HD~203\,064 = V1809~Cyg = BD~+43~3877 = ALS~11\,807.}                          
\textbf{SB1?.}
\citet{Alduetal82} identified this object as a SB1 and \citet{GiesBolt86} provided ephemerides for the radial velocity variations but warned that
their origin is likely to be pulsational rather than orbital. In GOSSS~I we give a spectral classification of O7.5~IIIn((f)).
The WDS lists a dim companion 3\farcs8 away but we do not see it in our AstraLux data and it does not appear in Gaia DR2 either. 68~Cyg
is a runaway star \citep{Maizetal18b}.

\subsection{Cepheus-Camelopardalis}

\paragraph{HD~206\,267~AaAb = BD~$+$56~2617~AaAb = ALS~11\,937~AaAb.}                              
\textbf{SB2+Ca.}
This system was classified as O6.5~V((f))~+~O9.5:~V by \citet{Burketal97}. In GOSSS~I we derived the similar classification of 
O6.5~V((f))~+~O9/B0~V that here we modify with a new analysis of our GOSSS data to O6~V(n)((f))~+~B0:~V. A LiLiMaRlin spectrum
yields the same spectral classification.
HD~206\,267~AaAb is the central object of a high-order multiple system. The two closest components (Aa and Ab) have a small 
$\Delta m$ and are separated by just 0\farcs1 according to the WDS. Our AstraLux data (Table~\ref{AstraLuxdata} and Fig.~\ref{AstraLux1})
confirms that and detects a significant clockwise motion at near constant separation consistent with the previous WDS results that yields
a combined preliminary orbital period of 153$\pm$10~a. A third, dim, object, B, is located 1\farcs7 away and should not contribute 
significantly to the 
LiLiMaRlin spectra. The seven-year span between our first and last epochs does not yield a significant relative motion for the Aa,B pair.
Finally, two bright companions, C and D, are found within 25\arcsec\
of the primary. Therefore, three components are likely to contribute to the spectral classifications here: the two spectroscopic ones in Aa and the 
quasi-stationary in Ab \citep{Raucetal18}, implying that one of the spectral types is a composite. \citet{Raucetal18} also note that there are no 
apparent eclipses caused by the inner binary despite its short period.
    
\paragraph{14~Cep = LZ~Cep = HD~209\,481 = BD~$+$57~2441 = ALS~12\,096.}                           
\textbf{SB2.}
\citet{Mahyetal11b} classified this system as O9~III~+~ON9.7~V. In GOSSS~I we derived a classification of O9~IV(n)~var~+~B1:~V: that
is revised here with a new analysis to O8.5~III~+~BN0~V. A LiLiMaRlin epoch yields the similar spectral types O9~III~+~BN0~V. 
Despite its short period, the system is not eclipsing. In the WDS catalog a very dim component is located 2\farcs8 away but we do not see 
it in our AstraLux images. 

\paragraph{ALS~12\,502.}                                                                           
\textbf{SB2E+Ca (previously SB2E).}
This eclipsing binary system \citep{Macietal04} was identified as a SB2 by \citet{Neguetal04}, who classified it is as O8:~+~O9:. 
In GOSSS data we clearly see double lines and assign it the spectral classification O9~III:(n)~+~O9.2~IV:(n). Using a LiLiMaRlin we derive 
the classification O9~IV(n)~+~O9~V(n). In both the GOSSS and LiLiMaRlin we see evidence for the presence of a third light contaminating 
the spectrum. ALS~12\,502 has no entry in the WDS catalog but our AstraLux images (Fig.~\ref{AstraLux3}) show a complex system, quite 
possibly a previously poorly studied cluster \citep{Solietal12}. One component is located 1\farcs6 away (Table~\ref{AstraLuxdata}) 
and is the possible third light (note that we did not align the slit to separate that component as our GOSSS spectrum was obtained before 
our AstraLux data), hence the SBS typing as SB2E+Ca. Note that there is a significant magnitude difference between the two AstraLux epochs,
likely caused by the primary being in different eclipse phases.

\paragraph{DH~Cep = HD~215\,835 = BD~$+$57~2607 = ALS~12\,597.}                                    
\textbf{SB2.}
Despite its short period, DH~Cep is not an eclipsing binary but appears to be an ellipsoidal variable.
\citet{Burketal97} classified this system as O5.5~III(f)~+~O6~III(f) and in GOSSS~II we classified it as O5.5~V((f))~+~O6~V((f)). Here we
find the same spectral classification using a LiLiMaRlin epoch and we note that there are small differences in spectral classifications between
epochs, prompting us to add var suffixes. The WDS catalog lists only two distant companions.
    
\paragraph{ALS~12\,688.}                                                                           
\textbf{SB2E.}
This eclipsing binary system \citep{Lewaetal09} was identified as a spectroscopic binary in GOSSS~III, where a classification of 
O5.5~V(n)((fc))~+~B was given. No spectroscopic orbit has been published. This target has no entry in the WDS catalog. 

\paragraph{AO~Cas = HD~1337 = BD~$+$50~46 = ALS~14\,758.}                                          
\textbf{SB2E.}
\citet{BagnGies91} classified this system as O9.5~III~+~O8~V. In GOSSS~II we obtained O9.2~II~+~O8~V((f)). Here we use a LiLiMaRlin epoch
to obtain a similar classification of O9.2~II~+~O7.5~V. The WDS catalog lists only three dim companions, the closest one 2\farcs8 away.
 
\paragraph{HD~108 = BD~+62~2363 = ALS~6036.}                                                       
\textbf{SB1?.}
\citet{Hutc75} reported this object to be a SB1 but \citet{Nazeetal08d} pointed out that the spectroscopic variations are instead caused
by magnetic effects. Indeed, HD~108 is an Of?p star \citep{Walbetal10a} classified as O8~fp in GOSSS~I. Its magnetic/rotational cycle is 
$\sim$50~a long and its variations are very slow so when GOSSS~I was published we had not been collecting data for enough time to see the star
in both of its two extreme states but now we have. We show in Fig.~\ref{GOSSS} a 2009 observation when HD~108 was in its O8.5~fp state
(\HeI{4471} unfilled, no \CIII{4650} emission, H$\beta$ in absorption) and a 2017 observation in its O6.5~f?p state (\HeI{4471} partially
filled, \CIII{4650} in emission, H$\beta$ in emission with absorption wings), leading to a combined classification of O6.5-8.5~f?p~var.
The WDS lists a dim companion 3\farcs2 away but we do not see it in our AstraLux data and it does not appear in Gaia DR2 either.

\paragraph{V747~Cep = BD~+66~1673 = ALS~13\,375.}                                                  
\textbf{SB1E (previously E).}
\citet{Majaetal08} identified this system as an eclipsing binary but no spectroscopic orbits or separate spectral classifications have ever
been published to our knowledge. In GOSSS~III we derived a combined spectral type of O5.5~V(n)((f)) and our LiLiMaRlin data show a clear line
motion between epochs, allowing us to identify it as a SB1 system. V747~Cep has no entry in the WDS catalog and our AstraLux 
images reveal no significant companion.

\paragraph{HD~12\,323 = BD~+54~441 = ALS~6886.}                                                    
\textbf{SB1.}
HD~12\,323 was identified as a SB1 by \citet{BoltRoge78}. In GOSSS~II we classified it as ON9.2~V.
This object has no WDS entry. It is a runaway star \citep{Maizetal18b}.

\paragraph{DN~Cas = BD~$+$60~470 = ALS~7142.}                                                      
\textbf{SB2E+Ca (previously SB2E+C).}
\citet{Hilletal06} classified this system as O8~V~+~B0.2~V. DN~Cas had not appeared in GOSSS before and here we classify it 
as O8.5~V~+~B0.2~V. A LiLiMaRlin epoch yields the same spectral classification. \citet{Bakietal16} used precise eclipse timings 
to deduce the existence of a low-mass third body with a $\sim$42~a orbit, which at the $\sim$2~kpc distance to the system from Gaia DR2 
yields a semi-major axis of $\sim$20~mas. This target has no entry in the WDS catalog but our AstraLux data (Table~\ref{AstraLuxdata} and
Fig.~\ref{AstraLux1}) reveals a previously unknown visual component (which we label B) at a distance of 1\farcs07, indicating that DN~Cas 
has at least four components. However, as the third and fourth bodies do not contribute significantly to the integrated spectrum or to the
radial velocity of the other two components, we type DN~Cas simply as SB2E+Ca\footnote{Properly speaking, we should classify the system as SB2e+C+Ca, but
we do not because the third body is likely a low-mass star instead of an intermediate-mass one.}. DN~Cas~B is slightly redder than A in $z-i$, as 
expected from a coeval, young companion. Note, however, that we do not have multiple AstraLux epochs, so we do not know if the measured magnitude 
differences are affected by eclipses or not (see below for the similar IU~Aur~AB case).

\paragraph{BD~$+$60~497 = ALS~7266.}                                                               
\textbf{SB2.}
\citet{RauwDeBe04} gave a classification of O6.5~V((f))~+~O8.5-O9.5~V for this system, which does not show eclipses despite its short 
period. In GOSSS~I we classified it as O6.5~V((f))~+~O8/B0~V and here we reclassify it as O6.5~V(n)((f))z~+~O9.5:~V based on better data. 
Using LiLiMaRlin data instead of GOSSS we derive slightly later spectral types, O7~V(n)((f))z~+~B0:~V(n).
BD~$+$60~497 has no entry in the WDS catalog and no significant companion is seen in our AstraLux images.

\paragraph{HD~15\,558~A = BD~$+$60~502~A = ALS~7283.}                                              
\textbf{SB2.}
This long-period eccentric SB2 system was classified as O5.5~III(f)~+~O7~V by \citet{DeBeetal06a}. 
In GOSSS~II we classified the composite spectrum as 
O4.5~III(f), i.e., with a significantly earlier primary. With a new GOSSS spectrogram we obtain here O4.5~III(fc)p~+~O8:, where the
secondary is quite weak and, hence, its spectral type is rather uncertain. As an anomaly in the spectrum, the primary appears to have not 
only the significant \CIII{4650} emission characteristic of the Ofc stars \citep{Walbetal10a} but also \CIV{4658} in emission, hence the
p suffix. HD~15\,558 is a high-order multiple system with 13 components in the WDS catalog. The only bright component within 15\arcsec\ 
is HD~15\,558~B, located 9\farcs9 away. We aligned the GOSSS slit to acquire spectra of both components simultaneously and we obtained a 
classification of B0~V for B. 

\paragraph{HD~16\,429~A = BD~$+$60~541~A = ALS~7374.}                                              
\textbf{SB1+Ca.}
This system was classified by \citet{McSw03} as O9.5~II~+~O8~III-V~+~B0~V?, with the primary corresponding to the visual Aa 
component and the other two spectroscopic components to the Ab component, located 0\farcs28 away. What makes this target special is that
the $\Delta m$ between Aa and Ab is larger than two magnitudes (Fig.~\ref{AstraLux2} and Table~\ref{AstraLuxdata}, note that we have reanalyzed the
data in \citealt{Maiz10a}), making the signal of the spectroscopic binary very weak in the integrated spectrum. \citet{McSw03} actually underestimated 
the magnitude difference between Aa and Ab and was only able to compute a SB1 orbit for Ab after subtracting the constant-radial-velocity A component, 
leading us to suspect that the spectroscopic components may be in Aa instead of in Ab. In GOSSS~II we were only able to give a composite 
O9~II-III(n)~Nwk spectral classification. We attempted to separate Aa and Ab using lucky spectroscopy but we were unable to do so. HD~16\,429 
is a high-order multiple with three additional components listed in the WDS catalog. HD~16\,429~B is an F star (\citealt{McSw03}, GOSSS~I, and
Fig.~\ref{GOSSS}) and the recent Gaia DR2 results give a large relative proper motion between A and B, indicating that it is a foreground object,
something that our measurements (Table~\ref{AstraLuxdata}) confirm.
We aligned the GOSSS slit with the other bright component, HD~16\,429~C, located 53\arcsec\ away and we determined that its spectral 
type is B0.7~V(n).  

\paragraph{HD~17\,505 A = BD~$+$59~552~A = ALS~7499~A.}                                            
\textbf{SB2+C.}
This target is a system with spectral types O6.5~III((f))~+~O7.5~V((f))~+~O7.5~V((f)) according to \citet{Hilletal06}, with the 
secondary and tertiary forming a near identical pair with a short 8.571\,0~d period and the primary being stationary in their data
(\citealt{Raucetal18} does not detect a radial velocity change in the tertiary either and the inner binary does not produce eclipses, hence the 
SB2+C status for HD~17\,505~A). In 
GOSSS~II we gave a composite spectral classification of O6.5~IIIn(f). We have obtained a new GOSSS spectrum using lucky spectroscopy to which 
we assign a classification of O6.5~IV((f))~+~O7~V((f)), but one of the two spectral types should be a composite. The 
HD~17\,505~system is a high-order multiple, with the WDS catalog listing 20 components (Fig.~\ref{AstraLux3}). 
Some of those are likely to be not bound to the central stars but instead be part of the IC~4848 cluster where this system is located. 
HD~17\,505~B is close enough to A (Table~\ref{AstraLuxdata}) to be likely bound but too distant to contribute significantly
to the flux in the echelle aperture (unless seeing is poor or the guiding system is not set up to properly follow A).
In GOSSS~I we gave HD~17\,505~B a spectral classification of O8~V, raising the number of O stars in the system to four, a classification we confirm here
with the new lucky spectroscopy data. The WDS lists a first epoch
1830 separation of the A,B pair larger than the current one by 1\arcsec, which would imply a relative proper motion of $\sim$5~mas/a or a change of
$\sim$35~mas in the 7 years between our two AstraLux observations, an effect we do not observe in our data, where the two relative positions 
are essentially coincident. The recent Gaia DR2 proper motions also yield a relative proper motion for A,B of less than 1~mas/a, indicating that
the separation for the first epoch in the WDS is likely incorrect.

\paragraph{HD~15\,137 = BD~$+$51~579 = ALS~7218.}                                                  
\textbf{SB1.}
This system is a low-amplitude single-lined spectroscopic binary \citep{McSwetal07}. In GOSSS~I we classified it as O9.5~II-IIIn.
HD~15\,137 has no entry in the WDS catalog and our AstraLux images show no significant companion.

\paragraph{CC~Cas = HD~19\,820 = BD~$+$59~609 = ALS~7664.}                                         
\textbf{SB2E.}
\citet{Hilletal94} derived spectral types of O8.5~III~+~B0~V for this system. In GOSSS~I we obtained a composite spectral
type of O8.5~III(n)((f)). Here we publish a GOSSS spectrogram where the companion leaves a weak signal that allows us to classify CC~Cas 
as O8.5~III(n)((f))~+~B. The WDS catalog lists only a very weak component 2\farcs8 away that is undetected in our AstraLux images.

\paragraph{HD~14\,633~AaAb = BD~$+$40~501 = ALS~14\,760.}                                          
\textbf{SB1+Ca?.}
This system is a low-amplitude single-lined spectroscopic binary \citep{McSwetal07}. In GOSSS~I we classified it as ON8.5~V. The AaAb 
nomenclature is used because \citet{Aldoetal15} detected a possible companion a short distance away but with a large $\Delta m$ uncertainty,
a case similar to that of HDE~229\,232~AB above.  Our AstraLux images show no apparent companion but the Ab companion should be too close for 
detection. Note that HD~14\,633~AaAb is WDS~02228$+$4124~CaCb, as WDS~02228$+$4124~A corresponds to the brighter late-type HD~14\,622. 
This target is a runaway star \citep{Maizetal18b}.

\paragraph{$\alpha$~Cam = HD~30\,614 = BD~$+$66~358 = ALS~14\,768.}                                
\textbf{SB1?.}
This target has a single SB1 orbit published \citep{ZeinMusa86} which we deem suspicious as it is eccentric ($e = 0.45$), with a short period 
(3.6784~d), and a primary that is a late-O supergiant (in GOSSS~I we classified it as O9~Ia), making it difficult for the orbit not to be quickly 
circularized or even for the companion to stay outside the primary. Furthermore, there are no detected eclipses, making it likely that the
radial velocity variations are caused by pulsations. This object has no entry in the WDS catalog
and no significant companions are seen in our AstraLux images. $\alpha$~Cam is a runaway star \citep{Maizetal18b}.

\paragraph{MY~Cam~A = BD~$+$56~864~A = ALS~7836~A.}                                                
\textbf{SB2E+Cas (previously SB2E).}
\citet{Loreetal14} classified this extremely fast and short-period eclipsing binary as O5.5~V~+~O7~V. In GOSSS~III we obtained a similar
classification of O5.5~V(n)~+~O6.5~V(n) which we refine here to O5.5~Vnz~+~O6.5~Vnz with a lucky spectroscopy observation. 
Using a LiLiMaRlin epoch we classify the system as 
O6~V(n)~+~O6.5~V(n)z. We add the var suffix in all cases, as the spectral types appear to change as a function of the orbital phase.
MY~Cam~A has no entry in the WDS catalog and with our AstraLux images we detect a previously unknown B component 0\farcs73 away 
(Table~\ref{AstraLuxdata} and Fig.~\ref{AstraLux1}) for which a possible slight motion is detected, but it needs to be confirmed with future
observations. Our lucky spectroscopy observation allowed us to spatially separate the spectrum of MY~Cam~B and give it a classification of 
B1.5:~V (Fig.~\ref{GOSSS}). The B component is slightly redder than A in $z-i$, as expected from the spectral type difference.

\subsection{Auriga}

\begin{figure}
\centerline{\includegraphics*[width=\linewidth]{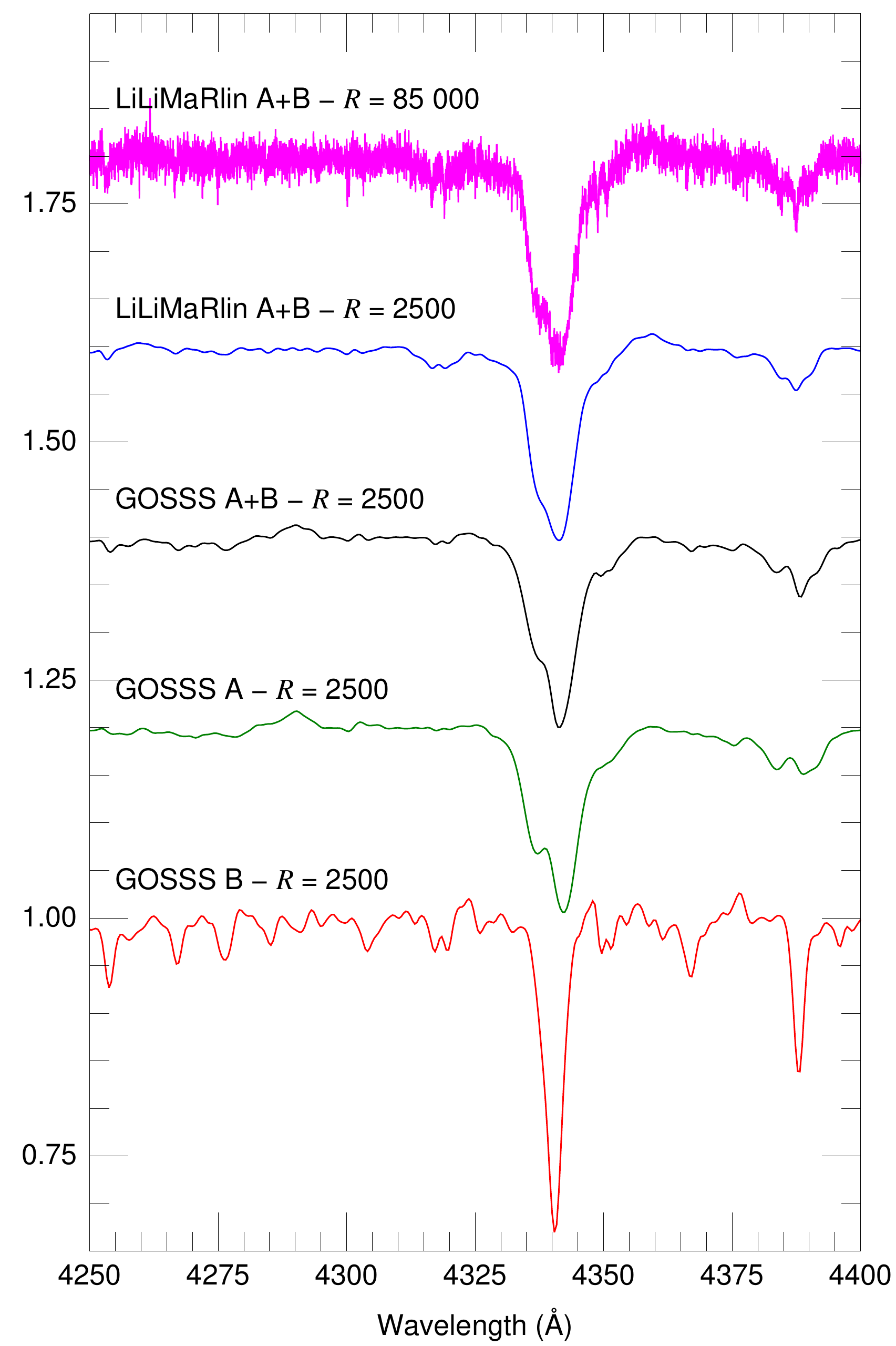}}
\caption{Plot equivalent to Fig.~\ref{HD_191_201_HeI_4922} for LY~Aur and the H$\gamma$ region (all plots are offset by 0.2 continuum 
         units with respect to the adjacent ones). Note that in this case there is a small phase difference between the GOSSS and 
         LiLiMaRlin epochs for the A orbit.}
\label{LY_Aur_Hgamma}
\end{figure}

\paragraph{LY~Aur~A = HD~35\,921~A = BD~$+$35~1137~A = ALS~8401~A.}                                
\textbf{SB2E+SB1as.}
\citet{Mayeetal13b} classified this system as O9~II~+~O9~III. In GOSSS~II we classified it as O9.5~II~+~O9~III. Here we present a new 
spectrogram obtained through lucky spectroscopy which we use to slightly revise the classification to O9.2~II(n)~+~O9.2~II(n). 
The visual B component is located 0\farcs6 away with a $\Delta m$ of 1.87~mag (\citealt{Maiz10a} and Fig.~\ref{AstraLux1}). 
Using lucky spectroscopy we aligned the slit to obtain spatially separated spectra for each component 
and derived a GOSSS spectral type of B0.2~IV for LY~Aur~B (Fig.~\ref{LY_Aur_Hgamma}, note that as A and B are separated in GOSSS, B
is not included in the star name), not far from the B0.5 prediction of \citet{Mayeetal13b}. According to those authors, the B 
component is a SB1, so LY~Aur~AB contains four stars, making it the only system in this paper with four objects within the high-resolution aperture 
that induce detectable radial velocity variations (the second component of LY~Aur~B is not seen in the spectra but if it were its radial velocity shifts 
should be larger than those of B itself).

\paragraph{IU~Aur~AB = HD~35\,652~AB = BD~$+$34~1051~AB = ALS~8369~AB.}                            
\textbf{SB2E+Ca.}
\citet{Ozdeetal03} classified this system as O9.5~V~+~B0.5~IV-V. IU~Aur~AB had not appeared in GOSSS papers before and here we classify 
it as O9.5~IV(n)~+~O9.7~IV(n). Using LilLiMaRlin we obtain a different spectral classification of O8.5 III(n)~+~O9.7~V(n). We believe the 
discrepancies between those spectral classifications are caused by [a] the known presence of a third light that is being added to the primary 
or the secondary depending on the phase of this very fast binary and [b] eclipse effects. Therefore, those spectral classifications should be
taken as provisional until a more in-depth multi-epoch study is performed. Nevertheless, we can already point out that it is already clear that
the secondary must be O9.7 or earlier based on the \SiIII{4552}/\HeII{4542} seen in different phases. For the third object \citet{Ozdeetal03} 
measured an orbit with a period of 293.3~d through light-time effects and estimated that the third light contributes a 24\% 
of the total light (which corresponds to a $\Delta m$ of 1.25). On the other hand, the WDS catalog lists a B companion 0\farcs1 away with a $\Delta m$ of 
1.36$\pm$1.01 mag discovered by Hipparcos. However, that component should have an orbit measured in centuries, implying there are two different 
companions. The question is which one of them is the main contributor to the third light, the outer B or the intermediate unnamed one with the 293.3~d
period? \citet{Ozdeetal03} favor the inner one based on the large $\Delta m$ uncertainty measured by Hipparcos for the outer one. Our AstraLux data
(Table~\ref{AstraLuxdata} and Fig.~\ref{AstraLux2}) shows that the A,B pair has a variable $\Delta z$ between 1.40 and 2.06 magnitudes, where the 
variability fits the known eclipses of the A component. Taking the upper value as the one that corresponds to the uneclipsed A yields a contribution of B
to the total light of 13$\pm$1\%. Therefore, the outer (B) component appears to be the dominant contribution to the third light but further analyses
are needed to determine whether the intermediate unnamed component also has a significant contribution. Our AstraLux images reveal the existence of two 
dim previously undetected companions 3\farcs6 and 4\farcs0 away (lower left quadrant of the IU~Aur~AB panel in Fig.~\ref{AstraLux2}), to which we 
assign the respective component classifications of C and D. Note that if future observations using e.g. interferometry resolve the intermediate component
it should be identified as Ab, leaving Aa for the SB2E primary. Finally, we note that the four inner components of IU~Aur have the same 
configuration as DN~Cas (but with a smaller $\Delta m$ between A and B), so the same comment about classifying it as SB2E+Ca or SB2E+C+Ca applies here.

\paragraph{HD~37\,737 = BD~$+$36~1233 = ALS~8496.}                                                 
\textbf{SB1.}
This target is a single-lined spectroscopic binary \citep{McSwetal07} that was classified as O9.5~II-III(n) in GOSSS~I. It has no entry in the 
WDS catalog and our AstraLux images show no significant nearby companion.

\paragraph{HD~37\,366 = BD~$+$30~968 = ALS~8472.}                                                  
\textbf{SB2.}
\citet{Boyaetal07a} classified this star as O9.5~V~+~B0-1~V. In GOSSS~I we gave a composite classification of O9.5~IV but with a new 
spectrogram we are able to give it a new classification as O9.5~V~+~B1:~V, which we also reproduce with a LiLiMaRlin spectrogram. 
The WDS catalog lists a dim companion 0\farcs6 away that we also see in our AstraLux data (Fig.~\ref{AstraLux1}). 

\subsection{Orion-Monoceros}
 
\paragraph{15~Mon~AaAb = S~Mon~AaAb = HD~47\,839~AaAb = BD~+10~1220~AaAb = ALS~9090~AaAb.}         
\textbf{SB1a.}
15~Mon~AaAb is a visual + SB1 binary whose orbit has been studied in different papers \citep{Giesetal93,Giesetal97,Cvetetal09,Cvetetal10,Toko18b} with diverging
results. The discrepancies arise because only a partial orbit has been observed (with a periastron around 1996), the radial velocity amplitude of the
spectroscopic orbit is small (with historical measurements having a relatively large scatter), and the orbit has a large eccentricity. A third component, B, 
is located 3\arcsec\ away from the Aa+Ab pair. In \citet{Maizetal18a} we used lucky spectroscopy to separate AaAb (O7~V((f))z~var) from B (B2:~Vn). 
Our AstraLux data resolves the three components and shows a significant orbital motion of Ab with respect to Aa in the lapse of ten years. Also, in a manner
similar to what we see for the HD~193\,322~Aa+Ab+B triple system, there is a significant change in the relative position of B with respect to Aa, in this case
more easily seen in position angle (Table~\ref{AstraLuxdata} and Fig.~\ref{AstraLux2}). The detected motion is compatible with the effect of the Aa,Ab orbit 
in the position of Aa. This system and its cluster, NGC~2264, will be analyzed in a separate paper (Ma{\'\i}z Apell\'aniz 2019, in preparation). 

\paragraph{$\delta$~Ori~Aa = Mintaka~Aa = HD~36\,486~A = BD~$-$00~983~A = ALS~14\,779~A.}          
\textbf{SB2E+Cas.}
\citet{Harvetal02} classified this object as O9.5~II~+~B0.5~III and noted the complexity of the system. It is composed of an inner 
eclipsing binary (Aa1+Aa2) with a 5.7~d period (to whose components belong their spectral types) with a visual companion (Ab) located 
0\farcs3 away (Fig.~\ref{AstraLux1}). All three stars are bright enough to contribute to the spectrum, though Aa1 is the brightest 
\citep{Pabletal15,Shenetal15}. In \citet{Maizetal18a} we applied the novel lucky spectroscopy technique to spatially separate Aa from 
Ab and obtain a GOSSS spectral classification of O9.5~IINwk for Aa and one of O9.7~III:(n) for Ab (as Aa and Ab are separated in GOSSS, Ab
is not included in the star name). We were unable to separate the spectroscopic components in Aa because the two epochs 
we obtained were not close to quadrature but we did observe some small variations between the two spectrograms that were indicative of 
orbit-induced radial belocity changes. Also, there were differences between the Aa spectra and the AaAb one we presented in GOSSS~I. Most notably, some lines 
were narrower, as Ab is a moderately fast rotator. Note that there is some confusion with the component nomenclature for this system. For
example, Simbad currently calls ``*del~Ori~B'' what we call $\delta$~Ori~Ab here (following the WDS catalog). There are two distant 
components of the system, one relatively bright (C) and one very faint (B). $\delta$~Ori~C has its own Henry Draper designation 
(HD~36\,485). 

\paragraph{HD 47\,129 = Plaskett~star = V640~Mon = BD~$+$06~1309 = ALS~9054.}                      
\textbf{SB2.}
The binarity of this object was discovered by \citet{Plas22}. \citet{Lindetal08} classified it as O8~III/I~+~O7.5~V/III and noted 
the variability and complexity of the spectra (see also \citealt{Struetal58}). 
Some of the variability of the spectrum was assigned to the magnetic field of the 
secondary by \citet{Grunetal13}. In GOSSS~I we classified the composite spectrum as O8~fp~var. Here we present a new GOSSS spectrogram
that yields O8~Iabf~+~O8.5:fp, where we do not assign a luminosity class to the secondary due to the peculiarity of the spectrum. The WDS
catalog lists three faint components within 1\farcs2, of which the most distant one appears in our AstraLux images (Fig.~\ref{AstraLux1})
    
\paragraph{HD~48\,099 = BD~$+$06~1351 = ALS~9098.}                                                 
\textbf{SB2.}
\citet{Mahyetal10} classified this system as O5.5~V((f))~+~O9~V. It shows no eclipses despite its short period. In GOSSS~III we found 
the same spectral classification with the only addition of a z~suffix to the primary. We also have a LiLiMaRlin spectrum at the correct 
phase and found the same spectral classification as GOSSS but with an added uncertainty to the spectral subtype of the secondary.
HD~48\,099 has no entry in the WDS catalog and appears single in our AstraLux data. 

\paragraph{HD~46\,149 = BD~$+$05~1282 = ALS~8983.}                                                 
\textbf{SB2.}
\cite{Mahyetal09} classified this system as O8~V~+~B0-1~V. In GOSSS~I we obtained a composite classification of O8.5~V. The target has no
entry in the WDS catalog and we see no obvious companion in our AstraLux data.

\paragraph{$\zeta$~Ori~AaAb = Alnitak~AaAb = HD~37\,742~AB = BD~$-$02~1338~AaAb = ALS~14\,793~AB.} 
\textbf{SB2.}
\citet{Hummetal13a} give spectral types of O9.5~Ib~+~B0.5~IV for this object
but it should be noted that their spectral type for the secondary is 
estimated, not directly measured, as the $\Delta m$ between the two components is large (2 magnitudes) and the radial velocity differences are
small. Another important point is that their spectroscopic secondary is the visual Ab, as $\zeta$~Ori~AaAb is one of the few spectroscopic
systems with a full interferometric orbit, with a seven-year period. $\zeta$~Ori~B is another bright component 2\farcs4 away
(Fig.~\ref{AstraLux2}), too far to significantly contribute to the LiLiMaRlin spectra. In
\citet{Maizetal18a} we applied lucky spectroscopy to obtain separate spectra for AaAb and B. Both are of O-type, with the composite AaAb
classified as O9.2~IbNwk~var (slightly earlier than the \citealt{Hummetal13a} classification for Aa) and B as O9.7~IIIn. 

\paragraph{$\sigma$~Ori~AaAb = HD~37\,468~AaAb = BD~$-$02~1326~AaAb = ALS~8473~AaAb.}              
\textbf{SB2a+Cas.}
\citet{SimDetal11b} classified this system as O9.5~V~+~B0.5~V, where the interferometric visual pair Aa,Ab \citep{Schaetal16} is also a SB2 
with a 143.198$\pm$0.005~d period \citep{SimDetal15a}. The aperture used by \citet{SimDetal11b} included the other bright visual companion, B, currently located 
0\farcs26 away and with a 159.9~a period, for which they provided a spectral classification of B0/1~V. 
Figure~\ref{AstraLux1} shows two AstraLux observations of $\sigma$~Ori where the clockwise motion of B with respect to AaAb is apparent over the 
ten years spanned. In \citet{Maizetal18a} we applied lucky spectroscopy to obtain spatially separated spectra for AaAb and B for the first time
(as AaAb and B are separated in GOSSS, B is not included in the star name). Here we present a new lucky spectroscopy observation with a crucial
difference with the previous one: it was obtained near the Aa,Ab periastron. This allows to spatially+kinematically separate the three components 
(Fig.~\ref{GOSSS}) and derive three spectral classifications with a single observation for the first time. Aa is O9.5~V, Ab is B0.2~V, and B is B0.2~V(n).
$\sigma$~Ori is a high-order multiple system with many entries in the WDS catalog and is located at the center of the cluster with the same name \citep{Caba14}. 

\paragraph{HD~46\,573 = BD~+02~1295 = ALS~9029.}                                                   
\textbf{SB1.}
\citet{Mahyetal09} identified it as a SB1. In GOSSS~I we classified it as O7~V((f))z.
HD~46\,573 has no WDS entry and appears single in our AstraLux data.

\paragraph{$\theta^1$~Ori~CaCb = HD~37\,022~AB = BD~$-$05~1315~CaCb = ALS~14\,788~AB.}             
\textbf{SB1?+Sa.}
As the main ionizing source of the Orion nebula, $\theta^1$~Ori~CaCb is one of the most studied O stars. It is also one of the
most complicated ones in terms of variability, with different periods from 15~d to 120~a having been proposed in the last two decades. Nowadays we
understand that the shortest 15.422$\pm$0.002~d period \citep{Stahetal96} is rotational and its associated variations are caused by magnetic effects
\citep{Donaetal02}. \citet{Weigetal99} discovered a close companion ($\theta^1$~Ori~Cb) for which an eccentric astrometric+SB1 orbit with a period of 
$\sim$11~a was computed later on by \citet{Krauetal09b}. As a further complication, \citet{Lehmetal10} proposed that the primary (Ca) is itself a SB1 with 
a 61.5~d period but this has not been confirmed and the 4:1 relationship between the rotation period of the primary and the alleged inner period could be 
a real resonance or a data artifact. Hence the SB1?+Sa typing. Note that the Ca+Cb pair is too close to be resolved with AstraLux. $\theta^1$~Ori~CaCb 
was classified in GOSSS~I as O7~Vp. In Fig.~\ref{GOSSS} we show two GOSSS spectra taken with the same setup eight days apart to illustrate the variations
associated with its magnetic cycle. The most obvious one is the transformation of \HeII{4686} from an absorption profile to an inverse P-Cygni one but
others are also visible, such as a change in the H$\beta$ profile caused by the similar apparition of an emission component that is not strong enough to 
emerge from the photospheric absorption but is sufficient to significantly change the shape and equivalent width of the line. Note also that \CIII{4650}
is in emission with an intensity similar to that of \NIII{4634} and variable, the most prominent defining characteristic of the Of?p phenomenon. Therefore,
here we add $\theta^1$~Ori~CaCb to the group of Galactic Of?p stars (becoming the sixth member of the club) and give it a spectral classification of
O7~f?p~var.

\paragraph{$\theta^2$~Ori~A = HD~37\,041 = BD~$-$05~1319~A = ALS~14\,789.}                         
\textbf{SB2 (previously SB1).}
\cite{SticLloy01} gave spectral types of O9~V~+~B0.5~V for this system but we should note that they only calculated a SB1 orbit and 
the secondary spectral type is an estimation, not a real measurement. In GOSSS~I we derived a composite spectral type of O9.5~IVp and 
here we present a new spectrogram where we obtain a separate classification of O9.2~V~+~B0.5:~V(n), which we believe to be the first time
a spectral type is measured for the secondary. A LiLiMaRlin spectrogram at a different orbital phase than the GOSSS one yields a
similar classification of O9.5~V~+~B0.2:~V. There is a faint visual Ab component listed in the WDS catalog with a $\Delta m$ of 3.2 mag and a separation of 
0\farcs4 which we also detect in our AstraLux images (Fig.~\ref{AstraLux1}). 
    
\paragraph{$\iota$~Ori = Hatysa = HD~37\,043 = BD~$-$06~1241 = ALS~14\,790.}                       
\textbf{SB2.}
\citet{Sticetal87} classified this eccentric SB2 system with apsidal motion as O9~III~+~B1~III. In GOSSS~I we derived a composite 
spectral classification of O9~IIIvar. Here we present a new spectrogram with separate spectral types of O8.5~III~+~B0.2:~V. With 
a LiLiMaRlin spectrogram we achieve a slightly better separation and we obtain a classification of O8.5~III~+~B0.2~V (the same as GOSSS
without the uncertainty in the spectral subtype of the secondary). The WDS catalog lists a weak visual companion 0\farcs1 away but it 
should not influence the spectral classification. 

\paragraph{HD~52\,533~A = BD~$-$02~1885~A = ALS~9251~A.}                                           
\textbf{SB1.}
In GOSSS~I we classified this single-lined spectroscopic binary \citep{McSwetal07} as O8.5~IVn. HD~52\,533 has nine components listed in 
the WDS catalog, including three nearby dim ones (Ab, B, and G) visible in Fig.~\ref{AstraLux2} (see also Table~\ref{AstraLuxdata}. 
The second brightest component listed is
C (=~BD~$-$02~1886), 22\arcsec\ to the north, which is a late-type star with a Gaia~DR2 proper motion very different from that of A, so
it is likely a foreground object.

\paragraph{HD~54\,662~AB = BD~$-$10~1892~AB = ALS~197~AB.}                                         
\textbf{SB2a.}
\citet{Boyaetal07a} classified this SB2 system as O6.5~V~+~O7-9.5~V. In GOSSS~II we give a composite spectral type of O7~Vz~var?. 
\citet{LeBoetal17} have resolved this spectroscopic binary into their visual orbit and have been able to compute an orbit with a period
of 2103.3~d, a value that agrees well with the previous value quoted in GOSSS~II based on OWN data (2119~d) and with the more recent 
determination by \citet{Mossetal18} of 2103.4~d. According to \citet{Tetzetal11} it is a runaway star based on its radial velocity 
(we did not detect it in \citealt{Maizetal18b} because its proper motion does not deviate significantly from the average values for O 
stars in that direction of the Galaxy). It has no entry in the WDS catalog and appears single in our AstraLux images.
    
\paragraph{HD~53\,975 = BD~$-$12~1788 = ALS~166.}                                                  
\textbf{SB2.}
\citet{Giesetal94} classified this object as O7.5~V~+~B2-3~V with a extreme magnitude difference between primary and secondary that
only allowed an estimate for the secondary spectral type but not an orbital measurement. In GOSSS~I we classified it as O7.5~Vz. 
The WDS catalog has no entry for this star and our AstraLux image does not reveal any significant companion.

\section{Analysis}

$\,\!\indent$ In this final section we summarize our results and present our first conclusions. In future papers we will present a more thorough analysis
once we include the southern systems and additional northern stars.

In this first MONOS paper we have presented updated spectroscopic classifications and visual multiplicity information for 92 multiple systems with 
$\delta > -20\degr$ that include at least one O+OBcc spectroscopic binary. The last similar analysis was that of M98, 
which had a different scope as it concentrated more on the astrometric than on the spectroscopic systems (as opposed to this paper), covered both
hemispheres, its sample selection was different, and no new spectral classifications were presented. 
Nevertheless, it is interesting to compare the information in common between both works. 
M98 identified 24 O+OBcc systems with $\delta > -20\degr$ and published SB1 or SB2 orbits\footnote{We select systems with an O in the spectroscopic status 
given in Table~1 of M98.}. We correct that number by including IU~Aur~AB (classified as 
B0.5~+~B0.5 in M98) and excluding SZ~Cam~AB and RY~Sct (which we classify as B stars) to yield 23 systems. The sample in this paper quadruples that 
number, a reflection of the important progress made in the study of spectroscopic binaries in the last two decades. Furthermore, if we look at those 23 systems
and compare the number of astrometric companions within 6\arcsec\ of the central source detected by M98 and compare it with the current number (including 
objects from the WDS catalog, Gaia DR2, and companions detected in this paper, see below) the number has grown from 12 to 28, an equally impressive advancement.
As we will show in future papers, there is still room for significant improvements in the case of new spectroscopic orbits. The same should be true for
the discovery of new astrometric companions, not only using lucky imaging but also more sophisticated techniques such as the ones used in the southern 
hemisphere by the SMaSH+ survey \citep{Sanaetal14}.

The value of our results lies mostly in the updated detailed information, such as the homogeneous revision of the spectral classification of the sample applying 
the criteria we have developed for GOSSS, that can be used for future global analyses of the multiplicity in O stars. We indeed plan to do such analyses in future 
papers ourselves but it is also worthy to point out some results for individual systems in this work. We present first-time O-type GOSSS spectral classifications
for 17 objects: HD~170\,097~A, QR~Ser, BD~$-$13~4923, HD~190\,967, HD~193\,611, HDE~228\,989, HDE~229\,234, Cyg~OB2-3~A, ALS~15\,133, ALS~15\,148, Cyg~OB2-1, 
ALS~15\,131, Cyg~OB2-20, Cyg~OB2-15, ALS~12\,502, DN~Cas, and IU~Aur~AB. For another six objects we present their first-time B-type GOSSS spectral 
classifications: HD~193\,322~B, Cyg~OB2-3~B, HD~15\,558~B, HD~16\,429~C, MY~Cam~B, and LY~Aur~B. 

We have identified one eclipsing system as SB2 for the first time, HD~170\,097~A, and derived its separate spectral classifications. 
To that one we can add a second 
eclipsing system in the sample, ALS~12\,688, that was identified as having double absorption lines for the first time in 
GOSSS~III. For another eclipsing binary, V747~Cep, we detect moving lines (SB1) for the first time using our LiLiMaRlin spectra. For two systems that 
have published SB1 orbits, BD~$-$16~4826 and $\theta^2$~Ori~A, we publish the first spectrograms with double lines (SB2) and derive their first true 
separate spectral classifications. For HD~168\,112~AB we also publish the first SB2 spectrograms and separate spectral classification.
We have also detected nine new astrometric components, one each around HD~190\,967, ALS~15\,133, Cyg~OB2-A11, Cyg~OB2-1, ALS~12\,502, 
DN~Cas, and MY~Cam~A and two around IU~Aur~AB.

The results of this paper will be added to GOSC at the time of publication, which will reach version 4.2. With them, GOSC will 
contain a total of 655 objects: 611 in the main catalog (Galactic O stars),
32 in supplement 2 (Galactic early-type stars), and 12 in supplement 3 (Galactic late-type stars).

It is not possible to do a full multiplicity analysis based on our 92 stars for two reasons: the sample is too small and is biased. The most obvious
bias is that we selected our stars from systems that had spectroscopic orbits previously published, so no single targets are included. Note, however, that
several SB1? targets are likely to be removed from the list once we check they are not really spectroscopic binaries. Another bias is
that objects with spectroscopic orbits tend to be bright (and, therefore, close to us), as multi-epoch high-resolution spectroscopy is expensive in observing 
time. Indeed, our 92 stars have an average $B$ magnitude of 9.4 with a range between 1.9 and 15.9. The other 428 O stars with GOSSS spectral classifications and
$\delta > -20\degr$ have an average $B$ magnitude of 10.5 with a range between 2.6 and 17.2. Nevertheless, we can still study the different proportions of double,
triple, and higher-order systems in a relatively unbiased way (other than the fact that many companions are likely to remain undetected, see above). To do such a study
we have compiled the statistics on: [1] $m_1$, the multiplicity as derived from the SBS, as defined above, which is the number of objects shown in the diagrams in
Fig.~\ref{sbstatus} and [2] $m_2$, the multiplicity derived from the number of objects within 6\arcsec\ from the SBS, the WDS catalog, Gaia DR2, low-mass objects detected 
through eclipse timings, and the new astrometric companions discovered here (being careful not to incur into double accounting). $m_1$ reflects the number of massive
and nearby (known) companions while $m_2$ adds more distant and less massive companions. $m_1$ should be relatively robust and, barring technological advances 
(e.g. an extreme version of lucky spectroscopy), we expect it not to change for most stars in the sample in the coming years, especially for objects with extensive 
previous coverage. $m_2$, on the other hand, should increase as new astrometric components are added or decrease as some of them are shown to be unbound to the central
object. The statistics for $m_1$ and $m_2$ are shown in Table~\ref{multstat}.

\begin{table}
\caption{Multiplicity statistics for the sample of 92 objects in this paper. The last row and column give the percentages for the total.}
\label{multstat}
\centerline{
\begin{tabular}{lrrrrrrr}
\hline
      & \multicolumn{6}{c}{$m_2$}                           &      \\
$m_1$ &      2 &      3 &      4 &      5 &      6 &      7 &   \% \\
\hline
2     &     36 &     25 &      4 &      3 &      1 &      1 & 76.1 \\
3     & \ldots &      6 &     12 &      2 &      1 & \ldots & 22.8 \\
4     & \ldots & \ldots &      1 & \ldots & \ldots & \ldots &  1.1 \\
\%    &   39.1 &   33.7 &   18.5 &    5.4 &    2.2 &    1.1 &      \\
\hline
\end{tabular}
}
\end{table}

Table~\ref{multstat} shows that most ($\sim$3/4) O-type spectroscopic multiple systems have just one massive nearby companion. On the other hand, only a minority of 
them have just one companion once we take into account distant and low-mass objects: about two fifths are double systems, another third are triple systems, and the rest 
are of a higher order. 
Massive stars prefer high-order multiplicity over simple binarity. 

\begin{acknowledgements}
We would like to thank Joel S\'anchez Berm\'udez, Michelangelo Pantaleoni Gonz\'alez, and the Calar Alto staff for their help with the AstraLux campaigns and
Miguel Penad\'es Ordaz for his help with the data compilation.
We also thank Sa{\'\i}da Caballero-Nieves for revising the data from \citet{Aldoetal15} for some targets and Brian Mason for providing us with the WDS detailed data and 
for his efforts maintaining the catalog.
Several authors acknowledge support from the Spanish Government Ministerio de Ciencia, Innovaci\'on y Universidades through different grants:
AYA2016-75\,931-C2-1/2-P (J.M.A., E.T.P., A.S., and E.J.A.), 
AYA2015-68\,012-C2-1/2-P (E.T.P., I.N., S.S.-D., J.L., and A.M.), 
AYA2016-79\,425-C3-2-P (J.A.C.), and
SEV2015-0548 (S.S.-D.).
R.H.B. acknowledges support from the ESAC Faculty Council Visitor Program. 
R.H.B. and J.I.A. were also supported by the Direcci\'on de Investigaci\'on y Desarrollo de la Universidad de La Serena through projects PR18\,143 and PR16\,142, respectively.
S.S.-D. acknowledges support from the Gobierno de Canarias grant ProID2017010115.
This research has made extensive use of the SIMBAD database (indeed, we would not have attempted if it did not exist), operated at CDS, Strasbourg, France. 
\end{acknowledgements}

\bibliographystyle{aa} 
\bibliography{general} 

\begin{landscape}
\begin{table}
\caption{Coordinates and spectral classifications for the sample in this paper sorted by GOSC ID. For the GOSSS spectral classifications we indicate which ones are {\bf mod}ified, {\bf new}, or correspond to {\bf vis}ual companions. For the alternate spectral classifications we give the reference. The information in this table is also available in electronic form at the GOSC web site (\url{http://gosc.cab.inta-csic.es}) and at the CDS via anonymous ftp to \url{cdsarc.u-strasbg.fr} (130.79.128.5) or via \url{http://cdsweb.u-strasbg.fr/cgi-bin/qcat?J/A+A/}.}
\label{spclas}
\centerline{
\scriptsize
\begin{tabular}{lccclllllllllll}
\hline
\hline
Name                  & GOSC ID            & R.A.         & Declination    & SBS        & \multicolumn{5}{c}{GOSSS classification}                          & \multicolumn{5}{c}{Alternate classification}                      \\
                      &                    & (J2000)      & (J2000)        &            & ST     & LC    & Qualifier   & Secondary (+tertiary)     & Ref.   & ST     & LC    & Qualifier   & Secondary (+tertiary)     & Ref.   \\
\hline
HD 164\,438           & 010.35$+$01.79\_01 & 18:01:52.279 & $-$19:06:22.07 & SB1        & O9.2   & IV     & \ldots     & \ldots                    & S14    & \ldots & \ldots & \ldots     & \ldots                    & \ldots \\
HD 167\,771           & 012.70$-$01.13\_01 & 18:17:28.556 & $-$18:27:48.43 & SB2        & O7     & III    & ((f))      & O8 III                    & S14    & O7     & III    & ((f))      & O8 III                    & new    \\
BD $-$16 4826         & 015.26$-$00.73\_01 & 18:21:02.231 & $-$16:01:00.94 & SB2        & O5.5   & V      & ((f))z     & \ldots                    & M16    & O5     & V      & ((f))z     & O9/B0 V                   & new    \\
HD 170\,097 A         & 015.48$-$02.61\_01 & 18:28:25.109 & $-$16:42:04.44 & SB2E       & O9.5   & V      & \ldots     & B1: V                     & new    & O9.5   & V      & \ldots     & B1: V                     & new    \\
QR Ser                & 016.81$+$00.67\_01 & 18:18:58.690 & $-$13:59:28.45 & SB2E       & O9.7   & III    & \ldots     & \ldots                    & new    & O9.5   & III    & \ldots     & B                         & S09    \\
V479 Sct              & 016.88$-$01.29\_01 & 18:26:15.045 & $-$14:50:54.33 & SB1        & ON6    & V      & ((f))z     & \ldots                    & M16    & \ldots & \ldots & \ldots     & \ldots                    & \ldots \\
HD 168\,075           & 016.94$+$00.84\_01 & 18:18:36.043 & $-$13:47:36.46 & SB2        & O6.5   & V      & ((f))      & \ldots                    & M16    & O6.5   & V      & ((f))      & B0-1 V                    & S09    \\
HD 168\,137 AaAb      & 016.97$+$00.76\_01 & 18:18:56.189 & $-$13:48:31.08 & SB2a       & O8     & V      & z          & \ldots                    & M16    & O7.5   & V      & z          & O8.5 V                    & new    \\
BD $-$13 4923         & 016.97$+$00.87\_01 & 18:18:32.732 & $-$13:45:11.88 & SB2        & O4     & V      & ((f))      & O7.5 V                    & new    & O4     & V      & ((f))      & O7.5 V                    & new    \\
MY Ser AaAb           & 018.25$+$01.68\_01 & 18:18:05.895 & $-$12:14:33.29 & SB2E+Sa    & O8     & Ia     & f          & O4/5 If + O4/5 V-III      & mod    & O8     & Iab    & f          & O4.5 If + O4: V-III       & new    \\
HD 168\,112 AB        & 018.44$+$01.62\_01 & 18:18:40.868 & $-$12:06:23.39 & SB2a       & O5     & IV     & (f)        & O6: IV:                   & mod    & O4.5   & III    & (f)        & O5.5 IV((f))              & new    \\
HD 166\,734           & 018.92$+$03.63\_01 & 18:12:24.656 & $-$10:43:53.04 & SB2E       & O7.5   & Ia     & f          & O9 Iab                    & mod    & O7.5   & Ia     & f          & O8.5 Ib(f)                & new    \\
HD 175\,514           & 041.71$+$03.38\_01 & 18:55:23.124 & $+$09:20:48.07 & SB2E+C     & O7     & V      & (n)((f))z  & B                         & M16    & O5.5   & V      & ((f))      & B0.5: V + O7.5 IV((f))    & new    \\
9 Sge                 & 056.48$-$04.33\_01 & 19:52:21.765 & $+$18:40:18.75 & SB1?       & O7.5   & Iab    & f          & \ldots                    & S11a   & \ldots & \ldots & \ldots     & \ldots                    & \ldots \\
Cyg X-1               & 071.34$+$03.07\_01 & 19:58:21.677 & $+$35:12:05.81 & SB1        & O9.7   & Iab    & p var      & \ldots                    & S11a   & \ldots & \ldots & \ldots     & \ldots                    & \ldots \\
HD 190\,967           & 072.33$+$01.81\_01 & 20:06:09.949 & $+$35:23:09.61 & SB2E       & O9.7:  & V      & \ldots     & B1.5 Iab                  & new    & O9.7:  & V      & \ldots     & B1.5 II                   & new    \\
HD 191\,201 A         & 072.75$+$01.78\_01 & 20:07:23.684 & $+$35:43:05.91 & SB2+Cas    & O9.5   & III    & \ldots     & B0 IV                     & S11a   & O9     & III    & \ldots     & O9 V                      & C71    \\
HD 191\,612           & 072.99$+$01.43\_01 & 20:09:28.608 & $+$35:44:01.31 & SB2        & O6-8   & \ldots & f?p var    & \ldots                    & S11a   & O6.5-8 & \ldots & f?p        & B0-2                      & H07    \\
HDE 228\,854          & 074.54$+$00.20\_01 & 20:18:47.219 & $+$36:20:26.08 & SB2E       & O6     & IV     & n var      & O5 Vn var                 & M16    & O6.5   & \ldots & \ldots     & O7.5                      & P52    \\
HDE 228\,766          & 075.19$+$00.96\_01 & 20:17:29.703 & $+$37:18:31.13 & SB2        & O4     & I      & f          & O8: II:                   & S11a   & O4     & I      & f          & O8-9 In                   & W73    \\
HD 193\,443 AB        & 076.15$+$01.28\_01 & 20:18:51.707 & $+$38:16:46.50 & SB2+Ca     & O9     & III    & \ldots     & \ldots                    & S11a   & O9     & III/I  & \ldots     & O9.5 V/III                & M13a   \\
BD $+$36 4063         & 076.17$-$00.34\_01 & 20:25:40.608 & $+$37:22:27.07 & SB1        & ON9.7  & Ib     & \ldots     & \ldots                    & S11a   & \ldots & \ldots & \ldots     & \ldots                    & \ldots \\
HD 193\,611           & 076.28$+$01.19\_01 & 20:19:38.748 & $+$38:20:09.18 & SB2E       & O9.5   & II     & \ldots     & O9.5 III                  & new    & O9.5   & III    & \ldots     & O9.7 III                  & new    \\
HDE 228\,989          & 076.66$+$01.28\_01 & 20:20:21.394 & $+$38:41:59.71 & SB2E       & O9.5   & V      & \ldots     & B0.2 V                    & new    & O9.5   & V      & \ldots     & B0 V                      & new    \\
HDE 229\,234          & 076.92$+$00.59\_01 & 20:24:01.298 & $+$38:30:49.55 & SB1        & O9     & III    & \ldots     & \ldots                    & new    & \ldots & \ldots & \ldots     & \ldots                    & \ldots \\
HD 192\,281           & 077.12$+$03.40\_01 & 20:12:33.121 & $+$40:16:05.45 & SB1        & O4.5   & IV     & (n)(f)     & \ldots                    & M16    & \ldots & \ldots & \ldots     & \ldots                    & \ldots \\
Y Cyg                 & 077.25$-$06.23\_01 & 20:52:03.577 & $+$34:39:27.51 & SB2E       & O9.5   & IV     & \ldots     & O9.5 IV                   & S11a   & O9.5   & IV     & \ldots     & O9.5 IV                   & new    \\
HDE 229\,232 AB       & 077.40$+$00.93\_01 & 20:23:59.183 & $+$39:06:15.27 & SB1+Ca?    & O4     & V:     & n((f))     & \ldots                    & M16    & \ldots & \ldots & \ldots     & \ldots                    & \ldots \\
HD 193\,322 B         & 078.10$+$02.78\_02 & 20:18:06.770 & $+$40:43:54.35 & \ldots     & B1.5   & V      & (n)p       & \ldots                    & vis    & \ldots & \ldots & \ldots     & \ldots                    & \ldots \\
HD 193\,322 AaAb      & 078.10$+$02.78\_01 & 20:18:06.990 & $+$40:43:55.46 & SB2a+Sa    & O9     & IV     & (n)        & \ldots                    & S11a   & O9     & V      & nn         & O8.5 III + B2.5: V:       & T11    \\
HD 194\,649 AB        & 078.46$+$01.35\_01 & 20:25:22.124 & $+$40:13:01.07 & SB2+Ca     & O6     & V      & ((f))      & O9.7: V                   & mod    & O6     & IV     & ((f))      & O9.5 V                    & new    \\
Cyg OB2-B17           & 079.84$+$01.16\_01 & 20:30:27.302 & $+$41:13:25.31 & SB2E       & O6     & Ia     & f          & O9: Ia:                   & M16    & O7     & Ia     & f          & O9 Iaf                    & S10    \\
Cyg OB2-3 B           & 079.97$+$00.98\_02 & 20:31:37.326 & $+$41:13:18.04 & \ldots     & B0     & IV     & \ldots     & \ldots                    & vis    & \ldots & \ldots & \ldots     & \ldots                    & \ldots \\
Cyg OB2-3 A           & 079.97$+$00.98\_01 & 20:31:37.509 & $+$41:13:21.01 & SB2E       & O8.5   & Ib     & (f)        & O6 III:                   & new    & O9     & III    & \ldots     & O6 IV:                    & K08    \\
ALS 15\,133           & 080.04$+$01.11\_01 & 20:31:18.330 & $+$41:21:21.66 & SB1        & O9.5   & IV     & \ldots     & \ldots                    & new    & \ldots & \ldots & \ldots     & \ldots                    & \ldots \\
Cyg OB2-A11           & 080.08$+$00.85\_01 & 20:32:31.543 & $+$41:14:08.21 & SB1        & O7     & Ib     & (f)        & \ldots                    & M16    & \ldots & \ldots & \ldots     & \ldots                    & \ldots \\
Cyg OB2-5 B           & 080.12$+$00.91\_02 & 20:32:22.489 & $+$41:18:19.45 & \ldots     & O7     & Ib     & (f)p var?  & \ldots                    & vis    & \ldots & \ldots & \ldots     & \ldots                    & \ldots \\
Cyg OB2-5 A           & 080.12$+$00.91\_01 & 20:32:22.422 & $+$41:18:18.91 & SB2E+Cas   & O6.5:  & Ia     & fe         & O7 Iafe                   & mod    & O5.5-6 & \ldots & \ldots     & O6.5                      & R99    \\
Cyg OB2-22 C          & 080.14$+$00.74\_01 & 20:33:09.598 & $+$41:13:00.54 & SB1E?      & O9.5   & III    & n          & \ldots                    & S11a   & \ldots & \ldots & \ldots     & \ldots                    & \ldots \\
Cyg OB2-22 B          & 080.14$+$00.75\_02 & 20:33:08.835 & $+$41:13:17.36 & SB1        & O6     & V      & ((f))      & \ldots                    & S11a   & \ldots & \ldots & \ldots     & \ldots                    & \ldots \\
Cyg OB2-41            & 080.15$+$00.79\_01 & 20:32:59.643 & $+$41:15:14.67 & SB1        & O9.7   & III    & (n)        & \ldots                    & M16    & \ldots & \ldots & \ldots     & \ldots                    & \ldots \\
ALS 15\,148           & 080.15$+$00.74\_01 & 20:33:13.265 & $+$41:13:28.74 & SB1        & O6.5:  & V      & \ldots     & \ldots                    & new    & \ldots & \ldots & \ldots     & \ldots                    & \ldots \\
Cyg OB2-9             & 080.17$+$00.76\_01 & 20:33:10.733 & $+$41:15:08.21 & SB2        & O4.5   & I      & f          & \ldots                    & S14    & O4     & I      & f          & O5.5 III(f)               & new    \\
Cyg OB2-1             & 080.17$+$01.23\_01 & 20:31:10.543 & $+$41:31:53.47 & SB1+Ca     & O8     & IV     & (n)((f))   & \ldots                    & new    & \ldots & \ldots & \ldots     & \ldots                    & \ldots \\
ALS 15\,131           & 080.19$+$00.81\_01 & 20:33:02.922 & $+$41:17:43.13 & SB1        & O7.5   & V      & ((f))      & \ldots                    & new    & \ldots & \ldots & \ldots     & \ldots                    & \ldots \\
Cyg OB2-20            & 080.19$+$01.10\_01 & 20:31:49.665 & $+$41:28:26.51 & SB1        & O9.7   & IV     & \ldots     & \ldots                    & new    & \ldots & \ldots & \ldots     & \ldots                    & \ldots \\
Cyg OB2-8 A           & 080.22$+$00.79\_01 & 20:33:15.078 & $+$41:18:50.51 & SB2        & O6     & Ib     & (fc)       & O4.5: III:(fc)            & S14    & O6     & Ib     & (fc)       & O4.5 III(fc)              & new    \\
Cyg OB2-70            & 080.23$+$00.71\_01 & 20:33:37.001 & $+$41:16:11.30 & SB1        & O9.5   & IV     & (n)        & \ldots                    & M16    & \ldots & \ldots & \ldots     & \ldots                    & \ldots \\
Cyg OB2-15            & 080.24$+$00.98\_01 & 20:32:27.666 & $+$41:26:22.08 & SB1        & O8     & III    & \ldots     & \ldots                    & new    & \ldots & \ldots & \ldots     & \ldots                    & \ldots \\
ALS 15\,115           & 080.27$+$00.81\_01 & 20:33:18.046 & $+$41:21:36.90 & SB1        & O8     & V      & \ldots     & \ldots                    & M16    & \ldots & \ldots & \ldots     & \ldots                    & \ldots \\
Cyg OB2-27 AB         & 080.29$+$00.66\_01 & 20:33:59.528 & $+$41:17:35.48 & SB2E+Ca    & O9.7   & V      & (n)        & O9.7 V:(n)                & M16    & O9.5   & V      & \ldots     & B0 V                      & K12    \\
\hline
\end{tabular}
}
\end{table}
\end{landscape}

\addtocounter{table}{-1}
\begin{landscape}
\begin{table}
\caption{(Continued).}
\centerline{
\scriptsize
\begin{tabular}{lccclllllllllll}
\hline
\hline
Name                  & GOSC ID            & R.A.         & Declination    & SBS        & \multicolumn{5}{c}{GOSSS classification}                          & \multicolumn{5}{c}{Alternate classification}                      \\
                      &                    & (J2000)      & (J2000)        &            & ST     & LC    & Qualifier   & Secondary (+tertiary)     & Ref.   & ST     & LC    & Qualifier   & Secondary (+tertiary)     & Ref.   \\
\hline
Cyg OB2-73            & 080.32$+$00.60\_01 & 20:34:21.930 & $+$41:17:01.60 & SB2        & O8     & V      & z          & O8 Vz                     & M16    & O8     & III    & \ldots     & O8 III                    & K09    \\
ALS 15\,114           & 080.54$+$00.73\_01 & 20:34:29.601 & $+$41:31:45.42 & SB2        & O7     & V      & ((f))      & O7 IV((f))                & mod    & O7     & V      & \ldots     & O9 V                      & K12    \\
Cyg OB2-29            & 080.55$+$00.80\_01 & 20:34:13.505 & $+$41:35:03.01 & SB1        & O7.5   & V      & (n)((f))z  & \ldots                    & M16    & \ldots & \ldots & \ldots     & \ldots                    & \ldots \\
Cyg OB2-11            & 080.57$+$00.83\_01 & 20:34:08.513 & $+$41:36:59.42 & SB1        & O5.5   & I      & fc         & \ldots                    & S11a   & \ldots & \ldots & \ldots     & \ldots                    & \ldots \\
LS III $+$46 11       & 084.88$+$03.81\_01 & 20:35:12.642 & $+$46:51:12.12 & SB2        & O3.5   & I      & f*         & O3.5 If*                  & M16    & \ldots & \ldots & \ldots     & \ldots                    & \ldots \\
HD 199\,579           & 085.70$-$00.30\_01 & 20:56:34.779 & $+$44:55:29.01 & SB2        & O6.5   & V      & ((f))z     & \ldots                    & S11a   & O6     & V      & ((f))      & B1-2 V                    & W01    \\
68 Cyg                & 087.61$-$03.84\_01 & 21:18:27.187 & $+$43:56:45.40 & SB1?       & O7.5   & III    & n((f))     & \ldots                    & S11a   & \ldots & \ldots & \ldots     & \ldots                    & \ldots \\
HD 206\,267 AaAb      & 099.29$+$03.74\_01 & 21:38:57.618 & $+$57:29:20.55 & SB2+Ca     & O6     & V      & (n)((f))   & B0: V                     & mod    & O6     & V      & (n)((f))   & B0: V                     & new    \\
14 Cep                & 102.01$+$02.18\_01 & 22:02:04.576 & $+$58:00:01.33 & SB2        & O8.5   & III    & \ldots     & BN0 V                     & mod    & O9     & III    & \ldots     & BN0 V                     & new    \\
ALS 12\,502           & 105.77$+$00.06\_01 & 22:34:45.972 & $+$58:18:04.64 & SB2E+Ca    & O9     & III:   & (n)        & O9.2 IV:(n)               & new    & O9     & IV     & (n)        & O9 V(n)                   & new    \\
DH Cep                & 107.07$-$00.90\_01 & 22:46:54.111 & $+$58:05:03.55 & SB2        & O5.5   & V      & (f)) var   & O6 V((f)) var             & mod    & O5.5   & V      & ((f)) var  & O6 V((f)) var             & new    \\
ALS 12\,688           & 107.42$-$02.87\_01 & 22:55:44.944 & $+$56:28:36.70 & SB2E       & O5.5   & V      & (n)((fc))  & B                         & M16    & \ldots & \ldots & \ldots     & \ldots                    & \ldots \\
AO Cas                & 117.59$-$11.09\_01 & 00:17:43.059 & $+$51:25:59.12 & SB2E       & O9.2   & II     & \ldots     & O8 V((f))                 & S14    & O9.2   & II     & \ldots     & O7.5 V                    & new    \\
HD 108                & 117.93$+$01.25\_01 & 00:06:03.386 & $+$63:40:46.75 & SB1?       & O6.5-8 & \ldots & f?p var    & \ldots                    & mod    & \ldots & \ldots & \ldots     & \ldots                    & \ldots \\
V747 Cep              & 118.20$+$05.09\_01 & 00:01:46.870 & $+$67:30:25.13 & SB1E       & O5.5   & V      & (n)((f))   & \ldots                    & M16    & \ldots & \ldots & \ldots     & \ldots                    & \ldots \\
HD 12\,323            & 132.91$-$05.87\_01 & 02:02:30.126 & $+$55:37:26.38 & SB1        & ON9.2  & V      & \ldots     & \ldots                    & S14    & \ldots & \ldots & \ldots     & \ldots                    & \ldots \\
DN Cas                & 133.88$-$00.08\_01 & 02:23:11.536 & $+$60:49:50.18 & SB2E+Ca    & O8.5   & V      & \ldots     & B0.2 V                    & new    & O8.5   & V      & \ldots     & B0.2 V                    & new    \\
BD $+$60 497          & 134.58$+$01.04\_01 & 02:31:57.087 & $+$61:36:43.95 & SB2        & O6.5   & V      & (n)((f))z  & O9.5: V                   & mod    & O7     & V      & (n)((f))z  & B0: V(n)                  & new    \\
HD 15\,558 A          & 134.72$+$00.92\_01 & 02:32:42.536 & $+$61:27:21.56 & SB2        & O4.5   & III    & (fc)p      & O8:                       & mod    & O5.5   & III    & (f)        & O7 V                      & D06    \\
HD 15\,558 B          & 134.73$+$00.93\_01 & 02:32:43.895 & $+$61:27:20.33 & \ldots     & B0     & V      & \ldots     & \ldots                    & vis    & \ldots & \ldots & \ldots     & \ldots                    & \ldots \\
HD 16\,429 C          & 135.67$+$01.16\_01 & 02:40:46.149 & $+$61:17:48.54 & \ldots     & B0.7   & V      & (n)        & \ldots                    & vis    & \ldots & \ldots & \ldots     & \ldots                    & \ldots \\
HD 16\,429 A          & 135.68$+$01.15\_01 & 02:40:44.951 & $+$61:16:56.04 & SB1+Ca     & O9     & II-III & (n) Nwk    & \ldots                    & S14    & O9.5   & II     & \ldots     & O8 III-V + B0 V?          & M03    \\
HD 16\,429 B          & 135.68$+$01.14\_01 & 02:40:44.805 & $+$61:16:49.43 & \ldots     & F      & \ldots & \ldots     & \ldots                    & vis    & \ldots & \ldots & \ldots     & \ldots                    & \ldots \\
HD 17\,505 A          & 137.19$+$00.90\_01 & 02:51:07.971 & $+$60:25:03.88 & SB2+C      & O6.5   & IV     & ((f))      & O7 V((f))                 & mod    & O6.5   & III    & ((f))      & O7.5 V((f)) + O7.5 V((f)) & H06    \\
HD 17\,505 B          & 137.19$+$00.90\_02 & 02:51:08.262 & $+$60:25:03.78 & \ldots     & O8     & V      & \ldots     & \ldots                    & vis    & \ldots & \ldots & \ldots     & \ldots                    & \ldots \\
HD 15\,137            & 137.46$-$07.58\_01 & 02:27:59.811 & $+$52:32:57.60 & SB1        & O9.5   & II-III & n          & \ldots                    & S11a   & \ldots & \ldots & \ldots     & \ldots                    & \ldots \\
CC Cas                & 140.12$+$01.54\_01 & 03:14:05.333 & $+$59:33:48.50 & SB2E       & O8.5   & III    & (n)((f))   & B                         & mod    & O8.5   & III    & \ldots     & B0 V                      & H94    \\
HD 14\,633 AaAb       & 140.78$-$18.20\_01 & 02:22:54.293 & $+$41:28:47.72 & SB1+Ca?    & ON8.5  & V      & \ldots     & \ldots                    & S11a   & \ldots & \ldots & \ldots     & \ldots                    & \ldots \\
$\alpha$ Cam          & 144.07$+$14.04\_01 & 04:54:03.011 & $+$66:20:33.58 & SB1?       & O9     & Ia     & \ldots     & \ldots                    & S11a   & \ldots & \ldots & \ldots     & \ldots                    & \ldots \\
MY Cam A              & 146.27$+$03.14\_01 & 03:59:18.290 & $+$57:14:13.72 & SB2E+Cas   & O5.5   & V      & nz var     & O6.5 Vnz var              & mod    & O6     & V      & (n) var    & O6.5 V(n)z var            & new    \\
MY Cam B              & 146.27$+$03.14\_02 & 03:59:18.345 & $+$57:14:13.15 & \ldots     & B1.5:  & V      & \ldots     & \ldots                    & vis    & \ldots & \ldots & \ldots     & \ldots                    & \ldots \\
LY Aur B              & 172.76$+$00.61\_02 & 05:29:42.600 & $+$35:22:29.90 & \ldots     & B0.2   & IV     & \ldots     & \ldots                    & vis    & \ldots & \ldots & \ldots     & \ldots                    & \ldots \\
LY Aur A              & 172.76$+$00.61\_01 & 05:29:42.647 & $+$35:22:30.07 & SB2E+SB1as & O9.2   & II     & (n)        & O9.7 II                   & mod    & O9     & II     & \ldots     & O9 III                    & M13b   \\
IU Aur AB             & 173.05$-$00.03\_01 & 05:27:52.398 & $+$34:46:58.23 & SB2E+Ca    & O9.5   & IV     & (n)        & O9.7 IV(n)                & new    & O8.5   & III    & (n)        & O9.7 V(n)                 & new    \\
HD 37\,737            & 173.46$+$03.24\_01 & 05:42:31.160 & $+$36:12:00.51 & SB1        & O9.5   & II-III & (n)        & \ldots                    & S11a   & \ldots & \ldots & \ldots     & \ldots                    & \ldots \\
HD 37\,366            & 177.63$-$00.11\_01 & 05:39:24.799 & $+$30:53:26.75 & SB2        & O9.5   & V      & \ldots     & B1: V                     & mod    & O9.5   & V      & \ldots     & B1: V                     & new    \\
15 Mon AaAb           & 202.94$+$02.20\_01 & 06:40:58.656 & $+$09:53:44.71 & SB1a       & O7     & V      & ((f))z var & \ldots                    & M18    & \ldots & \ldots & \ldots     & \ldots                    & \ldots \\
$\delta$ Ori Aa       & 203.86$-$17.74\_01 & 05:32:00.398 & $-$00:17:56.69 & SB2E+Cas   & O9.5   & II     & Nwk        & \ldots                    & M18    & O9.5   & II     & \ldots     & B0.5 III                  & H02    \\
HD 47\,129            & 205.87$-$00.31\_01 & 06:37:24.042 & $+$06:08:07.38 & SB2        & O8     & Iab    & f          & O8.5:fp                   & mod    & O8     & III/I  & \ldots     & O7.5 V/III                & L08    \\
HD 48\,099            & 206.21$+$00.80\_01 & 06:41:59.231 & $+$06:20:43.54 & SB2        & O5.5   & V      & ((f))z     & O9 V                      & M16    & O5.5   & V      & ((f))z     & O9: V                     & new    \\
HD 46\,149            & 206.22$-$02.04\_01 & 06:31:52.533 & $+$05:01:59.19 & SB2        & O8.5   & V      & \ldots     & \ldots                    & S11a   & O8     & V      & \ldots     & B0-1 V                    & M09    \\
$\zeta$ Ori AaAb      & 206.45$-$16.59\_01 & 05:40:45.527 & $-$01:56:33.26 & SB2        & O9.2   & Ib     & Nwk var    & \ldots                    & M18    & O9.5   & Ib     & \ldots     & B0.5 IV                   & H13    \\
$\sigma$ Ori AaAb     & 206.82$-$17.34\_01 & 05:38:44.765 & $-$02:36:00.25 & SB2a+Cas   & O9.5   & V      & \ldots     & B0.2 V                    & mod    & O9.5   & V      & \ldots     & B0.5 V                    & S11b   \\
$\sigma$ Ori B        & 206.82$-$17.34\_02 & 05:38:44.782 & $-$02:36:00.27 & \ldots     & B0.2   & V      & (n)        & \ldots                    & vis    & \ldots & \ldots & \ldots     & \ldots                    & \ldots \\
HD 46\,573            & 208.73$-$02.63\_01 & 06:34:23.568 & $+$02:32:02.94 & SB1        & O7     & V      & ((f))z     & \ldots                    & S11a   & \ldots & \ldots & \ldots     & \ldots                    & \ldots \\
$\theta^{1}$ Ori CaCb & 209.01$-$19.38\_01 & 05:35:16.463 & $-$05:23:23.18 & SB1?+Sa    & O7     & \ldots & f?p var    & \ldots                    & mod    & \ldots & \ldots & \ldots     & \ldots                    & \ldots \\
$\theta^{2}$ Ori A    & 209.05$-$19.37\_01 & 05:35:22.900 & $-$05:24:57.80 & SB2        & O9.2   & V      & \ldots     & B0.5: V(n)                & mod    & O9.5   & V      & \ldots     & B0.2: V                   & new    \\
$\iota$ Ori           & 209.52$-$19.58\_01 & 05:35:25.981 & $-$05:54:35.64 & SB2        & O8.5   & III    & \ldots     & B0.2: V                   & mod    & O8.5   & III    & \ldots     & B0.2 V                    & new    \\
HD 52\,533 A          & 216.85$+$00.80\_01 & 07:01:27.048 & $-$03:07:03.28 & SB1        & O8.5   & IV     & n          & \ldots                    & S11a   & \ldots & \ldots & \ldots     & \ldots                    & \ldots \\
HD 54\,662 AB         & 224.17$-$00.78\_01 & 07:09:20.249 & $-$10:20:47.64 & SB2a       & O7     & V      & z var?     & \ldots                    & S14    & O6.5   & V      & \ldots     & O7-9.5 V                  & B07    \\
HD 53\,975            & 225.68$-$02.32\_01 & 07:06:35.964 & $-$12:23:38.23 & SB2        & O7.5   & V      & z          & \ldots                    & S11a   & O7.5   & V      & \ldots     & B2-3 V                    & G94    \\
\hline
\end{tabular}
}
\begin{flushleft}References: B07:~\citet{Boyaetal07a}, C71:~\citet{ContAlsc71}, D06:~\citet{DeBeetal06a}, G94:~\citet{Giesetal94}, H02:~\citet{Harvetal02}, H06:~\citet{Hilletal06}, H07:~\citet{Howaetal07}, H13:~\citet{Hummetal13a}, H94:~\citet{Hilletal94}, K08:~\citet{Kimietal08}, K09:~\citet{Kimietal09}, K12:~\citet{Kobuetal12}, L08:~\citet{Lindetal08}, M03:~\citet{McSw03}, M09:~\citet{Mahyetal09}, M13a:~\citet{Mahyetal13}, M13b:~\citet{Mayeetal13b}, M16:~\citet{Maizetal16}, M18:~\citet{Maizetal18a}, P52:~\citet{Pear52}, R99:~\citet{Rauwetal99}, S09:~\citet{Sanaetal09}, S10:~\citet{Stroetal10}, S11a:~\citet{Sotaetal11a}, S11b:~\citet{SimDetal11b}, S14:~\citet{Sotaetal14}, T11:~\citet{tenBetal11}, W01:~\citet{Willetal01}, W73:~\citet{Walb73a}.\end{flushleft}
\end{table}
\end{landscape}

\end{document}